\documentclass{aa}  

\usepackage{graphicx}
\usepackage{txfonts}
\begin{document}

   \title{The fraction of bolometric luminosity absorbed by dust\\ in DustPedia\thanks{DustPedia is a collaborative focused research project supported by the European Union under the Seventh Framework Programme (2007- 2013) call (proposal no. 606824, P.I.\ J. I. Davies, {\tt http://www.dustpedia.com}). The data used in this work is publicly available at {\tt http://dustpedia.astro.noa.gr}.} 
  galaxies}

   \author{
S.~Bianchi\inst{1}
\and
P.~De~Vis\inst{2}
\and
S.~Viaene\inst{3,4}
\and
A.~Nersesian\inst{3,5,6}
\and
A.~V.~Mosenkov\inst{7,8}
\and
E.~M.~Xilouris\inst{5}
\and
M.~Baes\inst{3}
\and 
V.~Casasola\inst{1}
\and
L.~P. Cassar\`a\inst{5,9}
\and
C.~J.~R.~Clark\inst{10}
\and 
J.~I.~Davies\inst{10}
\and
I.~De~Looze\inst{3,11}
\and
W. Dobbels\inst{3}
\and
M.~Galametz\inst{12}
\and
F.~Galliano\inst{12} 
\and
A.~P.~Jones\inst{2}
\and
S.~Lianou\inst{12} 
\and
S.~C.~Madden\inst{12} 
\and
A. Tr\v{c}ka\inst{3}
}

   \institute{
INAF - Osservatorio Astrofisico di Arcetri, Largo E. Fermi 5,I-50125, Florence, Italy\\
              \email{sbianchi@arcetri.astro.it}
\and
Institut d'Astrophysique  Spatiale, UMR 8617, CNRS, Univerit\'e Paris Sud, Univerit\'e Paris-Saclay, Univerit\'e Paris Sud, Orsay, F-91405, France
\and
Sterrenkundig Observatorium, Universiteit Gent, Krijgslaan 281 S9, B-9000 Gent, Belgium
\and
Centre for Astrophysics Research, University of Hertfordshire, College Lane, Hatfield, AL10 9AB, UK
\and
National Observatory of Athens, Institute for Astronomy, Astrophysics, Space Applications and Remote Sensing, 
Ioannou Metaxa and Vasileos Pavlou GR-15236, Athens, Greece
\and
Department of Astrophysics, Astronomy \& Mechanics, Faculty of Physics, University of Athens, 
Panepistimiopolis, GR15784 Zografos, Athens, Greece
\and
Central Astronomical Observatory of RAS, Pulkovskoye Chaussee 65/1, 196140, St. Petersburg, Russia
\and 
St. Petersburg State University, Universitetskij Pr. 28, 198504, St. Petersburg, Stary Peterhof, Russia
\and
INAF - Istituto di Astrofisica Spaziale e Fisica Cosmica Milano, via Alfonso Corti 12, 20133, Milan, Italy
\and
School of Physics and Astronomy, Cardiff University, The Parade, Cardiff CF24 3AA, UK 
\and
Department of Physics and Astronomy, University College London, Gower Street, London WC1E 6BT, UK
\and
Laboratoire AIM, CEA/DSM - CNRS - Universit\'e Paris Diderot, IRFU/Service d'Astrophysique, CEA Saclay, 91191, Gif-sur- Yvette, France
}

   \date{}

 
  \abstract
   {}
   {We study the fraction of stellar radiation absorbed by dust, $f_\mathrm{abs}$, in 814 galaxies of different morphological types. The targets constitute the vast majority (93\%) of the DustPedia sample, including almost all large (optical diameter larger than 1\arcmin), nearby ($v\le$ 3000 km s$^{-1}$) galaxies observed with the {\em Herschel} Space Observatory.}
   {For each object, we model the spectral energy distribution from the ultraviolet to the sub-millimetre using the dedicated, aperture-matched DustPedia photometry  and the fitting code CIGALE. The value of $f_\mathrm{abs}$ is obtained from the total luminosity emitted by dust and from the bolometric luminosity, which are estimated by the fit.}
   {On average, 19\% of the stellar radiation is absorbed by dust in DustPedia galaxies. The fraction rises to 25\% if only late-type galaxies are considered. The dependence of $f_\mathrm{abs}$ on morphology, showing a peak for Sb-Sc galaxies, is weak; it reflects a stronger, yet broad, positive correlation with the bolometric luminosity, which is identified for late-type, disk-dominated, high-specific-star-formation rate, gas-rich objects. We find no variation of $f_\mathrm{abs}$ with inclination, at odds with radiative transfer models of edge-on galaxies. These results call for a self-consistent modelling of the evolution of the dust mass and geometry along the build-up of the stellar content. We also provide template spectral energy distributions in bins of morphology and luminosity and study the variation of $f_\mathrm{abs}$ with stellar mass and specific star formation rate. We confirm that the local Universe is missing the high $f_\mathrm{abs}$, luminous and actively star-forming objects necessary to explain the energy budget in observations of the extragalactic background light.}
   {}

   \keywords{dust, extinction -- infrared: galaxies -- galaxies: photometry -- galaxies: ISM -- galaxies: evolution}

   \maketitle
%

\section{Introduction}

A common refrain in the literature states that dust, despite representing only a tiny fraction of the total baryonic content of galaxies, has a profound effect on their physics and appearance.
Undoubtedly, dust has a major role in shaping their Spectral Energy Distribution (SED), by absorbing starlight in the ultraviolet (UV), optical and Near Infra-Red (NIR) ranges and re-emitting it at 
Mid and Far Infra-Red (MIR and FIR, respectively) and submillimetre (submm) wavelengths (for a review, see \citealt{GallianoARA&A2018}). Observations from the Infrared Astronomical Satellite (IRAS; \citealt{NeugebauerApJL1984}) provided 
the first means to attempt an estimate of the amount of radiation reprocessed by dust  in local galaxies.
\citet{SoiferAJ1991} used the total dust luminosity density from the 12 to 100 $\mu$m luminosity functions of the IRAS Bright Galaxy Sample and an 
estimate of the observed starlight luminosity density to assess that the infrared luminosity is about 30\% of that from late-type stars. Thus, the ratio between the dust and bolometric 
- i.e. stars + dust - luminosities (a quantity we name $f_\mathrm{abs}$ in this work) would be $\approx 24\%$.
\citet{XuA&A1995} derived $f_\mathrm{abs}$ in a more direct way, from 
a (very sparse) coverage of the UV-optical-FIR SED for each object in a sample of 135 UV-selected Late-Type Galaxies (LTGs). The average value for their sample is  $\langle f_\mathrm{abs} \rangle= 31\pm1 \%$. A similar approach 
was used by \citet{PopescuMNRAS2002} for 28 LTGs in the Virgo cluster, with a better coverage of the UV-optical SED and FIR data extending to 170 $\mu$m, from the Infrared Space Observatory (ISO; \citealt{KesslerA&A1996}). They 
obtained $\langle f_\mathrm{abs} \rangle=24 \pm 2 \%$\footnote{This value is obtained from the full sample of  \citet{PopescuMNRAS2002} as displayed in their Fig. 2. However, in the abstract and conclusion the
authors prefer to quote $\langle f_\mathrm{abs} \rangle=30\%$, the mean for spirals of Hubble type later than Scd.}. Thus, dust was found to be responsible for 1/3 to 1/4 of the total bolometric luminosity of a galaxy, 
a result in agreement with estimates from the Interstellar Radiation Field in our own Galaxy \citep{CoxA&AR1989}. 

The knowledge of $f_\mathrm{abs}$  is relevant for radiative transfer (RT) studies. Encoded in the quantity is information on the mass of the grains, 
on their space distribution with respect to the photon sources, and, of course, on their absorption/emission properties. RT fits to the surface-brightness distribution of edge-on 
spiral galaxies \citep{XilourisSub1998,BianchiA&A2007,DeGeyterMNRAS2014} have proved unable in most cases to predict the correct levels of FIR/submm emission \citep{BianchiA&A2000b,BaesA&A2010b,DeLoozeMNRAS2012}.
These, as we will see, for the typical dust-lane edge-ons correspond to $f_\mathrm{abs}\approx30-40\%$. Instead, the emission predicted in the FIR is about a factor 2-4 smaller than 
observed \citep[see][for the latest, state-of-the-art, RT modelling]{MosenkovA&A2016,MosenkovA&A2018}. Different solutions to this {\em energy balance} problem have been proposed, 
involving more complex geometries than those inferred from fits to optical images \citep{PopescuA&A2000,BianchiA&A2008,SaftlyA&A2015} or an
enhanced dust emissivity in the FIR/submm with respect to the standard Milky Way (MW) properties \citep{AltonA&A2004,DasyraA&A2005}. Yet, some edge-ons do not show the same problem
 \citep{DeGeyterMNRAS2015,MosenkovA&A2018} and one might wonder if the issue is related to a dependence of $f_\mathrm{abs}$ on other galactic properties.
 
A characterization of $f_\mathrm{abs}$ in the local Universe also impinges upon studies of galaxy evolution. Provided that the galaxy sample is representative enough of the local Universe population, 
the { \em luminosity-weighted}  $\langle f_\mathrm{abs} \rangle$ should be analogous to the $f_\mathrm{abs}$ that describes the wavelength-dependent specific luminosity density, or cosmic SED (CSED), at redshift zero. 
The integration of CSEDs from all epochs until the current one results in the Extragalactic Background Light (EBL). Since the first detection of the EBL in the FIR/submm from the satellite COBE 
\citep{PugetA&A1996,HauserApJ1998}, it has become clear that a strong evolution of the FIR/submm luminosity (i.e. an increase of $f_\mathrm{abs}$ for CSEDs at $z>0$) is required
\citep{FranceschiniMNRAS1998,FranceschiniA&A2001}. In contrast with the SEDs of local galaxies, the EBL implies that about half of the UV-optical photons have been absorbed by dust over cosmic times
\citep[i.e. $f_\mathrm{abs}\approx 50\%$; for the latest EBL estimates, see][]{DriverApJ2016,FranceschiniA&A2017}.

The earlier determinations of $f_\mathrm{abs}$ relied heavily on correction factors to derive the full luminosity over a certain spectral range from a few datapoints. In particular, the full energy output under the peak of 
dust emission up to 1000 $\mu$m was derived from the wavelengths accessible to IRAS or ISO by extrapolating fits to the available data or interpolating up to ground-based submm/mm observations 
available for a few objects. These corrections could increase the observed dust luminosities from 40\% \citep{SoiferAJ1991,XuA&A1995} to 100\% or
above \citep{PopescuMNRAS2002}. The need for these corrections has been obviated by the advent of the {\em Herschel} Space Observatory \citep{PilbrattA&A2010}, and in particular by the submm coverage 
provided by the instrument SPIRE \citep{GriffinA&A2010}. 

Using {\em Herschel} and ancillary data for the KINGFISH sample \citep{KennicuttPASP2011},  \citet{SkibbaApJ2011} derived the ratio between dust and stellar emission by directly 
integrating well sampled SEDs. Their tabulated results convert into $\langle f_\mathrm{abs}\rangle=29\pm2 \%$ for LTGs (52 objects).
Similar integrations have been carried out by \citet{DaviesMNRAS2012} for the FIR-selected Bright Galaxy Sample of the {\em Herschel} Virgo Cluster Survey, though neglecting the 
contribution to the SED of UV and MIR radiation. For the 69 LTGs in \citet{DaviesMNRAS2012} it is $\langle f_\mathrm{abs}\rangle=31\pm2\%$. 
These results have been confirmed by \citet{ViaeneA&A2016}, so far the most complete study in terms of spectral coverage of the SED and number of objects from a complete sample. 
For 239 spiral galaxies from the {\em Herschel} Reference Survey \citep[HRS; ][]{BoselliPASP2010}, \citet{ViaeneA&A2016} find an average $\langle f_\mathrm{abs}\rangle=32\pm1\%$, after fitting the observed 
SED with the MAGPHYS fitting tool \citep{DaCunhaMNRAS2008}.

We perform here a similar analysis on the galaxies of the DustPedia project \citep{DaviesPASP2017}, including both LTGs and Early-Type Galaxies (ETGs).
DustPedia includes the vast majority of nearby ($v < 3000$ km s$^{-1}$), large (with an optical diameter of at least 1$\arcmin$) galaxies detected in their stellar component at 3.4 $\mu$m and with available data from the
{\em Herschel} Science Archive \citep{RiedingerESA2009}. Its main advantage over HRS and other {\em Herschel}-based samples lies in the dedicated, aperture matched, photometry that was carried out from UV up to submm wavelengths
\citep{ClarkA&A2018}. Furthermore, DustPedia has more than twice the number of objects available in the HRS, allowing us to explore the dependence of $f_\mathrm{abs}$ on a larger dynamic range of other 
observables and physical quantities. Also, the photometric coverage is slightly better: for the work described in the remainder of the paper, a {\em mean} of 21 photometric datapoints are available for each object, 
while in \citet{ViaeneA&A2016}  the {\em maximum} number was twenty. Finally, DustPedia builds on the final
characterisation of the observations from the {\em Herschel} Space Observatory, whose post-operation support ended in 2017; thus it constitutes the ultimate database on the local Universe before the 
next generation FIR/submm space missions.

The paper is organised as follows: in Sect.~\ref{sec:data} we briefly describe the selection of the sample, the method used in determining $f_\mathrm{abs}$ and the other data used in the analysis. 
The variation of $f_\mathrm{abs}$ with morphological type is presented in Sect.~\ref{sec:results}, while Sect.~\ref{sect:compa} is devoted to a comparison with previous results. In  Sect.~\ref{sec:lumi} we
illustrate the dependence of $f_\mathrm{abs}$ on the bolometric luminosity, in particular for LTGs. In Sect.~\ref{sec:checks} we check that the results are not biased by the inclusion of edge-ons or AGN in the sample. 
An additional bonus of the analysis, the definition of SED templates, is shown in Sect.~\ref{sec:template}. The  $f_\mathrm{abs}$, SEDs and other properties of DustPedia galaxies are compared to those at
higher redshift  in Sect.~\ref{sec:evol}.
Finally, we summarise our findings in Sect.~\ref{sec:summary}.

\section{Data and analysis}
\label{sec:data}

The fraction of energy absorbed by dust in a galaxy can be defined as the ratio between the luminosity emitted by dust and the bolometric (attenuated starlight+dust, equivalent to unattenuated starlight) luminosity:
\[
f_\mathrm{abs} = \frac{L_\mathrm{dust}}{L_\mathrm{stars}+L_\mathrm{dust}} =  \frac{L_\mathrm{dust}}{L_\mathrm{bolo}} = \frac{\int L^\mathrm{dust}_\lambda d\lambda}{\int L_\lambda d\lambda},
\]
where the luminosity density $L_\lambda$ can be derived directly from the observed flux density $F_\lambda$ (the SED) under the common assumption $L_\lambda=4\pi D^2 F_\lambda$ ($D$ is the galaxy distance, whose knowledge is however irrelevant for $f_\mathrm{abs}$). The integration is carried over for  wavelengths $\lambda  > $ 0.0912  $\mu\mathrm{m}$, the ionising radiation at shorter $\lambda$ being absorbed preferentially by the atomic gas. We have modelled the SED of most DustPedia galaxies using the CIGALE (Code Investigating GALaxy Evolution; \citealt{NollA&A2009}) software package\footnote{Version 0.12.1, available at {\tt http://cigale.lam.fr}.}. We provide here a few details of the sample selection and modelling, while a full description is presented in a companion paper \citep{NersesianA&A2018}.

\subsection{Sample selection and photometry}
\label{sect:sample}

The DustPedia photometric database consists of aperture-matched flux estimates from the UV to the submm for a sample of 875 galaxies. In addition, IRAS and {\em Planck}  photometry is provided in a supplementary dataset. We used all the bands where most of the contribution is either from stars or dust, i.e.\ from FUV up to the {\em Planck} 850~$\mu$m band. We excluded the photometric entries that have significant contamination from a nearby galactic or extragalactic source, and those with serious artefacts in the imagery or insufficient sky coverage for a proper estimate of the target/sky levels. From the supplementary IRAS/{\em Planck} database, we excluded the flux density that might miss a significant contribution from extended emission \citep[for details on the database and flagging codes, see][]{ClarkA&A2018}. Finally, we pruned all objects that, after the flux selection, resulted in no estimate at all around the peak of dust emission for $60 \le \lambda / \mu\mathrm{m} \le 500$, and those with an insufficient coverage of the stellar SED, i.e.\ with no remaining fluxes for $0.35 \le \lambda / \mu\mathrm{m} \le 3.6$. 

{ In total, 51 objects were excluded because of a global contamination flag, in most cases due to a nearby galaxy. The exclusion might in principle bias our results against objects undergoing starbursting episodes due to strong interactions. 
Indeed, these objects lie on the high luminosity tail of the distribution, with an average  luminosity at 3.4 $\mu$m almost an order of magnitude higher than for the rest of the sample. 
Using the average results for higher-luminosity galaxies, however, we tested that the exclusion has not likely altered significantly the main conclusions of the work.
Another 10 objects were excluded because of insufficient SED coverage, after the photometry flagging. Among them is, for example,  M~104 {\em Sombrero} (NGC~4594) which has all its FIR/submm images flagged because of insufficient map coverage for a proper determination of the sky level.
}
The final sample is thus reduced to a total of 814 objects. Still,  94\% of these galaxies have more than 15 available photometric datapoints.

\begin{figure*}
\center
\includegraphics[scale=0.45,clip=true,trim=0 59 0 0]{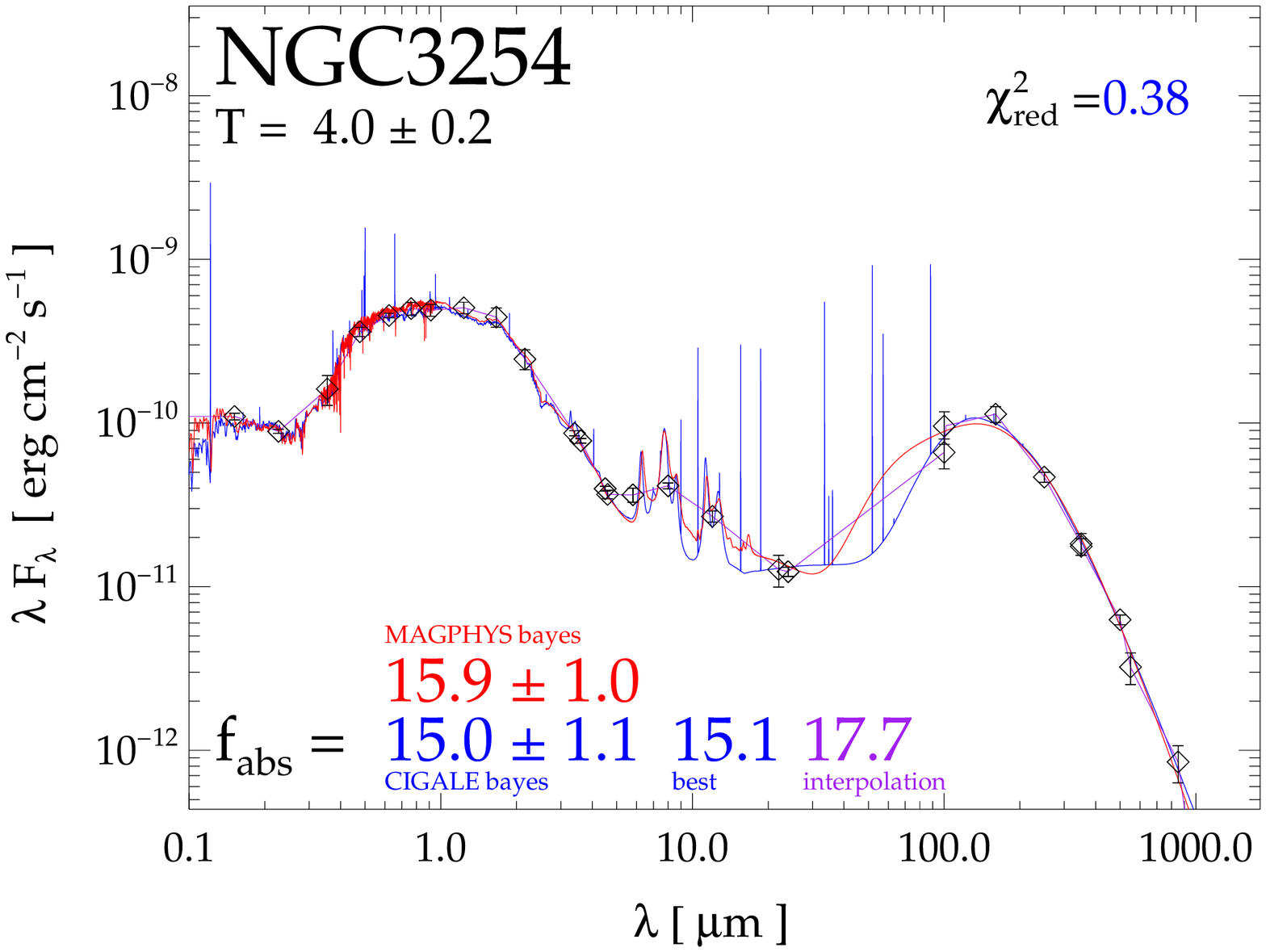}\includegraphics[scale=0.45,clip=true,trim=40 59 0 0]{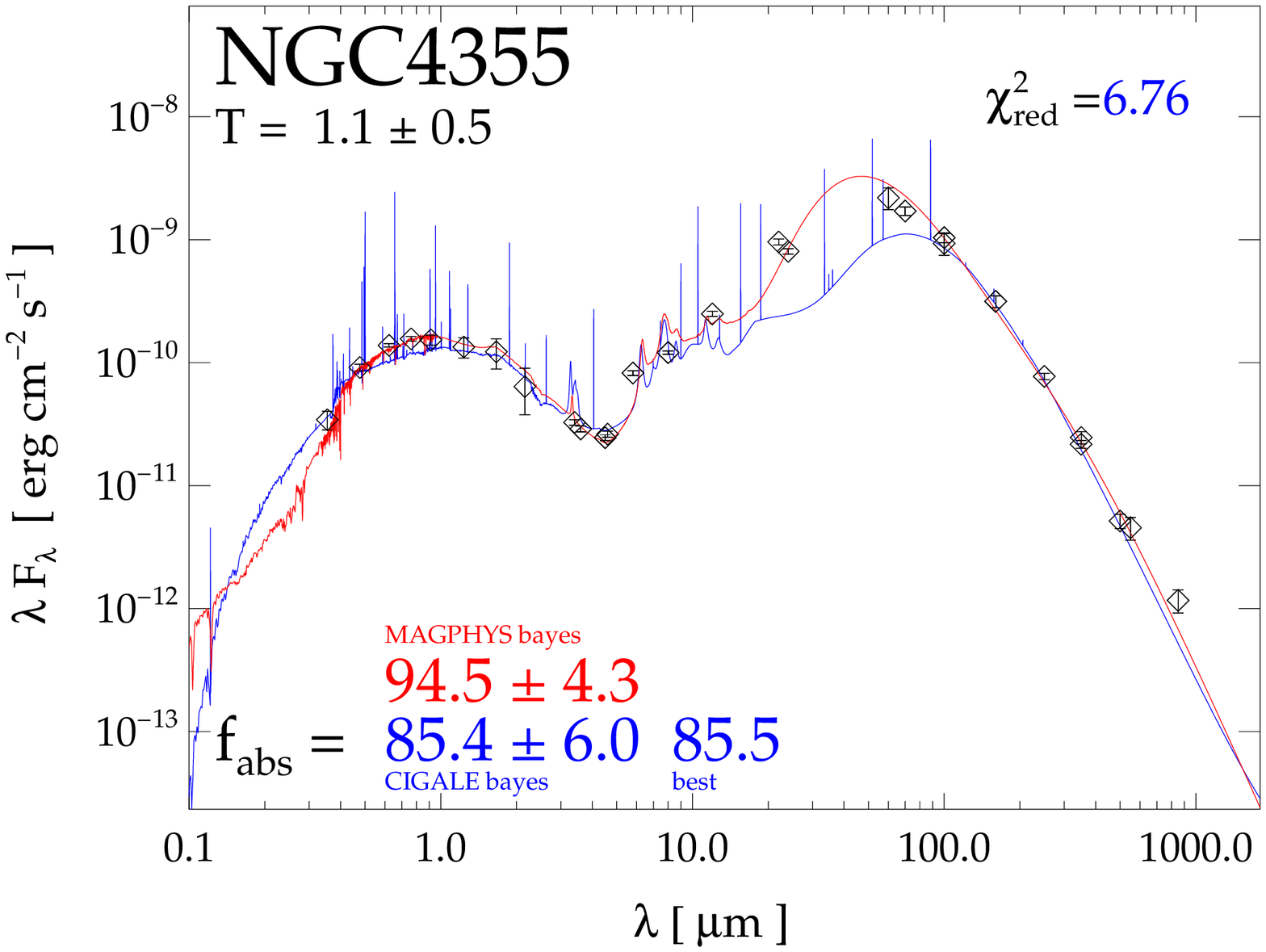}
\includegraphics[scale=0.45,clip=true,trim=0 0 0 0]{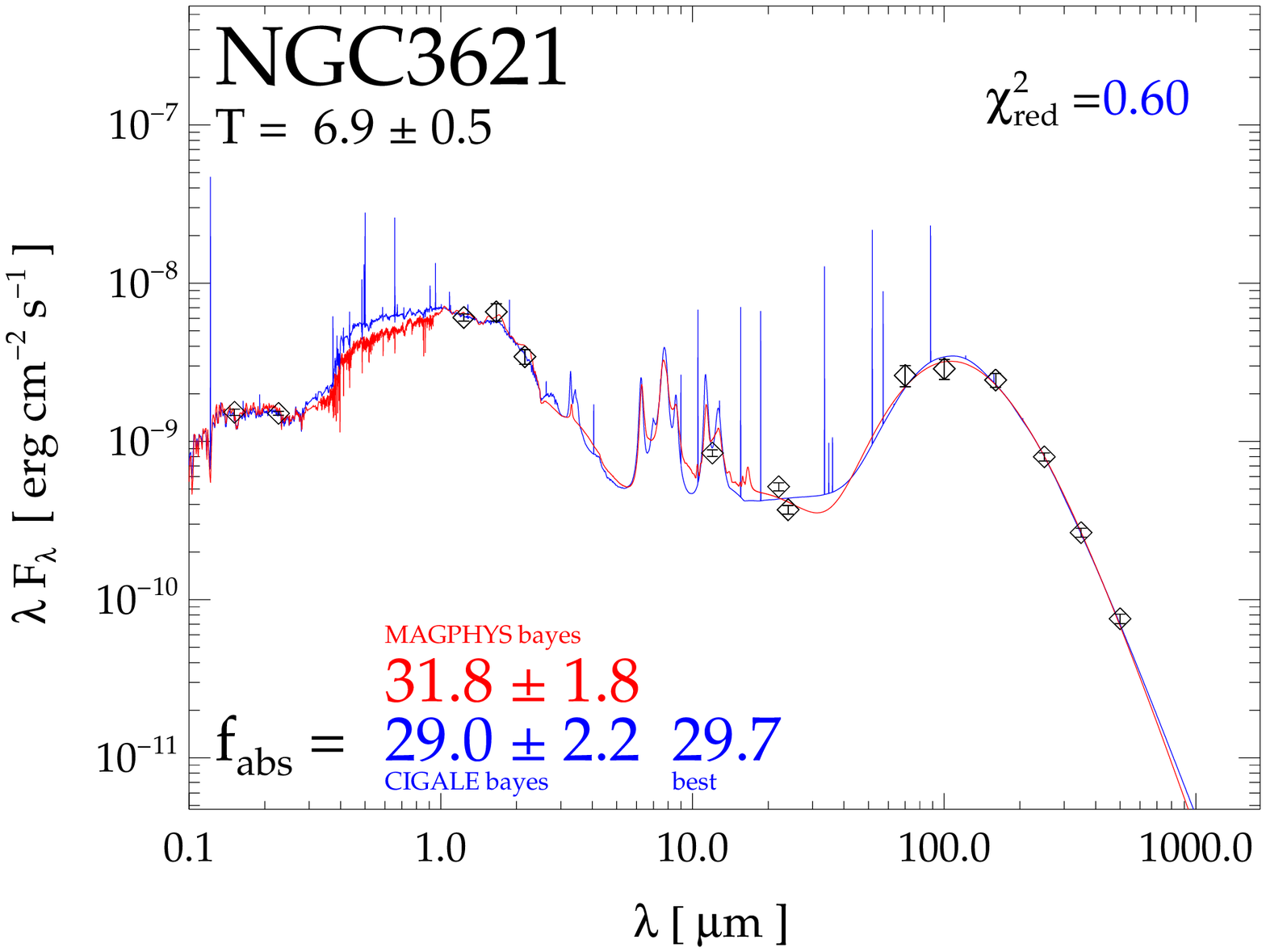}\includegraphics[scale=0.45,clip=true,trim=40 0 0 0]{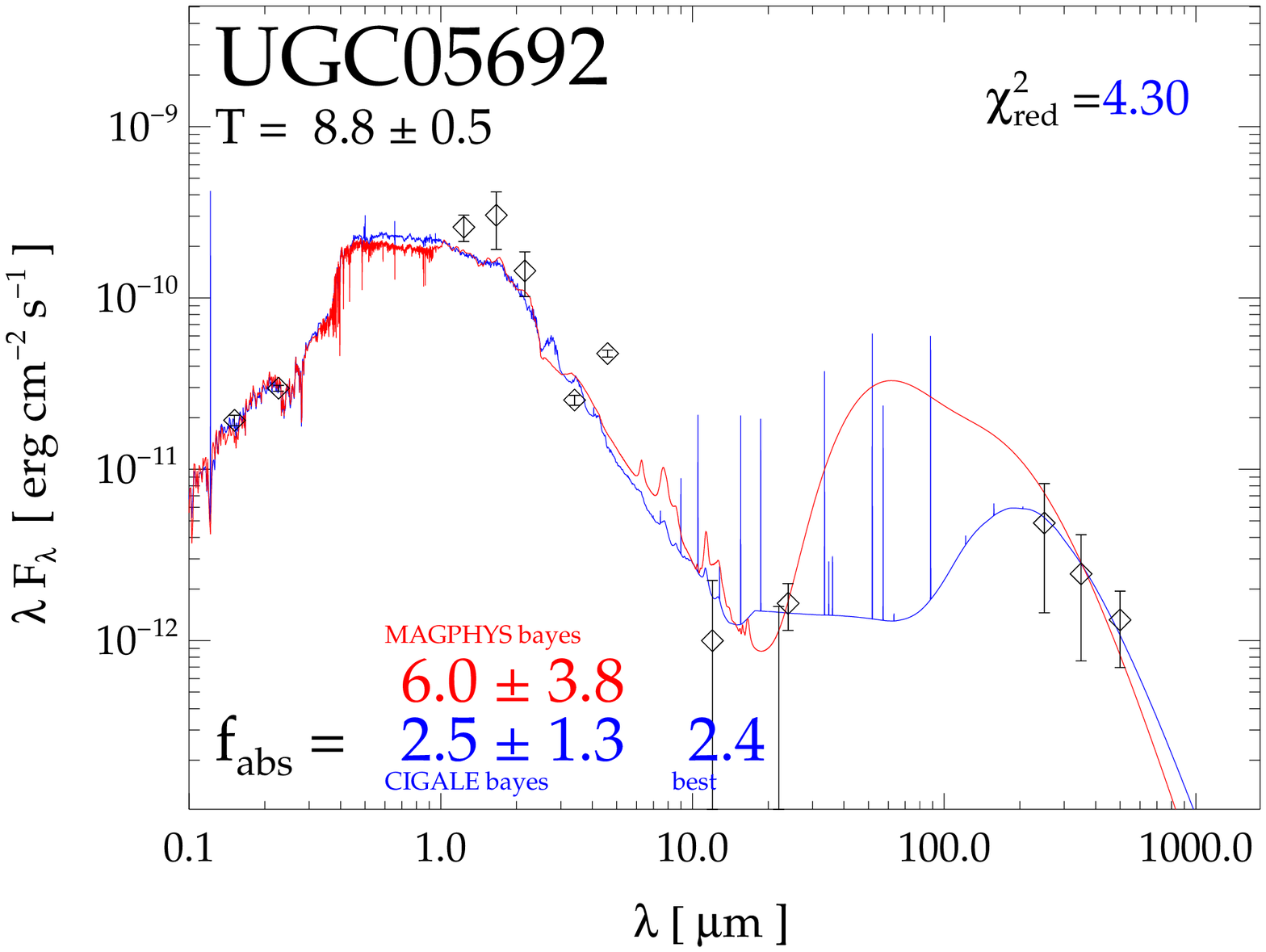}
\caption{Example SED fits for LTGs ($T>0.5$). Galaxies whose SEDs have a large band coverage ($\ge 25$ datapoints) are shown in the top row, those with a worse coverage ($\le 17$ datapoints) on the bottom row. Galaxies on the left have fits with $\chi^2$ at the peak of its distribution ($0.3 < \chi^2 <0.6$), those on the right a worse fit with  $\chi^2 >4$. The blue lines show CIGALE's best fit; both $f_\mathrm{abs}$ bayesian estimate (with uncertainty), used in the main text, and the estimate from best-fit parameters (i.e.\ the one describing the best fit itself) are reported. When available, we also show: in red, the MAGPHYS best fit (and bayesian estimate of $f_\mathrm{abs}$); in purple, the  piece-wise SED used for galaxies with very good data coverage (and the estimate of $f_\mathrm{abs}$ by direct integration of it). For the alternative estimates, see Sect.~\ref{app:compa}.}
\label{fig:ltgs_sedfits}
\end{figure*}

\begin{figure*}
\center
\includegraphics[scale=0.45,clip=true,trim=0 59 0 0]{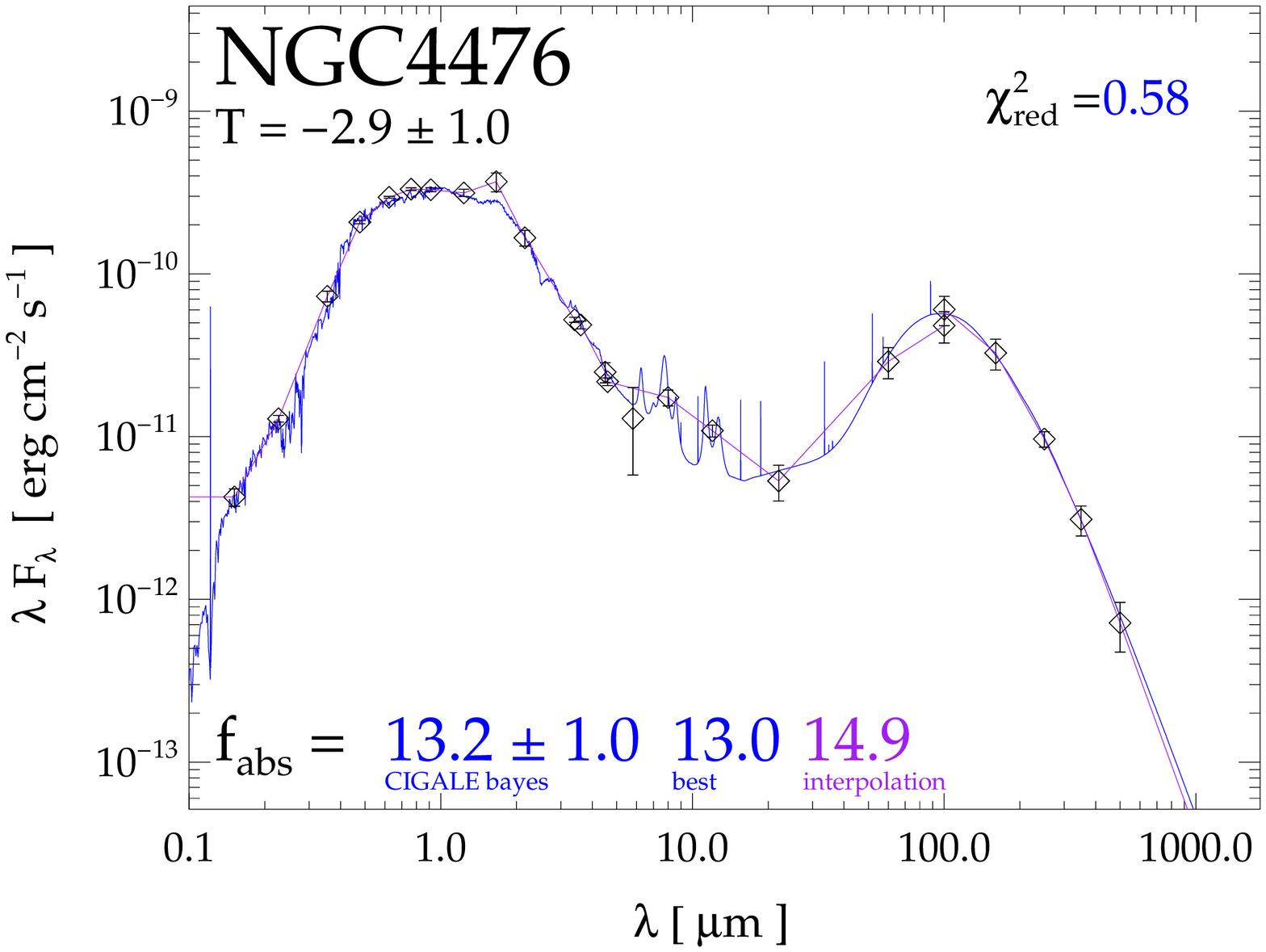}\includegraphics[scale=0.45,clip=true,trim=40 59 0 0]{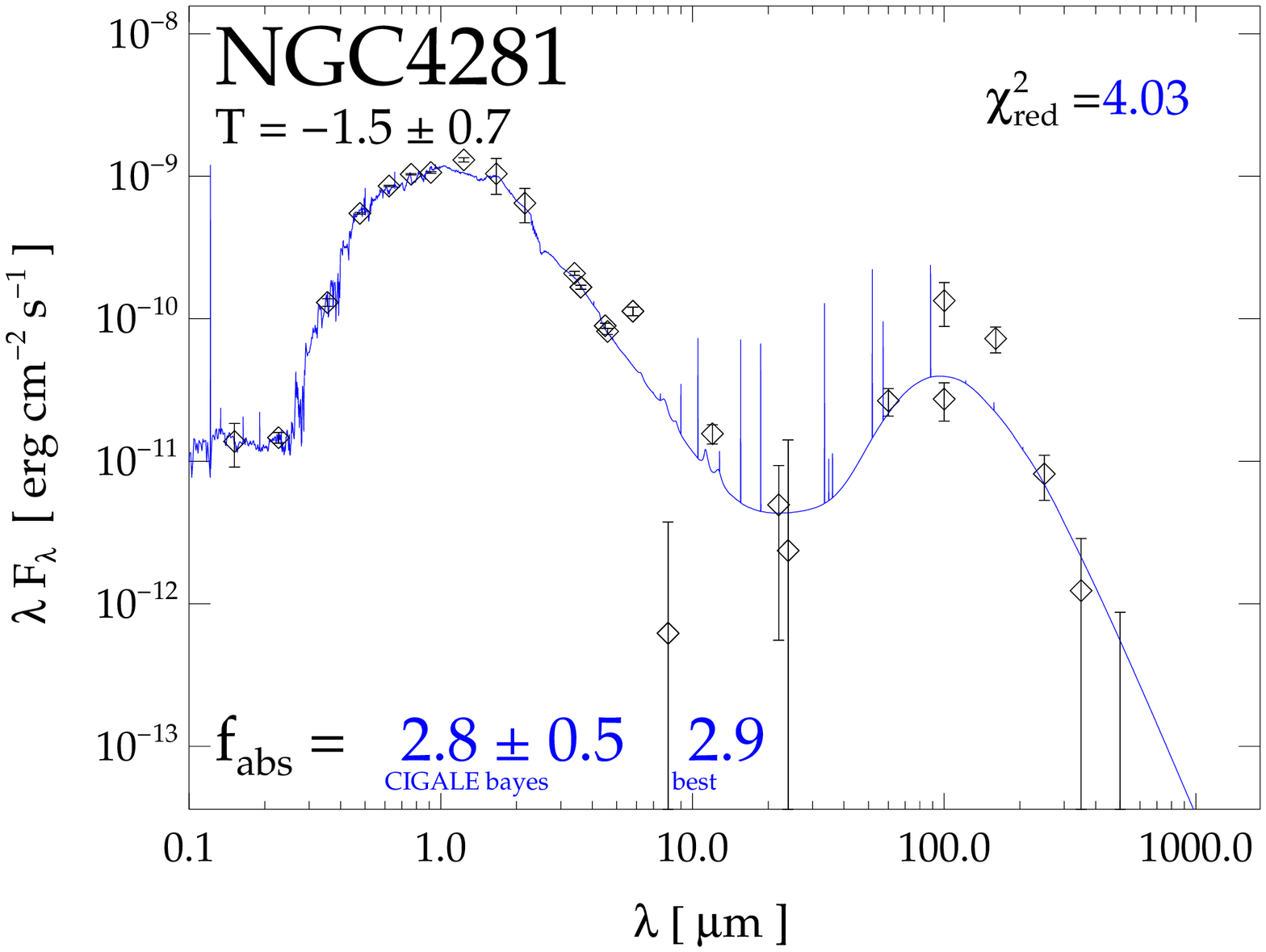}
\includegraphics[scale=0.45,clip=true,trim=0 0 0 0]{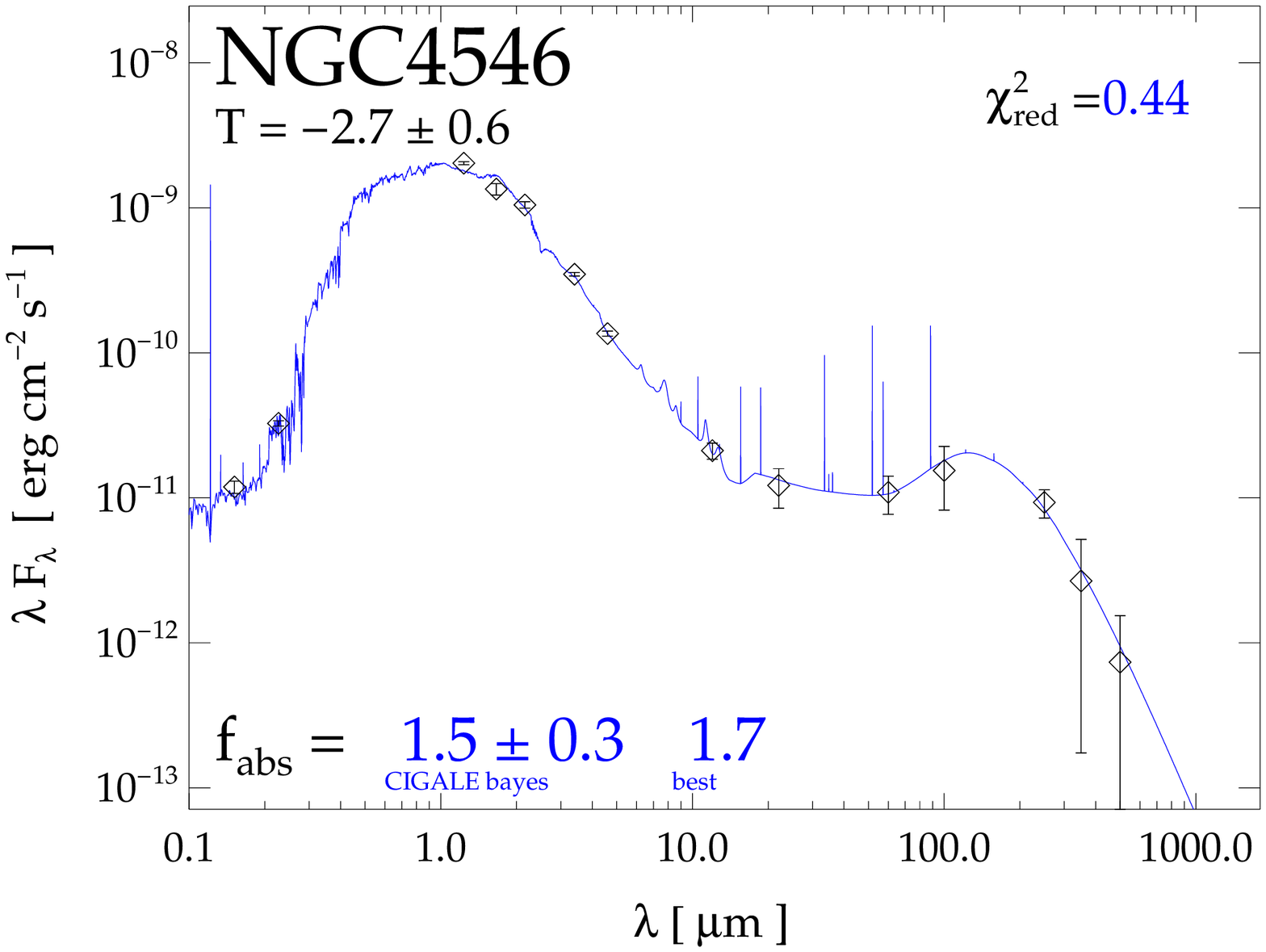}\includegraphics[scale=0.45,clip=true,trim=40 0 0 0]{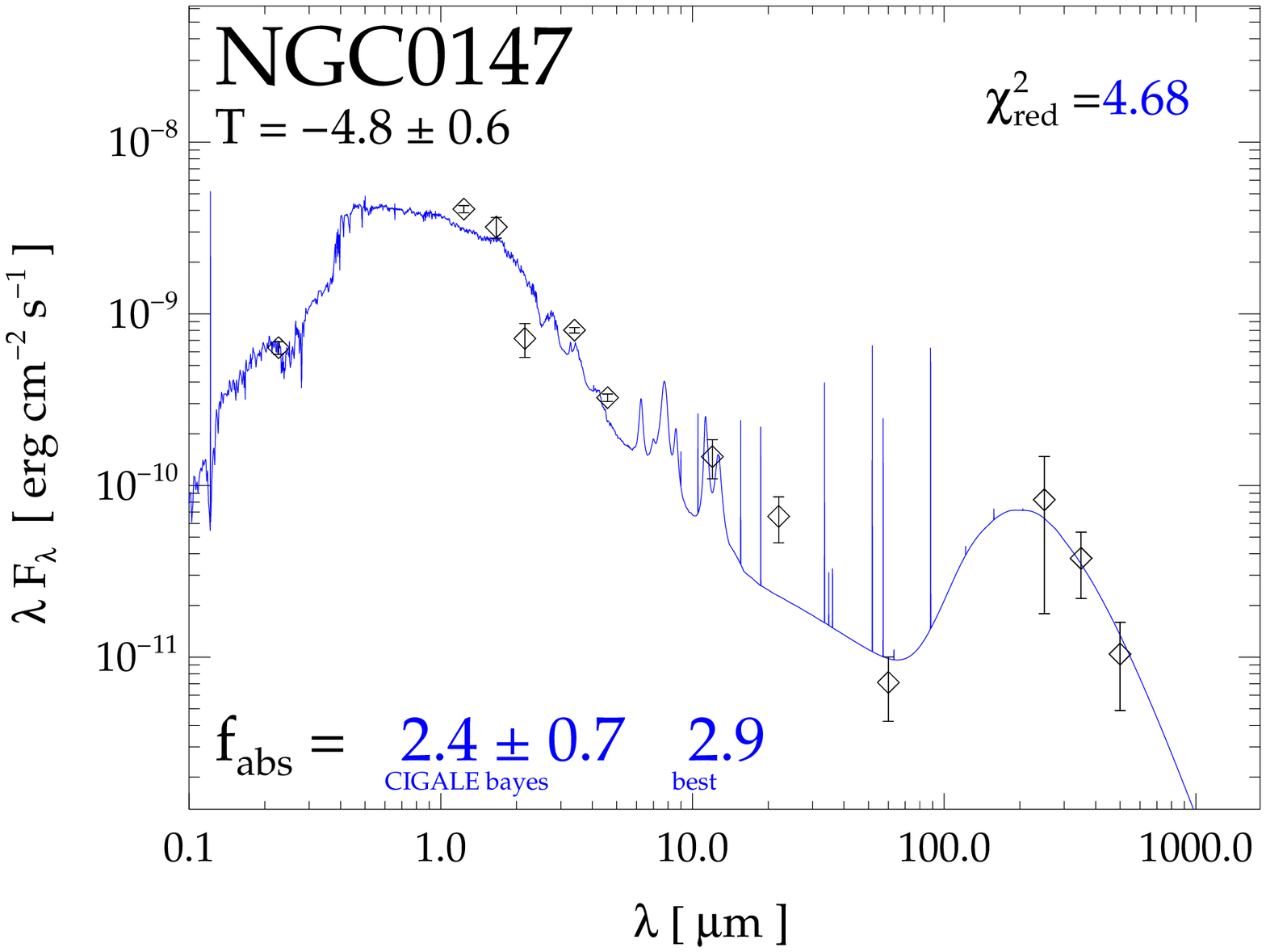}
\caption{Same as Fig.~\ref{fig:ltgs_sedfits} but for ETGs ($T<0.5$).}
\label{fig:etgs_sedfits}
\end{figure*}

\subsection{SED fitting}
\label{sec:cigale}

CIGALE models the SED of a galaxy by choosing a variety of modules for the stellar, gas and dust emission, and for dust attenuation; observations are compared with a grid of models defined by the various parameters describing each module; the code imposes the energy conservation between the amount of radiation absorbed and emitted by dust \citep{NollA&A2009,RoehllyProc2014}. For our work we have used: the {\em delayed and truncated} star formation history module, which was found to describe well both field and cluster spirals \citep{CieslaA&A2016}; a modified version of the standard starburst-like dust attenuation \citep{CalzettiApJ2000,BoquienA&A2016}; the \citet{BruzualMNRAS2003} single stellar population module of solar metallicity, coupled to the \citet{SalpeterApJ1955} initial mass function; the default set for the nebular emission module, based on CLOUDY templates \citep{FerlandPASP1998,BoquienA&A2016}, with ionization parameter 0.01 and all the ionising radiation absorbed by gas. The version of CIGALE we used includes a new dust emission module implemented by  \citet{NersesianA&A2018}: formally, this is similar to the available {\tt dl2014} module, that uses an updated version of the \citet{DraineApJ2007b} formalism (see also \citealt{CieslaA&A2014}). 
{
However, the dust emission templates are computed from 
 {\em The Heterogeneous dust Evolution Model for Interstellar Solids} \citep[THEMIS; ][]{JonesA&A2017}, DustPedia reference model, 
whose optical properties are firmly based on laboratory measurements of amorphous silicates, hydrocarbons and carbonaceous materials.  
}
By varying a set of 9 parameters, we built a grid of about 8$\times 10^7$ models. The complete grid specifications and an analysis of the performances of the new {\tt themis} dust emission module are presented in \citet{NersesianA&A2018}.

After minimizing the $\chi^2$ over the grid, CIGALE produces a best fit for the SED, for the various parameters describing the modules, and the physical quantities related to them. One could integrate numerically the fitted SED to obtain $L_\mathrm{dust}$, $L_\mathrm{bolo}$ and  $f_\mathrm{abs}$. The fitted SED might be considered as a reasonable, physically-motivated, interpolant between the datapoints;  for galaxies with a coarser SED coverage, it could be a substitute for the correction factors used in the earlier works. However, parameters derived by minimizing $\chi^2$  over a coarse  grid might be prone to degeneracies. Thus, we prefer the bayesian-like estimates computed by CIGALE from the likelihood of the models, which also allows for an estimate of the uncertainties. We used  the likelihood-weighted mean and standard deviation for $L_\mathrm{dust}$ and $L_\mathrm{bolo}$, and derived from these our estimates for $f_\mathrm{abs}$ and its uncertainty (which is mostly due to the uncertainty on $L_\mathrm{dust}$). 

Example SED fits are shown in Fig.~\ref{fig:ltgs_sedfits} and Fig.~\ref{fig:etgs_sedfits} for LTGs and ETGs, respectively. The figures show galaxies with different photometric coverage and with {\em good}  (i.e. at the peak of the $\chi^2_\mathrm{r}$ distribution, with  $0.3 < \chi_\mathrm{r}^2 < 0.6$) and {\em poor} ($\chi^2_\mathrm{r} > 4 $) fits. The $\chi^2_\mathrm{r}$ distribution is shown in \citet{NersesianA&A2018}. We find that only $\la 5$\% of the sample have  $\chi^2 _\mathrm{r}> 3$ (and $\la 3$\% with $\chi^2 _\mathrm{r}> 4$) . For the majority of these objects, the large  $\chi^2_\mathrm{r}$ values are the result of a few photometric points deviant from the general SED trend, and the CIGALE output can still be used to derive $f_\mathrm{abs}$, in particular when the bayesian-like estimates are used, as done here. In fact, the global trends we will show in the rest of the work change very little if the worse fits are excluded. Thus, we decided to retain the full sample. In Sect.~\ref{app:compa} we show the difference between the bayesian estimates we use and those from the best fit. We also compare the values of $f_\mathrm{abs}$ derived from CIGALE and those from other methods, in particular from the MAGPHYS fitting code \citep{DaCunhaMNRAS2008}, used by \citet{ViaeneA&A2016}.
{ 
In fact, for the limited scope of the current work, the dependence of the results on the chosen methodology (and dust model) is small.
}

From the CIGALE output, we also use in this work the bayesian estimates and uncertainties for the stellar mass ($M_\mathrm{stars}$) and for the star formation rate (SFR). For all the extensive quantities, we used the distances collected by \citet{ClarkA&A2018} and distributed together with the DustPedia photometry.

\subsection{Ancillary information}
\label{sec:morph_pam}
As a morphology indicator, we take the Hubble stage $T$ obtained from the HyperLEDA database \citep{MakarovA&A2014}\footnote{{\tt http://leda.univ-lyon1.fr/}. Values were downloaded in mid-2017 and are the same as those used by \citet{MosenkovA&A2018b}.}. HyperLEDA $T$'s might be non-integer, since, for most objects, they are the average over various estimates available in the literature. In the following, when we will refer to a specific morphology type defined by an integer value of $T$, we will use all objects in the range $[T-0.5,T+0.5)$ (e.g.\ the Sa sample defined by $T=1$ include all objects with $0.5 \le T < 1.5$). Also, an estimate of the uncertainty is derived by HyperLEDA from the literature scatter.

Alternative morphology indicators are taken from \citet{MosenkovA&A2018b}. They fitted a 2-D single S\'ersic (\citeyear{SersicBAAA1963,SersicBook1968}) surface brightness distribution to 3.4 $\mu$m images of most DustPedia galaxies. They divided the sample into disk-like galaxies, characterised by S\'ersic index $n < 2$ ($n=1$ is for exponential disks), and objects dominated by a spheroid (either spirals with dominant bulges or ellipticals) with larger values of $n$. While most late-type objects with $T>5$ are disk-like, objects classified in HyperLEDA as of earlier type show both disk-like and spheroid-like profiles. The S\'ersic indices $n$ is available for almost the entire sample (810 out of 814 objects). 

In order to test the effects of including edge-on galaxies in the analysis, we used the  inclinations estimated by  \citet{MosenkovA&A2018b}, which are more precise than those from HyperLEDA. They are available for most of the disk-like galaxies  (446 out of 453 objects with $n<2$). 
We also used the mass of atomic gas, $M_\ion{H}{i}$, together with  $M_\mathrm{star}$, as an indicator of the evolutionary stage of a galaxy. Through a literature search, $M_\ion{H}{i}$ has been retrieved for 709 objects (Casasola et al., De Vis et al., in prep.).

\section{Dependence on morphological type}
\label{sec:results}

\begin{figure*}
\includegraphics[width=\hsize]{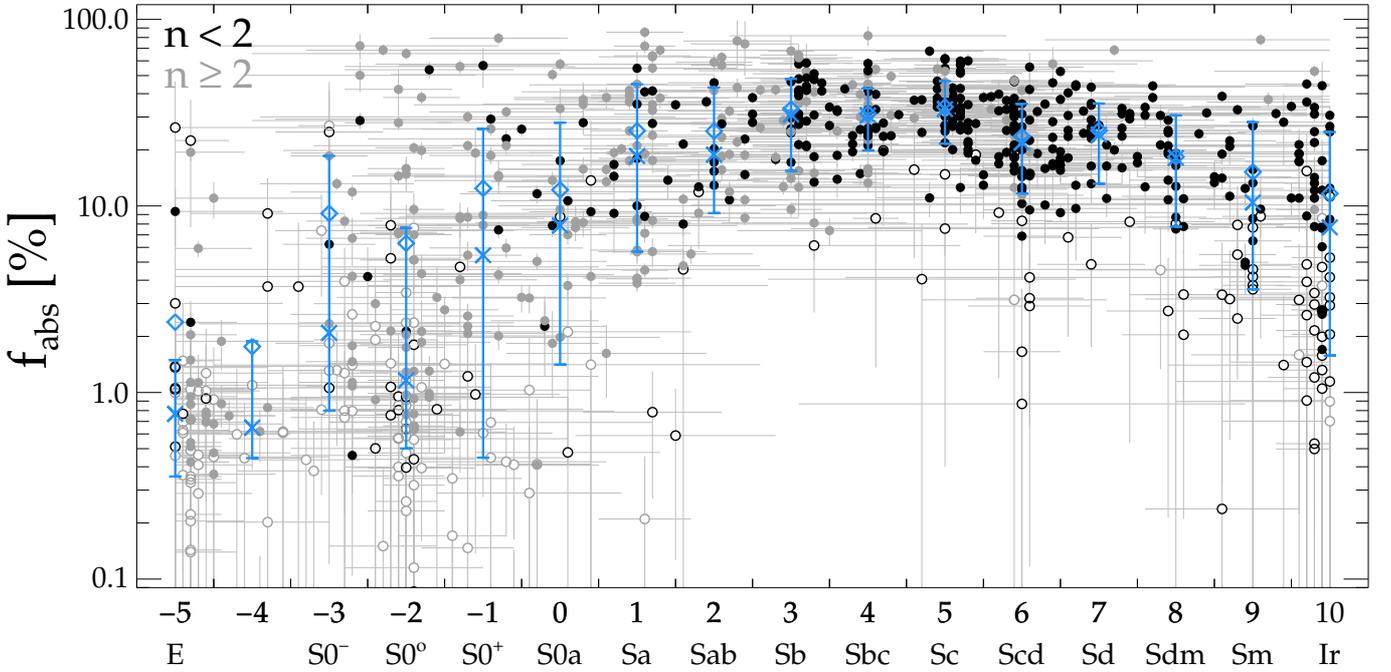}
\caption{$f_\mathrm{abs}$ vs Hubble stage $T$ for the DustPedia galaxies. For each bin in $T$, we plot the mean (diamonds), median (crosses) and the interval between the 16\% and 84\% percentiles (errorbars). Results on individual galaxies are also shown. Disk-dominated objects (those with fitted S\'ersic index $n <2$) are in black, spheroid-dominated ($n\ge2$) in gray. Open symbols are for galaxies with $f_\mathrm{abs}$ estimate below 2$\sigma$. Errorbars are shown in light gray for all objects.}
\label{fig:fabs_vs_T}
\end{figure*}

In Fig.~\ref{fig:fabs_vs_T} we show $f_\mathrm{abs}$ as a function of the Hubble stage $T$. We plot values (and error estimates on both axes) for each galaxy, using different colors for disk-dominated objects with S\'ersic index $n <2$ (black circles) and spheroid-dominated with  $n \ge2$ (gray circles). We also plot results for bins of width $\Delta T=1$ centred on each (integer) $T$ value (see Sect.~\ref{sec:morph_pam}): the values for the mean, standard deviation, median, lower 16\% and upper 84\% percentiles for each $T$ bin, as well as for broader bins in Hubble stage, are given in Table~\ref{tab:fabs_vs_T}. For the full sample, we obtain $\langle f_\mathrm{abs} \rangle = 19.1\pm0.6$\%. As expected, LTGs ($T\ge 0.5$) are usually richer in dust content and have a higher average,  $\langle f_\mathrm{abs} \rangle = 24.9\pm0.7$\%, than ETGs ($T < 0.5$), for which $\langle f_\mathrm{abs} \rangle = 7.4\pm0.8$\%. 

\begin{table*}
\caption{$f_\mathrm{abs}$ values for DustPedia galaxies, for bins of $T$}              
\label{tab:fabs_vs_T}      
\centering                                      
\begin{tabular}{c c c | c c | c c c}          
\hline\hline                        
$T$ & type & $N_\mathrm{obj}$ & mean & $\sigma$ & 16\% perc. & median & 84\% perc. \\    
\hline                                   
$                  -5$&                   E     & 51 & 2.4 & 5.4   & 0.4 & 0.8 & 1.5\\
$                  -4$&                         & 20 & 1.8 & 3.0   & 0.4 & 0.6 & 1.9\\
$                  -3$&              S0$^-$     & 34 & 9.1 &15.6   & 0.8 & 2.1 &18.6\\
$                  -2$&              S0$^o$     & 83 & 6.3 &13.4   & 0.5 & 1.2 & 7.6\\
$                  -1$&              S0$^+$     & 43 &12.4 &17.3   & 0.4 & 5.4 &25.9\\
$                   0$&                 S0a     & 37 &12.2 &14.3   & 1.4 & 7.9 &28.0\\
$                   1$&                  Sa     & 50 &25.3 &21.1   & 5.7 &18.5 &44.9\\
$                   2$&                 Sab     & 40 &25.2 &19.3   & 9.2 &18.8 &43.1\\
$                   3$&                  Sb     & 58 &33.0 &15.5   &15.4 &30.9 &48.0\\
$                   4$&                 Sbc     & 63 &31.2 &12.9   &19.9 &30.6 &42.7\\
$                   5$&                  Sc     & 70 &34.0 &13.6   &21.6 &33.1 &46.7\\
$                   6$&                 Scd     & 84 &23.6 &12.5   &11.7 &21.9 &35.3\\
$                   7$&                  Sd     & 46 &25.4 &12.7   &13.2 &24.1 &35.4\\
$                   8$&                 Sdm     & 32 &18.4 &10.7   & 7.8 &18.1 &30.6\\
$                   9$&                  Sm     & 36 &15.2 &15.2   & 3.6 &10.4 &28.1\\
$                  10$&                  Ir     & 67 &11.6 &12.3   & 1.6 & 7.7 &24.9\\
\hline
$         [-5.0,-3.5)$&                   E     & 71 & 2.2 & 4.8   & 0.4 & 0.7 & 1.9\\
$         [-3.5, 0.5)$&                  S0     &197 & 9.2 &15.0   & 0.6 & 2.3 &18.6\\
$         [ 0.5, 2.5)$&              Sa-Sab     & 90 &25.3 &20.2   & 7.1 &18.8 &43.1\\
$         [ 2.5, 5.5)$&               Sb-Sc     &191 &32.8 &13.9   &19.2 &31.1 &46.1\\
$         [ 5.5, 8.5)$&             Scd-Sdm     &162 &23.1 &12.4   &10.8 &21.2 &34.8\\
$         [ 8.5,10.0]$&               Sm-Ir     &103 &12.9 &13.4   & 2.1 & 8.6 &26.7\\
\hline
$         [-5.0, 0.5)$&                ETGs     &268 & 7.4 &13.5   & 0.5 & 1.4 &14.3\\
$         [ 0.5,10.0]$&                LTGs     &546 &24.9 &16.2   & 8.2 &23.6 &40.6\\
$         [-5.0,10.0]$&                 all     &814 &19.1 &17.4   & 1.1 &15.6 &36.2\\
\hline
\\
\end{tabular}
\end{table*}

For LTGs, a mild trend with $T$ can be seen, though with a large scatter: $\langle f_\mathrm{abs} \rangle$ is lower for $T=1-2$ objects ($\approx 25\%$), then it increases for  $T=3-5$. The peak  $\langle f_\mathrm{abs} \rangle $ is for Sc ($T=5$) galaxies, $34.0\pm 1.6$\%. For later types it decreases again, reaching a minimum for irregular galaxies ($T=10$), with  $\langle f_\mathrm{abs} \rangle = 12.3\pm1.6$. Despite the fact that the scatter in the $f_\mathrm{abs}$ distribution is generally large, it appears to be smaller - relative to the absolute value - for types $4 < T < 7$. In this range there is a scarcity of (extreme) outliers and the mean and median are very similar. This moderate uniformity might be due to the nature of the objects, most of which are disk-dominated, as shown by \citet{MosenkovA&A2018b}. An average, disk-like, galaxy of this type is NGC~3621 (Fig.~\ref{fig:ltgs_sedfits}, bottom left). Later-type objects are still predominantly disky, yet the scatter in $f_\mathrm{abs}$  increases, with a growing number of outliers on the lower side of the distribution. One such object is UGC~5692 (Fig.~\ref{fig:ltgs_sedfits}, bottom right). 

The scatter is larger also for the earlier LTGs with $T=1-3$, probably as a result of the increased number of spheroid-dominated with respect to disk-dominated objects. Though spheroid-dominated galaxies can have both low and high $f_\mathrm{abs}$ values, the 8  DustPedia objects with $f_\mathrm{abs} > 70\%$ are all  bulge-dominated. The most extreme case is NGC~4355 with $f_\mathrm{abs}=85\pm 6 \%$ (Fig.~\ref{fig:ltgs_sedfits}, top right\footnote{The fit for NGC~4355 is among the worst, because the upper end of the range we explored for the intensity of the interstellar radiation field  is not high enough \citep[for details, see][]{NersesianA&A2018}. The $f_\mathrm{abs}$ value we obtained is a lower limit to the true value. An alternative estimate  (see Sect.~\ref{app:magphys})  yields $f_\mathrm{abs}\approx 95\%$, a value which is still marginally consistent with ours, within $1.5\sigma$.}). The IR emission of this galaxy (also  known as NGC~4418) is powered by a compact starburst with a minor contribution from an AGN \citep{VareniusA&A2014}.  In Sect.~\ref{sec:checks} we will show that the inclusion of AGN-hosts in our sample does not bias the results.

Going to ETGs, the relative scatter is higher for lenticular galaxies: $f_\mathrm{abs}$ ranges from very low values, in most cases compatible with zero within the estimated uncertainty, up to the largest values observed for LTGs. The highest fractions of absorbed energy estimated in this morphology range are $f_\mathrm{abs}=79\pm 6$ and $72\pm 11 \%$ for NGC~1482 ($T=-0.8$) and NGC~1222 ($T=-2.6$), two spheroid-dominated objects undergoing merger-driven starburst episodes \citep{VagshetteNewA2012,YoungMNRAS2018}. An object with an {\em average}  $f_\mathrm{abs}$ is NGC~4476 shown in Fig.~\ref{fig:etgs_sedfits} (top left). As for Sa galaxies, in these morphology bins the $f_\mathrm{abs}$ distribution is not Gaussian, the mean value being  larger than the median. Also, there is no clear difference in $f_\mathrm{abs}$ between the disk-dominated galaxies and the rest of spheroid-dominated objects for these morphologies. For elliptical galaxies, instead, the scatter in $f_\mathrm{abs}$ is reduced and the number of objects with $f_\mathrm{abs}$ larger than a few percent is smaller. We will discuss the results on ETGs and the significance of the lowest $f_\mathrm{abs}$ values later.

\begin{figure*}
\sidecaption
\includegraphics[width=12cm]{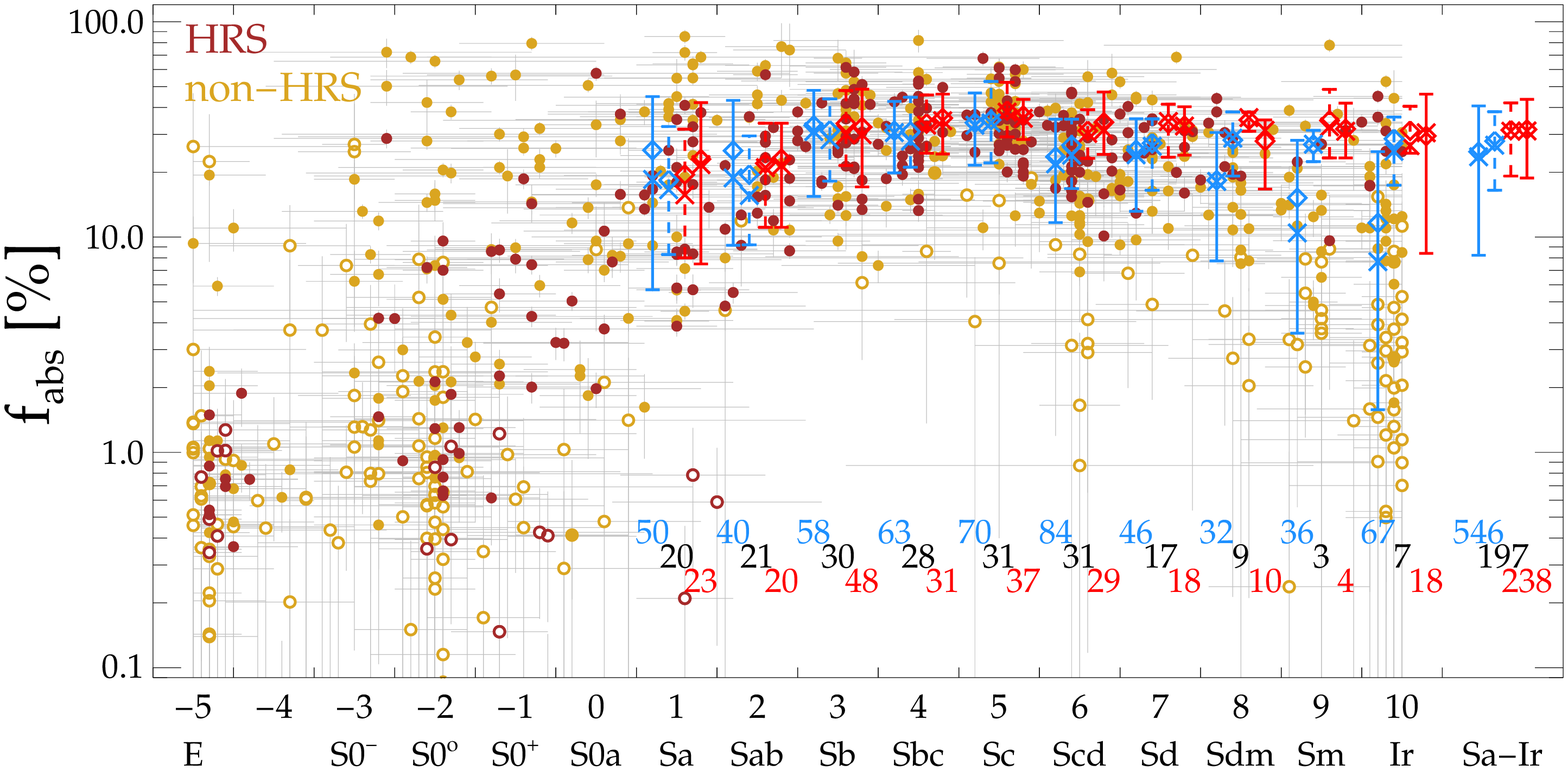}
\caption{Same as Fig.~\ref{fig:fabs_vs_T}, but showing average results for: the DustPedia LTG sample (blue symbols with solid errorbars); the \citet{ViaeneA&A2016} analysis on HRS LTGs (red symbols with solid errorbars); the current analysis and the $f_\mathrm{abs}$ values of  \citet{ViaeneA&A2016} for a sample of galaxies in common between the two works (blue and red symbols with dashed errorbars, respectively). For the common sample, the DustPedia morphology is used. We also show results for each galaxy from the current analysis: brown symbols are for HRS objects included in DustPedia, yellow for additional DustPedia objects. The number of galaxies used in each analysis and for each bin is indicated. The Im and Pec classes of \citet{ViaeneA&A2016} are combined into the Ir, $T=10$, bin. 
}
\label{fig:fabs_vs_T_hrs}
\end{figure*}

The subjective nature of the traditional morphology classification is reflected by the large uncertainty in the Hubble stage $T$, shown by the horizontal error bars in Fig.~\ref{fig:fabs_vs_T}. On average, the uncertainties are of order of the bin width. There are, however, objects with uncertainties in $T$ spanning over several bins. This mixing might be responsible for part of the scatter in $f_\mathrm{abs}$. Using the two-sample Kolmogorov-Smirnov (KS) test we found that in several cases there is a significant probability that the samples in nearby $T$ bins are drawn from the same distribution. For example, there is a probability of  74\% that the $T=-1 $ and 0 distributions are compatible with each other.
In a few cases, the probability is large even for bins that are more separated: it is 93\% for $T=0$ and 10, and 56\% for $T=2$ and 6. Guided by Fig.~\ref{fig:fabs_vs_T} and by the results of the KS tests, we divided the sample into six larger morphology samples (see Table~\ref{tab:fabs_vs_T}). The mean $f_\mathrm{abs}$  value is the lowest for E galaxies; it increases from 9 to 25\% going from S0 to Sa-Sab; it reaches a maximum for Sb-Sc with $\langle f_\mathrm{abs} \rangle \approx 33$\%;  it decreases again for later types, to 23\% for Scd-Sm  and 13\% for Sm-Ir. The distribution of $f_\mathrm{abs}$ in these larger samples are significantly different from each other, at a probability level $P<0.001$. The only exception are the  Sa-Sab  and Scd-Sdm bins, for which the KS test gives a probability of 10\% of coming from the same distribution. This, and the KS results on some of the unit $T$ bins, indicates that $f_\mathrm{abs}$ depends less on $T$ than on other physical properties of the samples. 

\section{Comparison with previous estimates}
\label{sect:compa}

If we limit the comparison to {\em Herschel}-based works, we find that our result is marginally consistent, at a $2\sigma$ level, with \citet{SkibbaApJ2011}: for their whole late-type sample it is $\langle f_\mathrm{abs} \rangle= 29\pm2\%$ vs our $24.9\pm0.7$\%. It is also consistent when only earlier spirals are included: it is  $\langle f_\mathrm{abs} \rangle = 34\pm3\%$ for their Sa-Sc ($0.5 \le T < 5.5$) and $30\pm1\%$ in our sample. As in \citet{SkibbaApJ2011}, we find that in later type spirals dust absorbs a smaller fraction of stellar radiation. This is apparently at odds with the earlier ISO-based results of \citet{PopescuMNRAS2002}, where later types have higher $f_\mathrm{abs}$. However, this might be simply due to the limited number of objects in their morphology bins and to the large corrections needed to take into account the submm spectrum not sampled by ISO. Yet, the average value for their full sample, $\langle f_\mathrm{abs} \rangle= 24\pm2\%$, is consistent with our estimate. For ETGs, \citet{SkibbaApJ2011} have $\langle f_\mathrm{abs} \rangle = 25\pm8\%$. Though their limited sample (10 objects vs our 268) might be biased towards ISM-rich objects,  the result is still marginally consistent at a $2\sigma$ level, with our estimate, $\langle f_\mathrm{abs} \rangle \approx 7.4\pm0.8\%$.

A slightly higher estimate is obtained for LTGs by \citet{DaviesMNRAS2012}, $\langle f_\mathrm{abs} \rangle = 31\pm2\%$, at about $3\sigma$ from our determination. This might be due to a series of causes: their sample is selected at $500 \mu$m, biasing towards objects with higher dust emission fluxes; it is a cluster sample, including only Virgo galaxies (though there is no apparent change in the SED of cluster and field galaxies; Davies et al., in prep.); the integration procedure in their work is coarser and excludes a few wavelength ranges with significant stellar and dust emission. Nevertheless, the average value for the 7 ETGs where they measured  $f_\mathrm{abs}$ is $8\pm2\%$, consistent with our estimate. 

On the higher side is also the $\langle f_\mathrm{abs} \rangle$ estimate of \citet{ViaeneA&A2016}. For their sample of 239 LTGs from HRS, they have $\langle f_\mathrm{abs} \rangle = 31.6\pm0.8\%$, $6\sigma$ apart from the value found here.  Since \citet{ViaeneA&A2016} is our reference work, for the completeness and dimension of their sample and the similarity of their method to ours, we investigated in details the reason for the discrepancy between the two analyses.

We first restricted the sample to the galaxies in common between \citet{ViaeneA&A2016} and this analysis: because of the DustPedia selection rules, of the requirements imposed in this work (Sect.~\ref{sec:data}), and of the differences between the morphology classification used by HRS and DustPedia, they reduce to 197 objects.  For them, we obtain $\langle f_\mathrm{abs} \rangle = 31.4\pm0.8$\% using the original estimates from \citet{ViaeneA&A2016}, very close to the value derived on their full sample. Instead, it is $\langle f_\mathrm{abs} \rangle = 27.9\pm0.8$\% using the results of our fits. The difference is thus reduced to $3\sigma$, showing that part of the disagreement with  \citet{ViaeneA&A2016} is due to the larger number and diversity of objects in DustPedia.

In Fig.~\ref{fig:fabs_vs_T_hrs} we plot the results of the averages described above as a function of $T$. For either the full HRS sample of \citet{ViaeneA&A2016}, or the common HRS-DustPedia sample, there is little dependence of $\langle f_\mathrm{abs} \rangle$ on $T$ for the later type spirals with $T\ga6$. Instead, DustPedia includes many more low $f_\mathrm{abs}$ objects in this morphology range, as can be seen by comparing the DustPedia-only datapoints with those of the DustPedia-HRS  (yellow vs brown circles, respectively). The result is the decline of $\langle f_\mathrm{abs} \rangle$ with $T$ we already described, together with the increased scatter in the distribution.

\begin{figure*}
\center
\includegraphics[height=8.5cm,clip=true,trim=5 15 0 0]{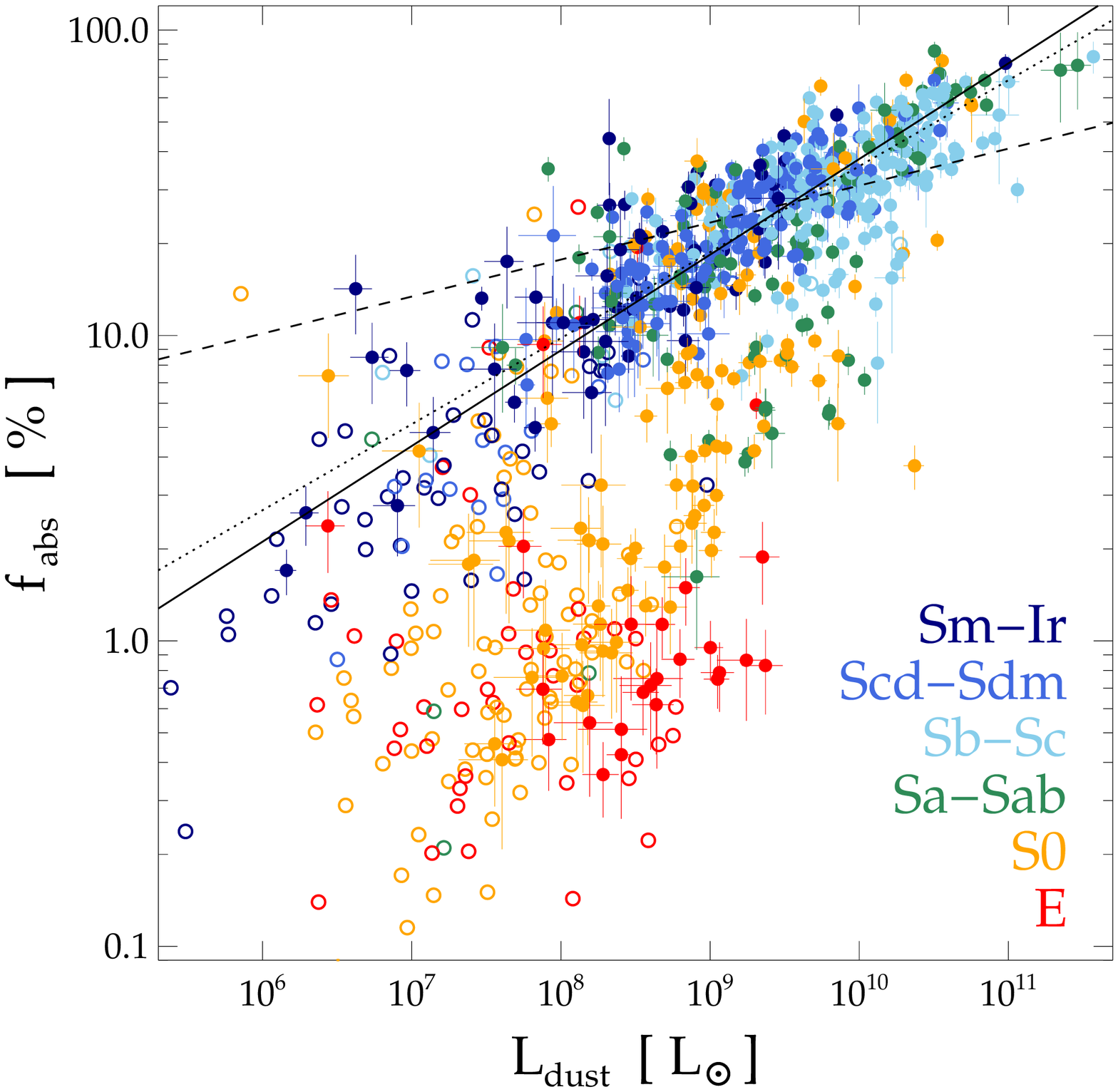}\includegraphics[height=8.5cm,clip=true,trim=82 15 0 0]{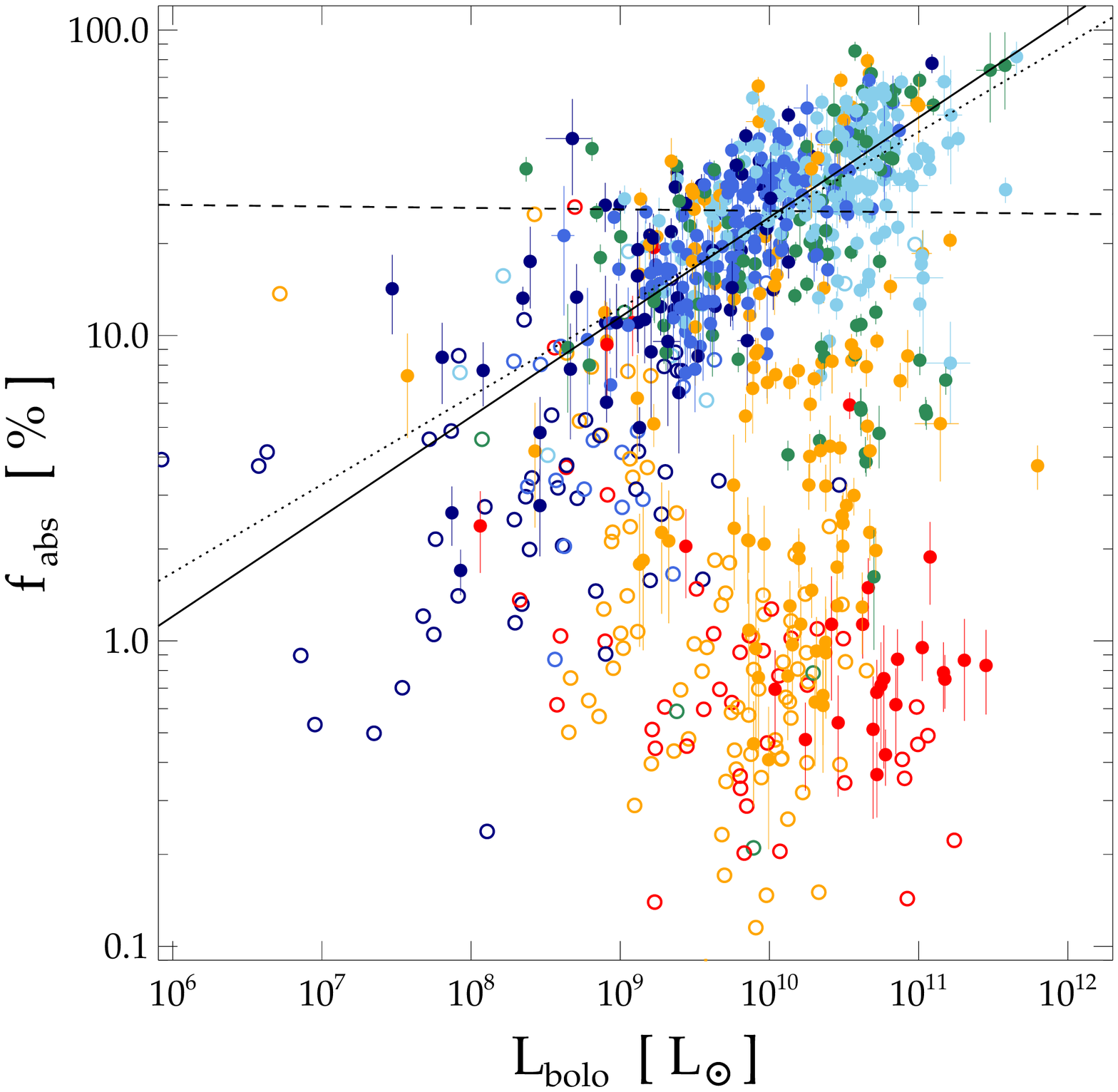}
\caption{$f_\mathrm{abs}$ vs $L_\mathrm{dust}$ (left panel) and vs $L_\mathrm{bolo}$ (right panel). Different colours refer to different morphological types. Open symbols without errorbars show $f_\mathrm{abs}$ estimates below 2$\sigma$. Solid lines are the linear fits to datapoints for types later than Sb. Dashed lines refer to the intrinsic correlations produced by the uncertainties in the luminosity, under the assumption of no correlation of  $f_\mathrm{abs}$ vs $L_\mathrm{bolo}$; dotted lines are the same, but assuming the same $f_\mathrm{abs}$ vs $L_\mathrm{bolo}$ correlation as observed (see text for details).
}
\label{fig:corr_lum}
\end{figure*}

A detailed comparison between the photometry and models of \citet{ViaeneA&A2016} and those of the current work on the common DustPedia-HRS sample have shown that various reasons might be responsible for the residual discrepancy. A first explanation might reside in the different SED fitting code used by  \citet{ViaeneA&A2016}, MAGPHYS \citep{DaCunhaMNRAS2008}.  Compared to DustPedia, more galaxies of the \citet{ViaeneA&A2016} sample do not have photometric constraints between 25 and 100 $\mu$m.
This, coupled with the tendency of MAGPHYS to fill the gap with an unconstrained warm dust component (see Sect.~\ref{app:magphys} and UGC~5692 in Fig.~\ref{fig:ltgs_sedfits}) tends to bias $f_\mathrm{abs}$ to larger values. The bias is stronger for the few later type galaxies in common between the two samples, several of which showing an unconstrained SED on the Wien side of the thermal peak. Other, more systematic, differences are due to photometry: the fluxes in \citet{ViaeneA&A2016} were not corrected for Galactic extinction \citep{DeVisMNRAS2017a};  the FUV and NUV fluxes from HRS are smaller, and the WISE 12 $\mu$m higher, than those from DustPedia \citep[for details, see][]{ClarkA&A2018}. All these differences concur, to various extent, in biasing $f_\mathrm{abs}$ to (slightly) larger values. 

In conclusion, our results are different from those of \citet{ViaeneA&A2016} in part because of the larger sample and diversity in DustPedia. When the same sample is used, there is a residual difference due to the methodology and photometry. The uniformity of the DustPedia photometry from the UV to the submm, combined with the more physical dust emission templates we have used in CIGALE, reassures us on the $f_\mathrm{abs}$ estimates done in the current work.

\section{Dependence on luminosity}
\label{sec:lumi}

In Fig.~\ref{fig:corr_lum} we show $f_\mathrm{abs}$ vs $L_\mathrm{dust}$ (left panel) and vs $L_\mathrm{bolo}$ (right panel). For LTGs, a positive trend (though with a large scatter) is clearly present: Sb-Sc objects appear to be clustered on the bright end of the trend and Sm-Ir objects on the faint one, though without a clear morphology division between intermediate luminosities. This indicates that luminosity is a more important factor than morphology in the determination of $f_\mathrm{abs}$. The stronger dependence on luminosity rather than on morphology might also be the reason for the non-null result of the KS test between the Sa-Sab and  Scd-Sdm distributions: while different in morphology, the two samples share a similar luminosity range. However, there are also significative departures from the trend: already for types Sb-Sc, but more pronouncedly for Sa-Sb,  a good fraction of bright galaxies have a lower $f_\mathrm{abs}$.  Going to ETGs,  the majority of lenticular galaxies fall in the same low-$f_\mathrm{abs}$, high-luminosity region, though several of them still share the main LTGs trend.The vast majority of ellipticals, instead, have low $f_\mathrm{abs}$ and shows no correlation (with the exception of a few cases on the LTGs trend, possibly due to misclassified lenticulars).

\subsection{The trend for LTGs}

\begin{table*}
\caption{$f_\mathrm{abs}$ values for DustPedia galaxies later than Sb, for bins of  $L_\mathrm{bolo}$. For $L_\mathrm{bolo}$  we also give the mean and standard deviation within each bin. }              
\label{tab:fabs_vs_L}      
\centering                                      
\begin{tabular}{c c c | c c | c c c}          
\hline\hline
$\log_{10} L_\mathrm{bolo}/L_\odot$ & $\langle L_\mathrm{bolo}/L_\odot\rangle$& $N_\mathrm{obj}$ & mean & $\sigma$ & 16\% perc. & median & 84\% perc. \\    
\hline
$[ 7.5, 8.5)$&$ 1.6\pm 0.9\times 10^{ 8}$& 30& 5.1& 4.6& 1.2& 3.1& 8.6\\
$[ 8.5, 9.5)$&$ 1.7\pm 0.8\times 10^{ 9}$&104&13.1& 8.7& 3.8&11.9&21.3\\
$[ 9.5,10.5)$&$ 1.2\pm 0.8\times 10^{10}$&216&27.4&11.1&16.4&27.1&37.3\\
$[10.5,11.5)$&$ 6.2\pm 3.4\times 10^{10}$& 97&38.6&14.6&25.0&38.3&54.3\\
\hline
\end{tabular}
\end{table*}

\begin{figure}
\center
\includegraphics[height=8.5cm,clip=true,trim=5 15 0 0]{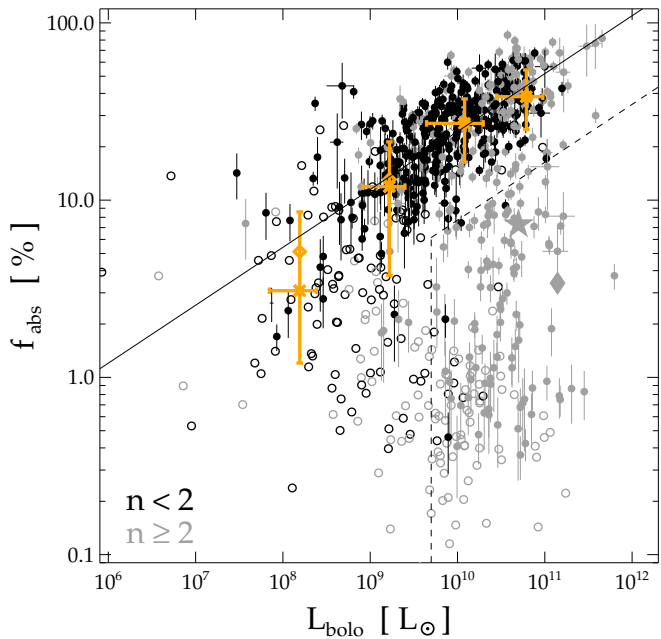}
\caption{
$f_\mathrm{abs}$ vs  $L_\mathrm{bolo}$. Different colors are used for disk- and bulge-dominated galaxies. Open symbols without errorbars show $f_\mathrm{abs}$ estimates below 2$\sigma$. 
We plot in orange the mean (diamonds), median (crosses) and the interval between the 16\% and 84\% percentiles (errorbars), for each of the $L_\mathrm{bolo}$ bins of Table~\ref{tab:fabs_vs_L}.
The fit to the main trend is indicated, together with the area for the selection of the low-$f_\mathrm{abs}$ tail. 
The star indicates M31 and the diamond NGC~4594 (see text for details).
}
\label{fig:corr_lum_disk}
\end{figure}

We quantify the strength of the correlation using the Kendall's correlation measure $\tau_K$. For almost all the $\tau_K$ measures presented in this work, the probability for the null hypothesis is small, at the $P<0.001$ level. 
For LTGs, we find $\tau_K = 0.47$ for  $f_\mathrm{abs}$ vs  $L_\mathrm{bolo}$ and 0.63 for  $f_\mathrm{abs}$ vs  $L_\mathrm{dust}$. The correlations improve to $\tau_K = 0.54$ and 0.66, respectively, if only galaxies later than Sb ($T\ge 2.5$) are considered. In the following, we use this further morphological selection to define the trend, thus removing the low-$f_\mathrm{abs}$, high-luminosity, Sa-Sab objects.  We grouped the galaxies into four bins in $L_\mathrm{bolo}$ (see Table~\ref{tab:fabs_vs_L} and  Fig.~\ref{fig:corr_lum_disk}).  The mean  steadily increases from $\langle f_\mathrm{abs} \rangle = 5.1\pm0.8\%$ for $7.5 \le \log_{10} (L_\mathrm{bolo}/L_\odot) < 8.5$ (where most of galaxies are disks of type Sm-Ir) to $38.6\pm1.5 \%$ for $\log_{10} (L_\mathrm{bolo}/L_\odot) \ge 10.5$ (which are predominantly Sb-Sc). The scatter is reduced with respect to that for $T$ bins, confirming that  $f_\mathrm{abs}$ has a stronger dependence on $L_\mathrm{bolo}$ than on morphology.  Yet, the scatter is still large for the larger luminosities, since some high-luminosity Sb-Sc galaxies stay on the  low-$f_\mathrm{abs}$, high-luminosity tail more common for the (removed) Sa-Sab objects. Linear fits (in log-log space) to the trends for $T\ge 2.5$ yield
\[
\log_{10} (f_\mathrm{abs} \left[\%\right]) = -(1.9\pm0.2) + (0.33\pm 0.02) \log_{10}(L_\mathrm{bolo}/L_\odot),
\]
and
\[
\log_{10} (f_\mathrm{abs}\left[\%\right]) = -(1.6\pm0.1) + (0.31\pm 0.01) \log_{10}(L_\mathrm{dust}/L_\odot),
\]
with errors  estimated with a Monte Carlo bootstrap procedure, fitting a thousand random representations of the dataset (the fits are shown as solid lines in Fig.~\ref{fig:corr_lum}). Using the same procedure we found  that datapoints where  $f_\mathrm{abs}$ is smaller than $2\times$ the estimated uncertainties (open symbols in Fig.~\ref{fig:corr_lum}, corresponding to $L_\mathrm{dust} \la 10^{8.5} L_\odot$), when considered together with their uncertainties, do not bias the Kendall's correlation measure (or the fit). The trend is also not significantly affected by the inclusion of the worse fits: $\tau_K$ changes  from 0.54 to 0.53 if only objects with $\chi^2_r <2$ are considered (432 out of the 456 galaxies of type later than Sb).

In Fig.~\ref{fig:corr_lum_disk} we plot $f_\mathrm{abs}$ vs $L_\mathrm{bolo}$ and distinguish among disk-dominated and spheroid-dominated galaxies, based on their S\'ersic index. Apparently, most of the (high signal-to-noise) $n<2$ objects align along the trend already discussed (with $\tau_K=0.51$), while $n\ge2$ are present both on the trend and on the lower $f_\mathrm{abs}$ locus of ETGs. Thus, the use of the alternative morphology indicator has no impact in refining the sample and reducing the scatter in the observed trends. We also experimented with different cuts (and different morphology indicators, such as the bulge-to-total ratio, available for $\approx 64$\% of the sample; \citealt{MosenkovA&A2018b}), but we were not able to find a better split between galaxies following the trend and galaxies outside of it.

If we consider only objects later than Sb, and thus following the main trend, we find that disk dominated objects with $n<2$ have a lower $\langle f_\mathrm{abs} \rangle$ than those with $n\ge2$: it is $23.5\pm0.7$\% vs $32\pm2$\%. However, this is the result of the bias in luminosity, since bulge-dominated objects have on average larger $L_\mathrm{bolo}$ (this is shown  in Fig.~\ref{fig:corr_lum_disk} by the larger number of gray symbols at the upper end of the luminosity range, and their  lack at the lower end). When the analysis is conducted for smaller luminosity bins, such as those in Table~\ref{tab:fabs_vs_L}, there is little difference between objects with $n$ below and above 2: for example, if we select all objects later than Sb with $10.5 \le \log_{10} (L_\mathrm{bolo}/L_\odot) < 11.5$, it is $\langle f_\mathrm{abs} \rangle=39\pm 2 $\% for $n<2$ and   $38\pm 3$\% for $n\ge 2$. The same is true for the third luminosity bin in Table~\ref{tab:fabs_vs_L} (there are few bulge-dominated systems in the first bins for significative statistics). 

\subsection{Significance of $f_\mathrm{abs}$ vs $L_\mathrm{bolo}$}

Trends similar to that shown for  $f_\mathrm{abs}$ vs $L_\mathrm{dust}$ have been found in the past. 
{ The FIR luminosity was found to correlate with FIR to optical or UV ratios, using IRAS \citep{SoiferAJ1989,WangApJ1996} and {\em Spitzer} data \citep{DaleApJ2009}.
Using {\em Herschel} observations of}
KINGFISH galaxies, \citet{SkibbaApJ2011} had shown a correlation consistent with our own.  In particular, they noted  the reduced fraction of absorbed radiation  for later type dwarf and irregular galaxies with respect to earlier type spirals.

On the other hand, the correlation of  $f_\mathrm{abs}$ vs $L_\mathrm{bolo}$ has received less attention (though correlation with a related quantity, the stellar mass, had been noted; see later in Sect.~\ref{sect:mstar}).
This correlation is potentially more interesting, as it denotes that dust absorption depends on the star-formation history of a galaxy. Indeed, we test in the following that the higher $\tau_K$ for $f_\mathrm{abs}$ vs  $L_\mathrm{dust}$ does not imply that the correlation is more physically significant than that with $L_\mathrm{bolo}$. The reason for this is that the luminosities shown in the x-axis are used to derive  $f_\mathrm{abs}$ on the y-axis; spurious correlations are produced by the uncertainties in the luminosities and by the intrinsic scatter in $f_\mathrm{abs}$. 

Following \citet{DeVisMNRAS2017a}, we investigated the significance of the correlations using Monte Carlo procedures. In a first test, we assumed the null hypothesis, i.e. that $f_\mathrm{abs}$ is actually independent of the luminosity. For each object, we generated a mock value of $L_\mathrm{bolo}$ assuming a Gaussian distribution centred on the measured value and with scatter provided by its uncertainty. We then extracted a random value of  $f_\mathrm{abs}$ (again assuming a Gaussian distribution, and the mean and scatter of the whole sample); from these, a new $L_\mathrm{dust}$ was produced, to which we added a Gaussian scatter using as uncertainty a value randomly chosen from those associated with the 20 nearest, true, estimates of the dust luminosity. Finally, a mock value of  $f_\mathrm{abs}$ was derived from the ratio of the luminosities and linear fits (in log-log space) were derived. In the two panels of Fig.~\ref{fig:corr_lum} we show with dashed lines the average slopes obtained after iterating the procedure a thousand times. A faint, negative,  correlation is found for $f_\mathrm{abs}$ vs $L_\mathrm{bolo}$; negative, because $L_\mathrm{bolo}$ appears in the denominator of the $f_\mathrm{abs}$  estimate, and faint because the error estimates for $L_\mathrm{bolo}$ are small. Instead,  a significant spurious trend for $f_\mathrm{abs}$ vs $L_\mathrm{dust}$ is present. In a second test, we verified that this bias is responsible for the larger $\tau_K$ for $f_\mathrm{abs}$ vs  $L_\mathrm{dust}$. The procedure is almost identical to the previous one, with one exception: for each object, the random value of $f_\mathrm{abs}$ is now derived by using the mean and scatter over the 20 objects which are nearest in the true estimates of the bolometric luminosity. We measured $\tau_K$ for the two relations and repeated the procedure a thousand times. The average trends are shown in the two panels of Fig.~\ref{fig:corr_lum} with dotted lines. The average correlations on the mock data are close to the fits to the true data (solid line), as much as the $\tau_K$ values (0.5 and 0.65 for the mock $f_\mathrm{abs}$ vs $L_\mathrm{bolo}$ and $f_\mathrm{abs}$ vs $L_\mathrm{dust}$, respectively). For $f_\mathrm{abs}$ vs $L_\mathrm{bolo}$  this is not unexpected, since in this second test we draw the mock $f_\mathrm{abs}$  from the observed correlation. For $f_\mathrm{abs}$ vs $L_\mathrm{dust}$, the test shows that the larger $\tau_K$ is simply produced by the larger uncertainties in $L_\mathrm{dust}$ and the scatter in $f_\mathrm{abs}$.

\subsection{Outliers and ETGs}

Moving from LTGs to lenticulars, a larger fraction of bright bulge-dominated galaxies move below the main trend. To investigate their nature, we selected galaxies with $L_\mathrm{bolo}\ge 5\times 10^9 L_\odot$ and set an arbitrary upper limit of $f_\mathrm{abs}$ at 3$\times$ below the fit to the main trend (dashed line in Fig.~\ref{fig:corr_lum_disk}). While only 3\% of Sb-Sc galaxies above the luminosity limit are included in the selection, the fraction grows to 30\% for Sa-Sab galaxies and increases further to  78\% for S0. Iconic objects that would fall in that locus of the plot are M~31, the {\em Andromeda} galaxy  ($T=3$,  $L_\mathrm{bolo} = 5\times 10^{10} L_\odot$ and $f_\mathrm{abs} = 7\%$, estimated from the fit of \citealt{ViaeneA&A2017}) and M~104  ($T=1.1\pm 3$, $L_\mathrm{bolo} = 1.4\times 10^{11} L_\odot$ and $f_\mathrm{abs} =3.4\%$ from \citealt{SkibbaApJ2011}; similar estimates from the fits of \citealt{DeLoozeMNRAS2012}). M31 is not part of DustPedia, while M~104 was excluded from the sample (see Sect.~\ref{sect:sample}):
both objects are shown with different markers in Fig.~\ref{fig:corr_lum_disk}. As these two galaxies are characterised by a large bulge and a ring-like dust distribution, the different $f_\mathrm{abs}$ behaviour of the other objects in the selection might in part be related to the different geometry of the stellar and dust distributions and to their evolution with respect to the objects along the trend. We will discuss this issue further in Sect.~\ref{sec:simple}.

A good fraction of ETGs have uncertain dust luminosities below $\approx 10^{8} L_\odot$. Indeed, 41\% of S0 and 65\% of ellipticals have $f_\mathrm{abs}$ determinations below $2\sigma$. Above that limit, some of the detections - and $f_\mathrm{abs}$ determinations - might also be spurious. In particular for bright ellipticals, the target area is defined by optical images when instead the dust emission, when detected, comes from a much smaller area, often unrelated to the stellar distribution \citep{SmithApJ2012b,DiSeregoA&A2013}. The aperture-matched photometry in FIR/submm bands might in some cases be dominated by strong cirrus feature or background sources in the large target area, rather than true dust emission from the source. From a visual check of the FIR imaging of  DustPedia ETGs, it is found that emission from the target itself is present in about a third of the ellipticals where $f_\mathrm{abs}$ is estimated to be above $2\sigma$  (8 out of 26 objects). An example of a spurious  $f_\mathrm{abs}\approx2\%$ determination is Andromeda's dwarf spheroidal NGC~147 (Fig.~\ref{fig:etgs_sedfits}), where the uncertain Herschel-SPIRE detection in the FIR is due to the MW cirrus foreground \citep{DeLoozeMNRAS2016}. Only a handful of ellipticals with true FIR emission detection have estimated $f_\mathrm{abs} \la 2\%$ at a $>2\sigma$ level. One of them is NGC~4487 (M87), where the FIR emission is dominated by synchrotron radiation \citep{BaesA&A2010} and not by dust. Since we only model the dust component, one might wonder if these low $f_\mathrm{abs}$ estimates should be considered as simple upper limits.
The contamination is present also for lenticulars, though to a lesser extent: about 78\% of  the $f_\mathrm{abs}\ge 2\sigma$ estimates are associated with true detections of FIR emission in the target (81 out of 104 galaxies). Nevertheless, the exclusion of the spurious $f_\mathrm{abs}$ estimates does not produce a significant change in the statistics of Table~\ref{tab:fabs_vs_T}, for both lenticulars and ellipticals.

Finally, we note that $f_\mathrm{abs}$ in part depends on the colour of the radiation heating the dust. For the same stellar and dust distributions, ETGs and earlier-type spirals with typically redder stellar SEDs suffer less absorption than bluer LTGs, because of the dust attenuation law. This translates into different trends for $f_\mathrm{abs}$ for galaxies where the younger stellar population dominates over the older one \citep{NersesianA&A2018}.

\section{Checking the assumptions}
\label{sec:checks}

Our estimate of $f_\mathrm{abs}$ from the observed SED relies on two main assumptions: the emission from both dust and stars is isotropic; stars are the only source of the radiation heating dust. We show in this Section that neither the inclusion of edge-on galaxies nor AGN bias the results discussed so far.

\subsection{Dependence on inclination}

Isotropic emission is generally assumed in most estimates of luminosities. The assumption is certainly valid for objects showing a spherical symmetry for both dust and stars. For disk galaxies, it is generally assumed to be valid for the dust emission, since the ISM is optically thin in MIR to submm bands. However, it might be invalid in highly inclined objects, where a larger fraction of the stellar output could be absorbed along the disk plane. 

RT models of edge-on galaxies (see Sect.~\ref{app:rt}) suggest that the estimate of $f_\mathrm{abs}$ decreases by $\approx$ 5-15\% when passing from the edge-on to the face-on inclination; and $L_\mathrm{bolo}$ increases by $\approx$ 30-40\%. Both effects are mainly due to the reduction of $L_\mathrm{star}$, as estimated from the output SED, in the edge-on case. Instead, we find little differences with inclination for DustPedia galaxies. When selecting disk-dominated objects, it is $\langle f_\mathrm{abs}\rangle\approx24\pm2\%$ {\em both} for the edge-ons (75 objects with $i=90^\circ$) and the  face-ons (69 objects with $i<40^\circ$, the limit chosen from the modelling of Sect.~\ref{app:rt} to encompass little variations of the estimate with $i$). The KS tests shows that these two samples have 95\% probability of coming from the same distribution. As a reference, the whole disk-dominated sample (446 galaxies) has $\langle f_\mathrm{abs}\rangle=21.6\pm0.7$\%. There might be a little bias towards face-on galaxies, which on average are 30\% brighter than edge-on and the full sample - as indeed expected from the models; yet the estimates for the two limiting inclinations are the same within the errors.

The different behaviour with respect to inclination between the real estimates and the simulation results might in principle be due to a selection bias: 
the {\em average} galaxy was modelled from the mean results of \citet{DeGeyterMNRAS2014}, which selected edge-ons with a prominent dust lane.
The same is true for other edge-ons analysed via RT fitting.
DustPedia includes 10 edge-on galaxies with detailed fits\footnote{They are: NGC~891  \citep{XilourisSub1998}, 
NGC~5907 \citep{XilourisSub1998,MosenkovA&A2018}, 
NGC~5529 \citep{XilourisSub1998,BianchiA&A2007,MosenkovA&A2018},
NGC~4013 \citep{XilourisSub1998,BianchiA&A2007,DeGeyterA&A2013,MosenkovA&A2018},
IC~2531 \citep{XilourisSub1998,MosenkovA&A2016},
NGC~4217 \citep{BianchiA&A2007,MosenkovA&A2018}, 
NGC~4302, NGC5746 \citep{BianchiA&A2007}, 
NGC~4565 \citep{DeLoozeMNRAS2012} and
IC~2461 \citep{DeGeyterMNRAS2014}. 
}. They are all disk-dominated galaxies of type Sb-Sc, according to our criteria. For them, we obtain $\langle f_\mathrm{abs} \rangle=36\pm4\%$, higher than the average on the full disk sample. However, these galaxies are brighter than average, with $10^{10} \la L_\mathrm{bolo}/L_\odot \la 10^{11}$.  If, among disk-dominated galaxies in the same luminosity range, we select the face-on cases (40 objects with $i<40^\circ$) it is  $\langle f_\mathrm{abs} \rangle=32\pm2\%$.
Thus, there is no appreciable difference between the estimates of $f_\mathrm{abs}$ on edge-on galaxies with well-defined dust lanes and face-on galaxies of analogous global properties. Results are identical if we select face-ons with $L_\mathrm{bolo}$  larger by up to 50\%, to mimic the behaviour expected from RT models.

Since we found no evidence for a dependence of $f_\mathrm{abs}$ on inclination, we considered edge-on galaxies together with the rest of the sample. 
In principle, a more isotropic behaviour of $f_\mathrm{abs}$ might be expected in galaxies where the energy budget is dominated by starlight from young 
clusters absorbed locally within parental clouds, an effect which is not captured by the smooth stellar and dust distributions adopted in
the modelling of edge-on galaxies. A reduced $f_\mathrm{abs}$ variation might be further concealed by the large observed scatter. However,
edge-on DustPedia galaxies {\em do} show a difference with respect to face-ons: for a similar FIR  SED, the stellar UV-to-NIR SED is redder
{ (with a difference in NUV-r colour of $\approx$1.8 magnitudes, as derived from the average SED of the galaxies analysed with RT fitting 
and that of face-ons in the same luminosity range).
}
Yet, for both inclinations, the energy exiting the galaxy as direct, attenuated, starlight, $L_\mathrm{stars}$, is the same, 
{ relatively to $L_\mathrm{dust}$ 
}
and the resulting  $f_\mathrm{abs}$ changes little. Investigating whether these effects are due to simplistic assumption in RT modelling of 
edge-ons, or to unforeseen selection rules in choosing their face-on counterparts, is beyond the scope of this paper. We hope to shed some light on 
these issues with dedicated RT modelling of the largest DustPedia galaxies (Verstocken et al., in prep., Nersesian et al., in prep.) and 
of galaxies in large samples of DustPedia-like objects from cosmological simulations (Tr\v{c}ka et al. in prep.).

\subsection{AGN}
In a galaxy hosting a luminous AGN, the dust torus heated by the hard radiation from the accretion disk might dominate the MIR emission over that produced by the diffuse dust and stellar component. At least for the case of the bright AGNs seen at high-z, reprocessed MIR radiation from the torus could contribute significantly to the FIR-submm SED emission \citep{SchneiderA&A2015,DurasA&A2017}. Even though such extreme objects are not likely to be present in a local sample, the host's dust might be directly exposed to the hard radiation from the nucleus, if its opening cone is not aligned with the disk axis.  Thus, the AGN contribution might bias the estimate of $f_\mathrm{abs}$ high with respect to the rest of the objects where stars are the dominant radiation source. 

We have used the 90\%-confidence criterium of \citet{AssefApJS2018}, based on the WISE 3.4 and 4.6 $\mu$m flux density, to ascertain which DustPedia galaxy might host a AGN. We find that 19 objects comply with that criterium. Among them are the record-holder NGC~4355 we already discussed; and NGC~1068, for which we are building detailed RT simulations to study the contribution of direct AGN radiation to dust heating (Viaene et al., in prep.). Several of these galaxies have high $f_\mathrm{abs}$. However, they share the same trend as the other objects: for example, if we select the hosts of type later than Sb (6 objects), we find $\langle f_\mathrm{abs} \rangle=43\pm16\%$ for an average $L_\mathrm{bolo}\approx1\times 10^{11} L_\odot$, a result compatible to that of the most luminous galaxies of the same type (See Table~\ref{tab:fabs_vs_L}). This shows that the presence of an AGN does not alter significantly the energy balance in the SED. 

For these reasons, we did not exclude AGN hosts from the analysis. A similar conclusion was reached by \citet{ViaeneA&A2016}.

\section{Template SEDs}
\label{sec:template}

\begin{figure*}
\center
\includegraphics[scale=0.45,clip=true,trim=0 59 0 3]{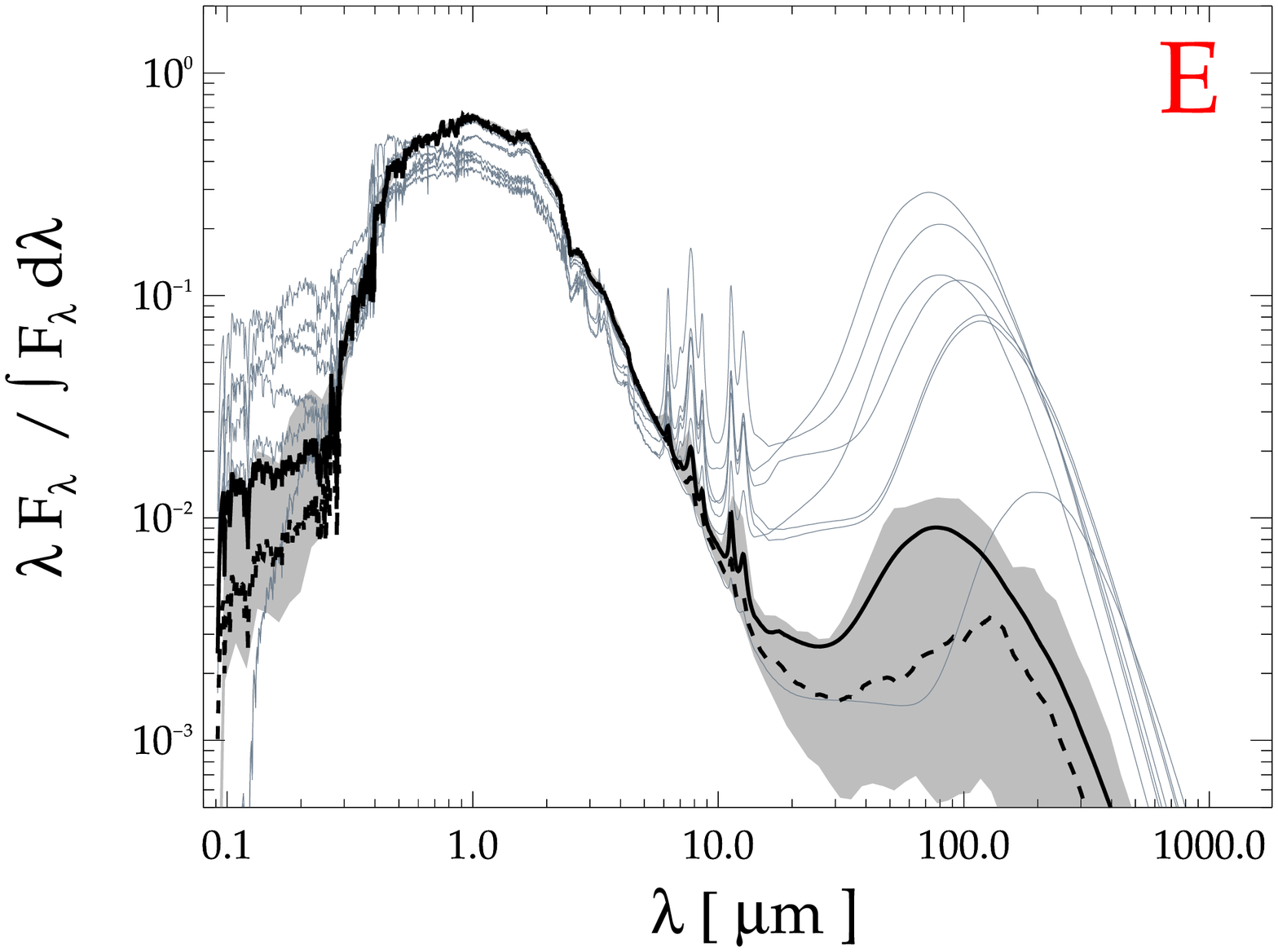}\includegraphics[scale=0.45,clip=true,trim=86 59 0 3]{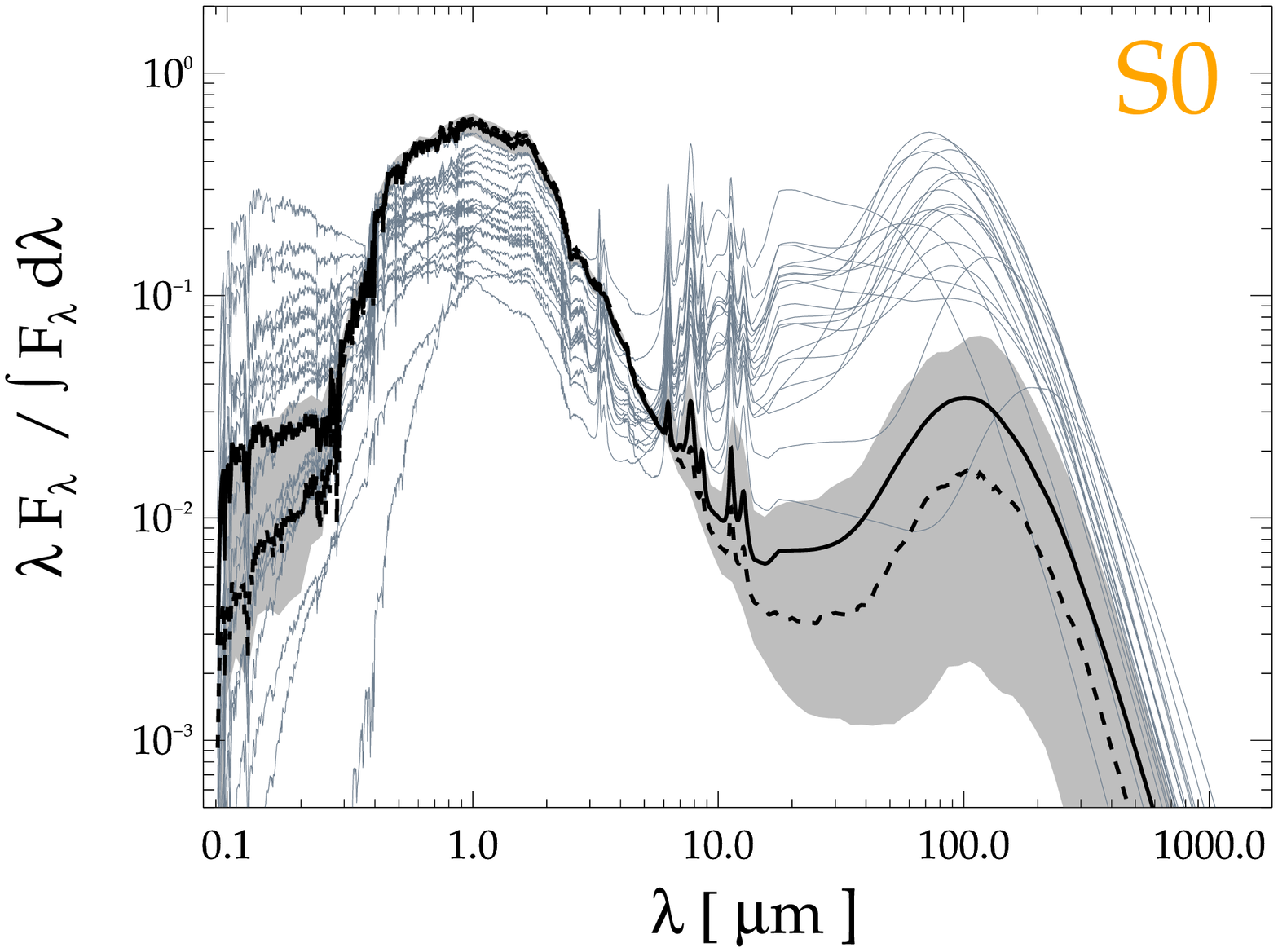}
\includegraphics[scale=0.45,clip=true,trim=0 59 0 3]{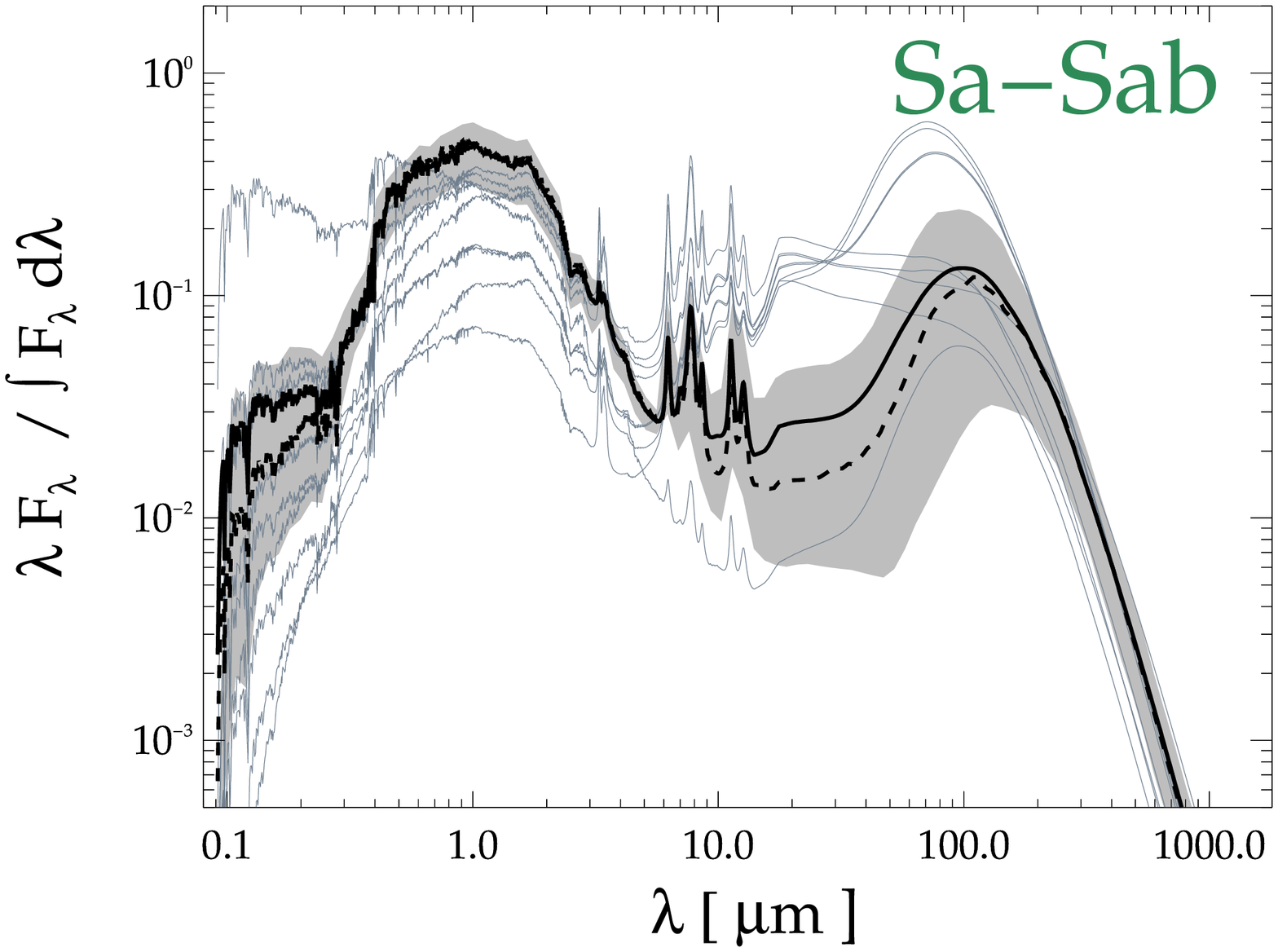}\includegraphics[scale=0.45,clip=true,trim=86 59 0 3]{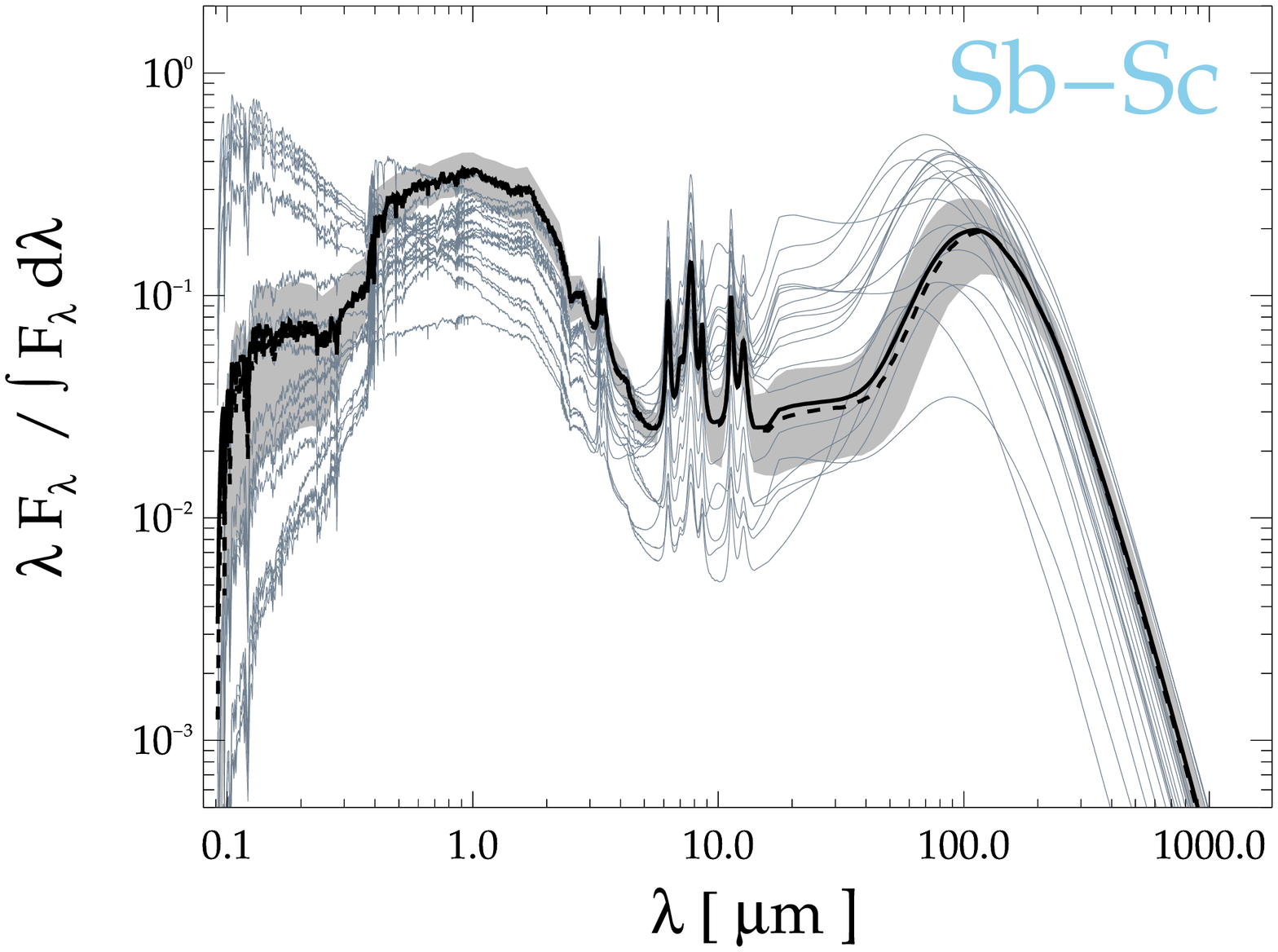}
\includegraphics[scale=0.45,clip=true,trim=0 0   0 3]{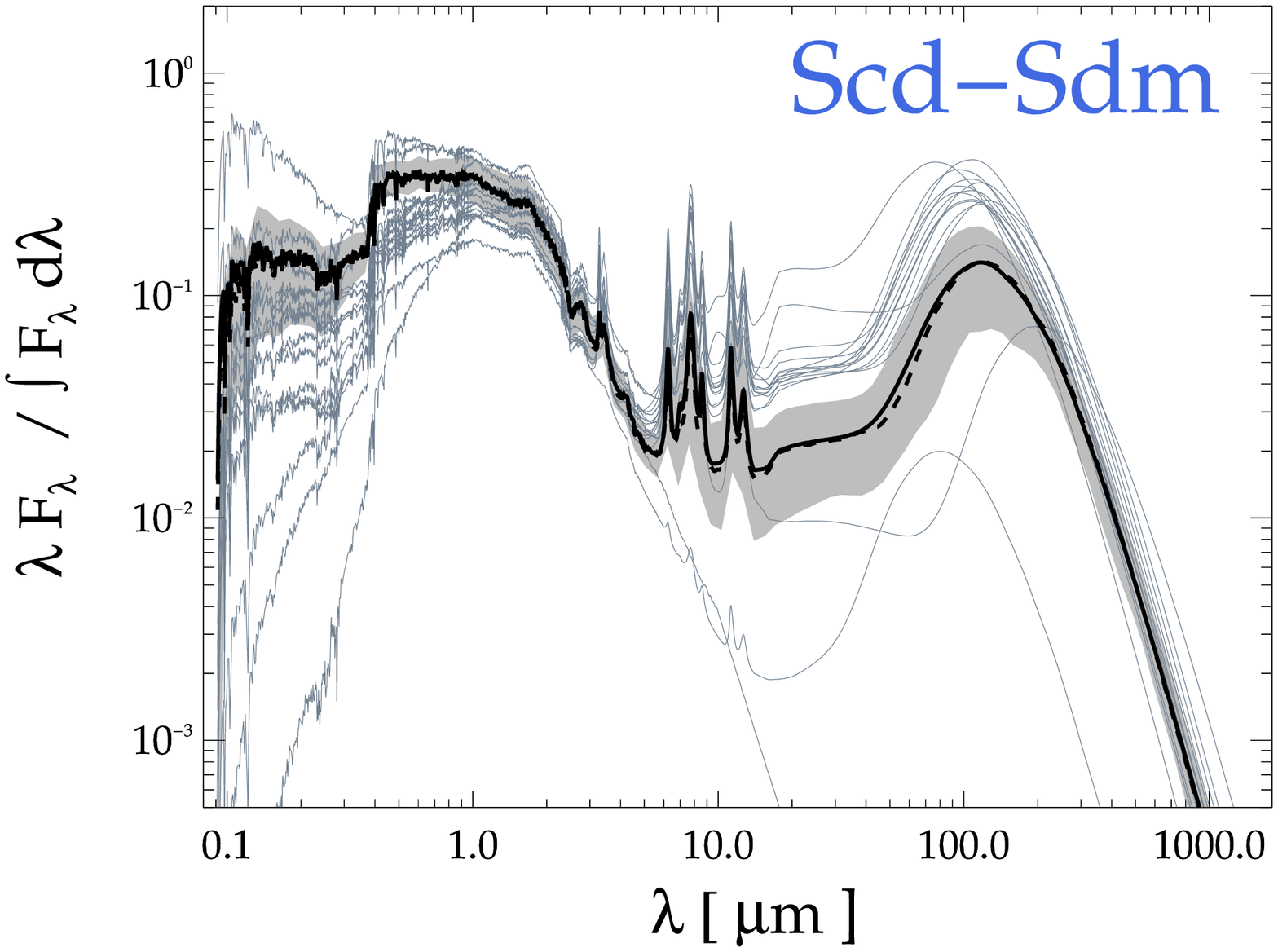}\includegraphics[scale=0.45,clip=true,trim=86 0 0 3]{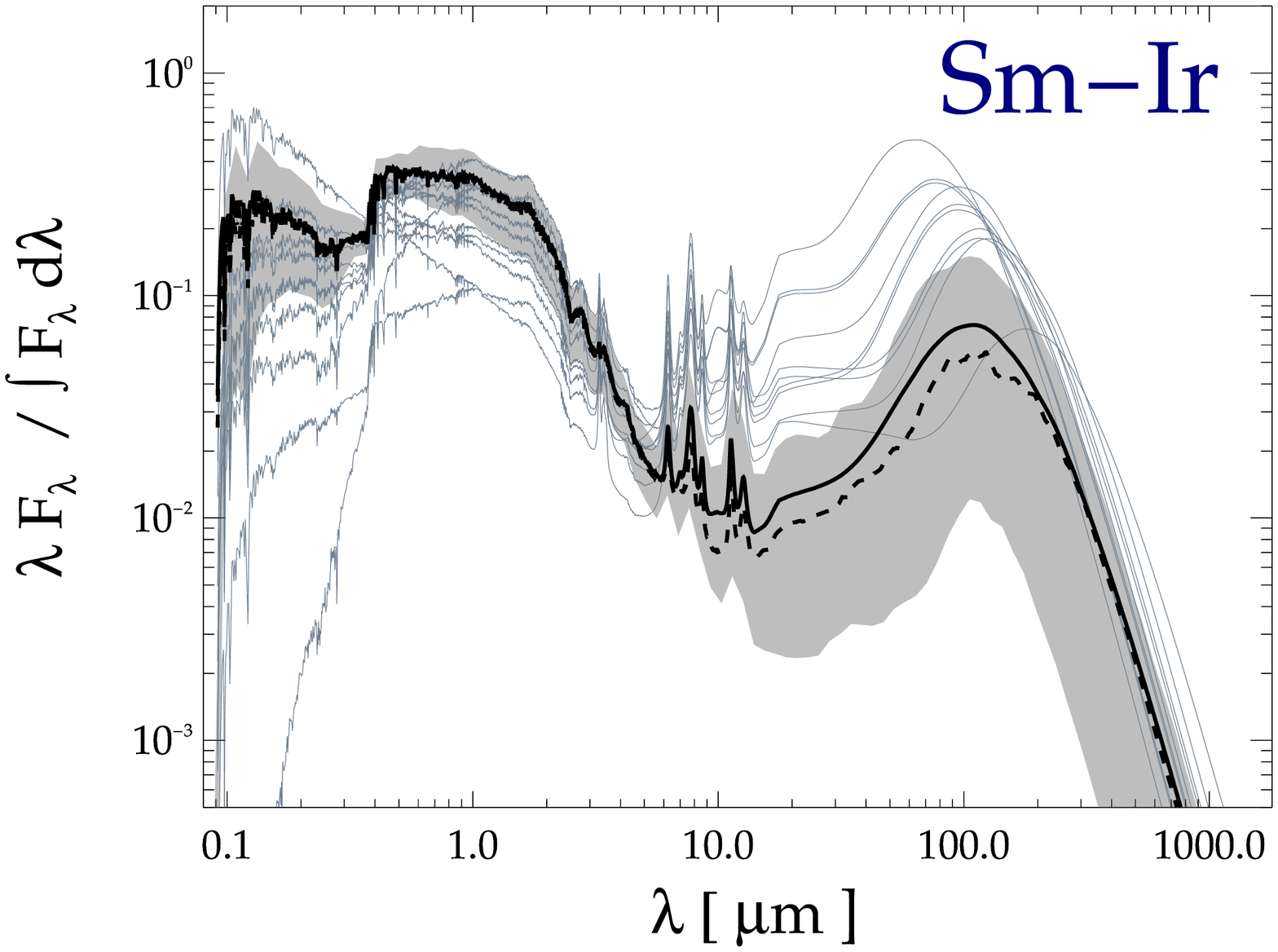}
\caption{Average (solid black lines) and median (dashed black lines) SEDs for the broad morphology bins defined in Table~\ref{tab:fabs_vs_T}. For each subsample, we excluded from the template generation the 10\% objects showing the most deviant SEDs (shown as gray lines). The shaded area delimits the 16 and 84\% percentiles.}
\label{fig:sed_T}
\end{figure*}

Several templates of the stellar and dust SEDs have  been produced in the past and used as tools to predict a galaxy's energy output and to study its evolution over cosmic time \citep[see, e.g., ][and references therein]{SmithMNRAS2012b,CieslaA&A2014}. As local benchmarks, we provide here templates based on our SED-fitting, which take advantage on the larger sample, broader spectral coverage and more homogeneous photometry of DustPedia galaxies with respect to previous work in the literature.

We derived spectral templates for the broad morphology bins of Table~\ref{tab:fabs_vs_T}, and for the bins in $L_\mathrm{bolo}$ defined in Table~\ref{tab:fabs_vs_L}; in the second case, we only use the galaxies of type later than Sb, which show a better defined trend of  $f_\mathrm{abs}$ vs $L_\mathrm{bolo}$, as we discussed in Sect.~\ref{sec:lumi}. For each galaxy subsample, templates were produced in this way: first,  the SED of each galaxy was normalised to its  $L_\mathrm{bolo}$, to have all outputs on a similar scale; second, for each of the wavelength bins of CIGALE's best fit, the mean, median and percentiles of the specific flux densities were derived.  The process was re-iterated once to remove the 10\% of the sample with the most deviant SEDs from the average\footnote{The templates are available at {\tt http://dustpedia.astro.noa.gr}.}.

\begin{figure*}
\sidecaption
\includegraphics[width=12cm]{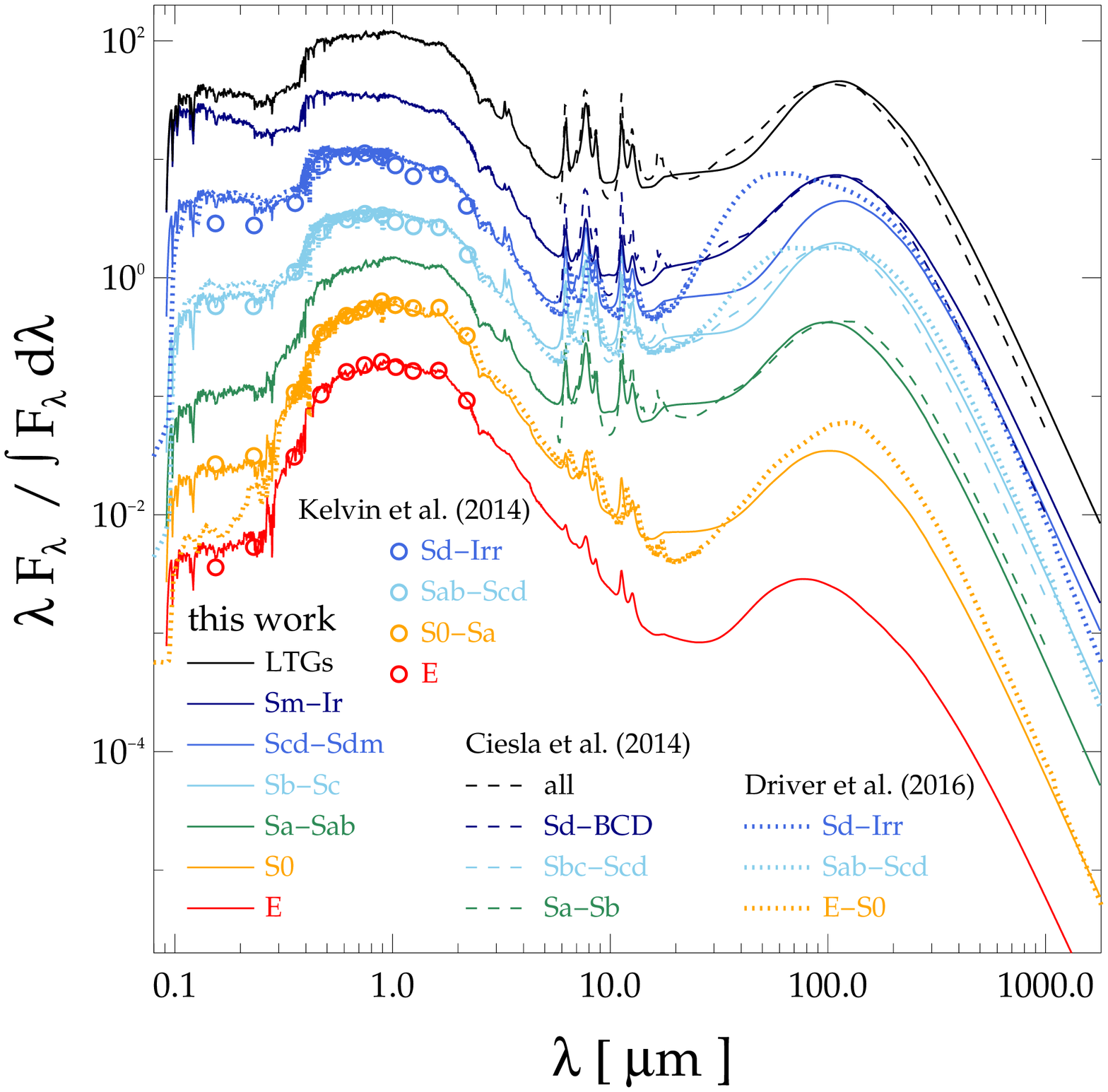}
\caption{The average templates of Fig.~\ref{fig:sed_T}, shifted by an arbitrary factor for clarity. Also plotted are the templates of \citet{CieslaA&A2014} (normalised to the same dust luminosity as ours) and the local CSEDs of \citet{KelvinMNRAS2014} and \citet{DriverMNRAS2016} (normalised to $\lambda=0.75\mu$m, i.e. the $i$-filter). We assigned the same colours to analogous morphology bins, even though the definition of the bins differs in the various references.
}
\label{fig:sed_T_compa}
\end{figure*}

Fig.~\ref{fig:sed_T} shows the average SEDs for each morphology bin. As expected from the $f_\mathrm{abs}$  analysis, the scatter around each template is large. In particular for ETGs, the average template is significantly different from the median, being biased by a relatively large number of objects with strong FIR emission. As an indication of the scatter, the panels of  Fig.~\ref{fig:sed_T} also include the individual SEDs of the 10\% of each subsample which are most deviant from the average (which were however excluded from the template creation). Apparently, the scatter around the peak of optical-NIR radiation (at $\lambda \approx 1 \mu$m) is smaller than that at the peak of FIR radiation (at $\lambda \approx 160 \mu$m): the former ranges between 10 and 40\%, for ellipticals and Sm-Ir, respectively, the latter has a minimum at about 30\% for Sb-Sc galaxies and goes over 100\% for ETGs. This might indicate that, in each morphology bin, there is a much larger variety in the dust emission spectrum than in the stellar; however, the reduced scatter in the optical results also from the normalization of each spectrum on $L_\mathrm{bolo}$, which is dominated in most cases (and in particular for ETGs), by the stellar output.

The average templates for each morphology bin  are shown together in Fig.~\ref{fig:sed_T_compa}. A clear trend is seen in the UV/optical/NIR, with the SED becoming bluer from earlier to later types. This is a well known fact \citep[see, e.g.][]{StratevaAJ2001}. In the figure we compare our templates with
the CSEDs for different morphological types derived by \citet{KelvinMNRAS2014} from 3727 galaxies ($0.025\le z \le 0.06$) in the volume-limited Galaxy And Mass Assembly (GAMA) survey. The CSEDs, obtained by integrating luminosity functions for the different Hubble types at different UV/optical/NIR bands, are very similar in shape to our determinations. This reassures us on the representativity of the DustPedia galaxies, despite the complexity of its sample definition.

In the FIR, our SEDs do not present a monotonous change with morphology, as foreseen from what is shown in Fig.~\ref{fig:fabs_vs_T}: the relative contribution of dust to the bolometric luminosity strongly increases from ETGs up to Sb-Sc; then it lowers more gradually for the later types. The peak of dust emission for ellipticals is shifted to shorter wavelengths, implying warmer dust (though we should keep in mind the skewness of the SED distribution in Fig.~\ref{fig:sed_T}). 
In analogy with Wien's displacement law and assuming modified blackbody emission and a typical MW dust absorption cross section (see Sect.~\ref{app:direct}), from the maximum of $\lambda\times F_\lambda$ we find $T_\mathrm{peak}=34$~K for ellipticals.
Bluer FIR colors and hotter dust for ellipticals have been noted earlier \citep{BoselliA&A2010,DaviesMNRAS2012,AuldMNRAS2013} and attributed to the more intense radiation field in these galaxies, or to the additional contribution of electron collisional heating in the hot gas \citep{BocchioA&A2013}. For LTGs, the peak progressively (but moderately) moves to longer wavelengths from Sa-Sab ($T_\mathrm{peak}=26$~K, similar also to that for S0) to Sb-Sc ($T_\mathrm{peak}=24$~K) and  to Scd-Sdm ($T_\mathrm{peak}=22$~K). This might reflect a dependence of the dust temperature on $L_\mathrm{bolo}$ (and thus on the intensity of the radiation field), whose average decreases going from Sa-Sab and Sb-Sc to Sc-Sdm.  A dependence of the dust temperature on the infrared luminosity had been reported earlier for galaxies in large FIR surveys \citep{SymeonidisMNRAS2013,MagnelliA&A2014}. For the earlier type spirals (and the lenticulars) the hotter dust might result from the higher intensity of the heating radiation field in the presence of a strong bulge  \citep{EngelbrachtA&A2010}. The later type bin (Sm-Ir), instead, does not follow the trend of LTGs and shows a bluer peak ($T_\mathrm{peak}=24$~K). In fact, dwarf galaxies tend to have warmer dust than more massive ones \citep[see, e.g.,][]{RemyRuyerA&A2015}.

\begin{figure*}
\center
\includegraphics[scale=0.45,clip=true,trim=0 59 0 3]{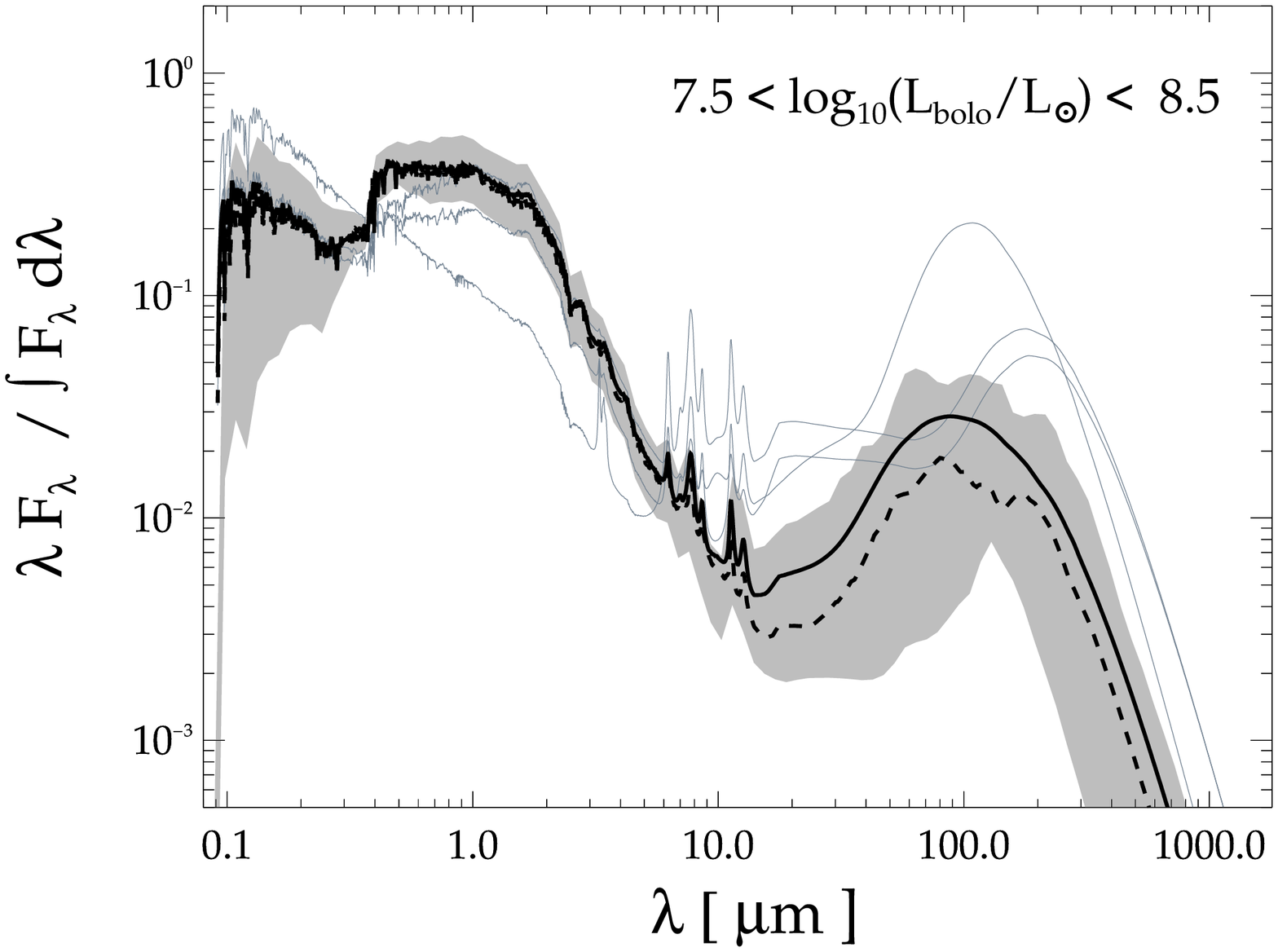}\includegraphics[scale=0.45,clip=true,trim=86 59 0 3]{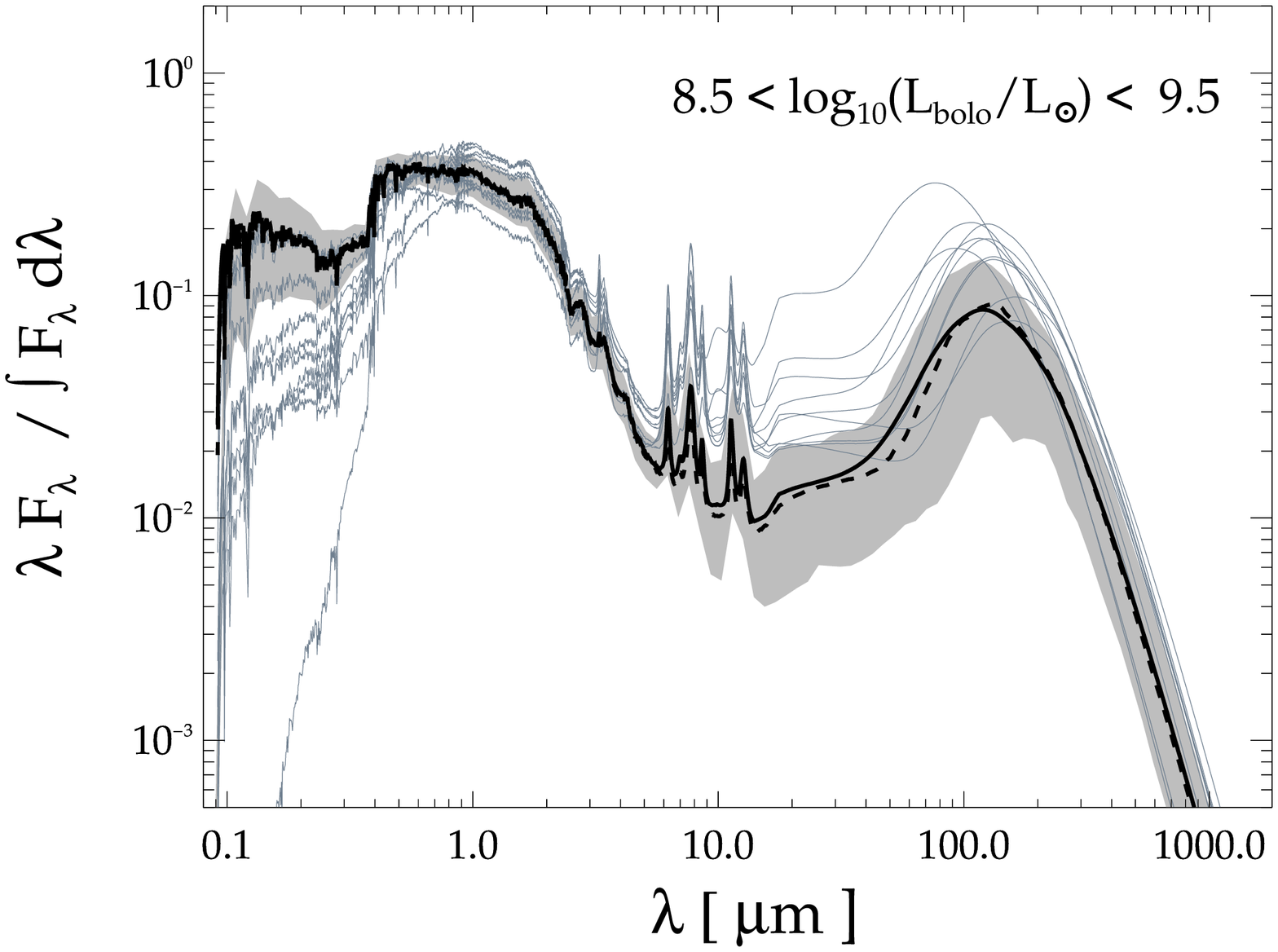}
\includegraphics[scale=0.45,clip=true,trim=0 0   0 3]{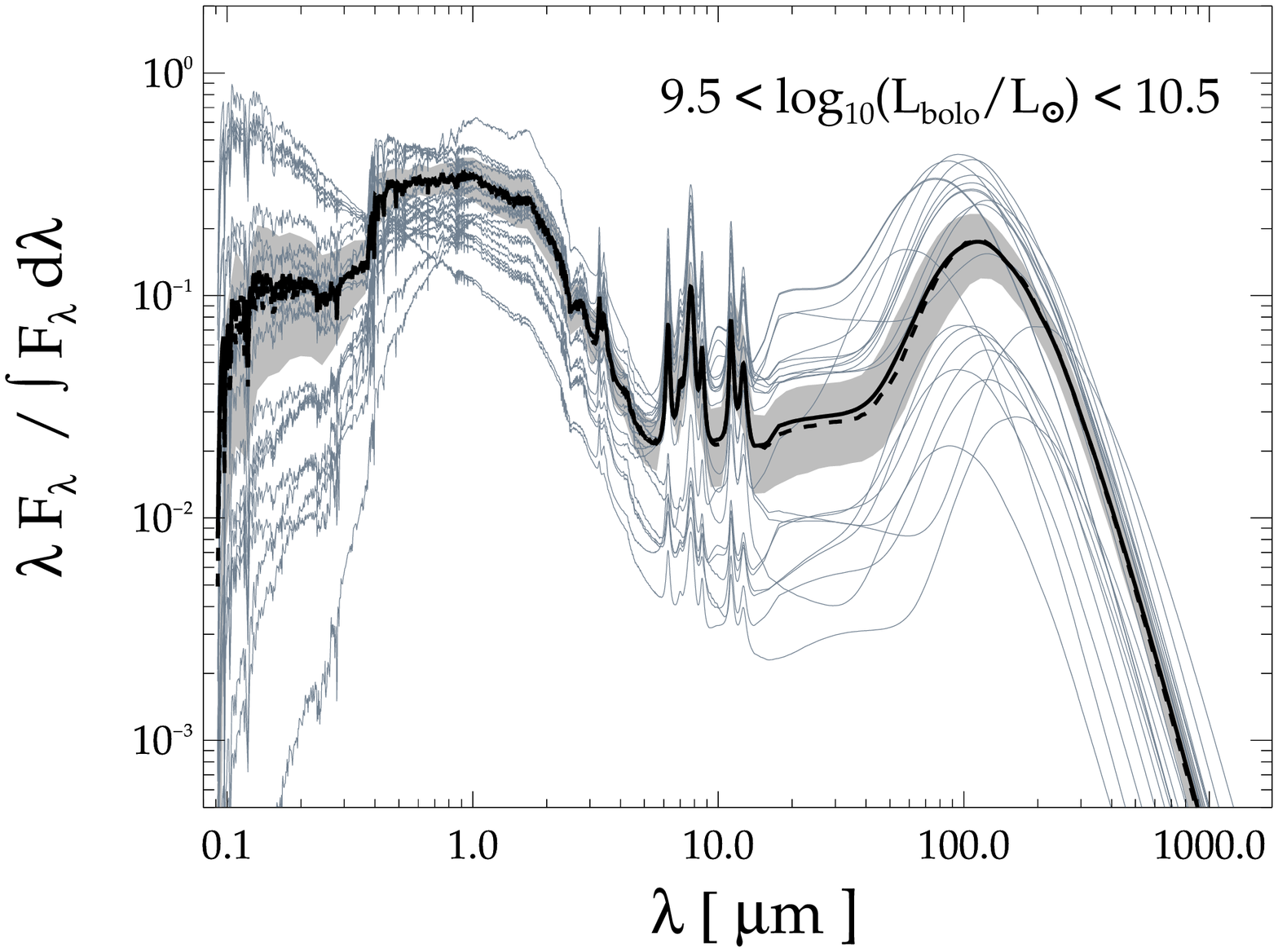}\includegraphics[scale=0.45,clip=true,trim=86 0 0 3]{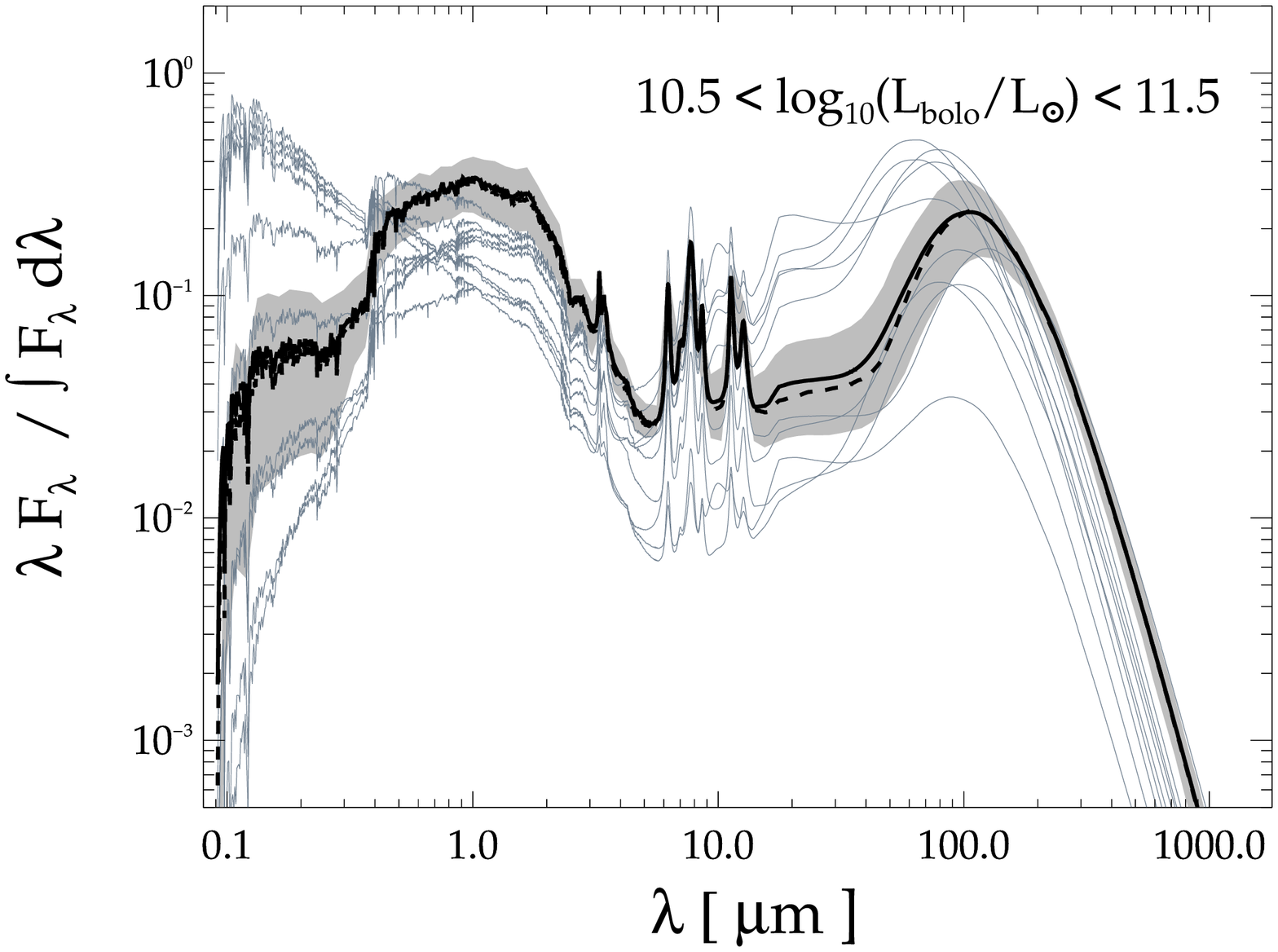}
\caption{Average (solid black lines) and median (dashed black lines) SEDs for four bins in $L_\mathrm{bolo}$ defined in Table~\ref{tab:fabs_vs_L} for galaxies later than Sb. For each subsample, we excluded from the template generation the 10\% objects showing the most deviant SEDs (shown as gray lines). The shaded area delimits the 16 and 84\% percentiles.}
\label{fig:sed_L}
\end{figure*}

We compared our templates with those derived  by \citet{CieslaA&A2014} for the gas-rich HRS galaxies. They fitted the dust emission SED only, using CIGALE and the \citet{DraineApJ2007} model. Their templates for various morphological bins are shown in Fig.~\ref{fig:sed_T_compa}, normalised to $L_\mathrm{dust}$. We also compare their average template for the full sample with an average template derived for all DustPedia LTGs (black lines). Despite the differences in the sample, photometry, and dust models, the templates are rather consistent, with little variations. In particular, the template for their earlier-type bin peaks at longer wavelengths than ours, and thus does not show the trend with the later types we discussed; however, this is probably due to their inclusion of Sb into the bin, while our analysis of the mean $f_\mathrm{abs}$ values made us conclude that objects of that morphology are similar to those of Sbc and Sc. Our templates are also consistent, within the differences in modelling and morphology bin definition, with those obtained by \citet{DriverMNRAS2016} by median stacking FUV-to-submm fits to the SED of GAMA galaxies with $z<0.06$. The major difference in this case is in the broader FIR SED and its peak at shorter wavelength; this is however due to the lack of MIR data and an unconstrained warm dust component introduced by the fitting tool they use, MAGPHYS (Sect.~\ref{sect:compa} and~\ref{app:magphys}).

\begin{figure*}
\sidecaption
\includegraphics[width=12cm]{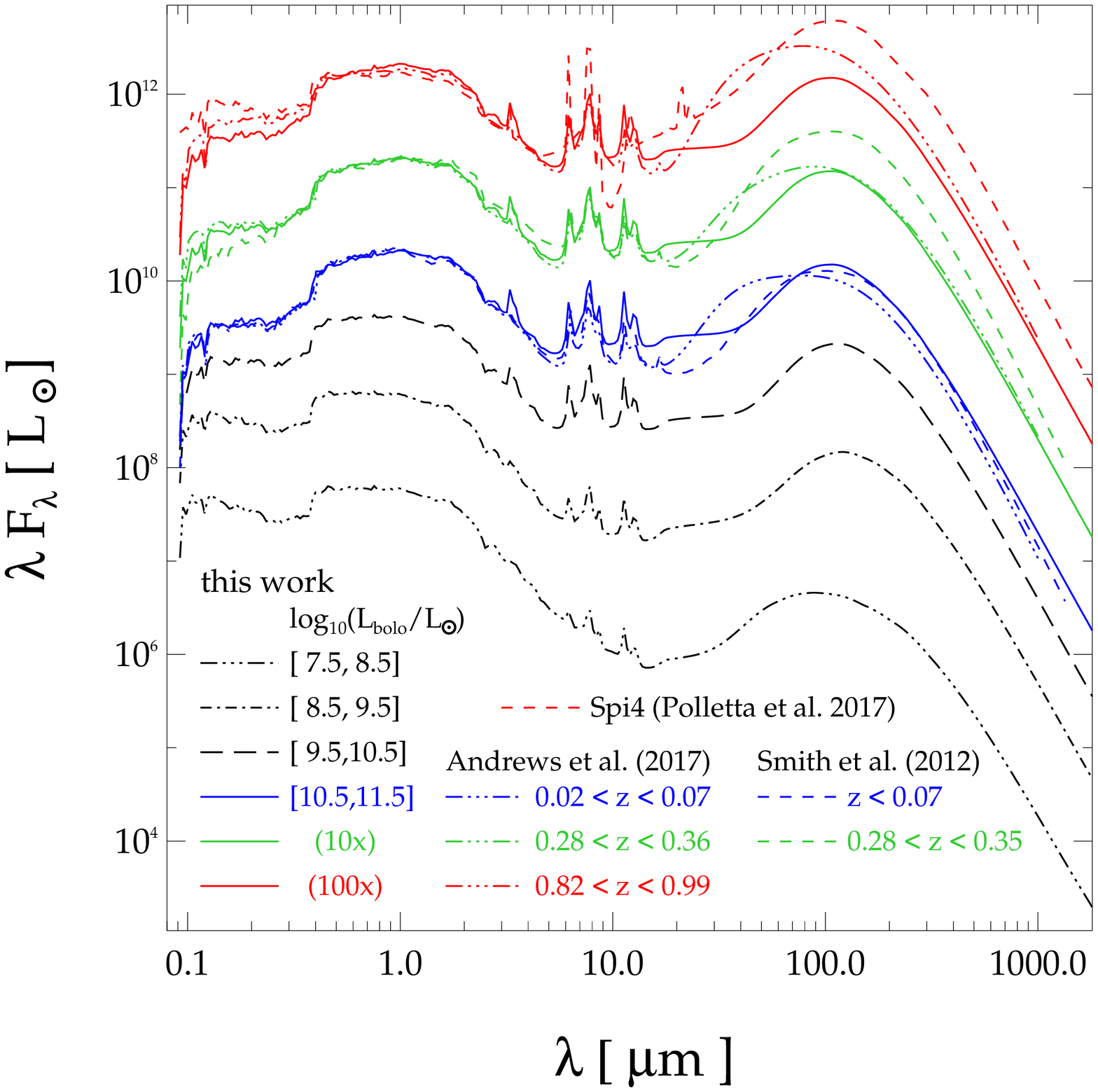}
\caption{The average templates of Fig.~\ref{fig:sed_L},  scaled to the mean $L_\mathrm{bolo}$ of each subsample. The highest-$L_\mathrm{bolo}$ template is shown three times, at the original scale (solid blue line) and multiplied by arbitrary factors (solid green and red lines). Also shown are two templates at different redshifts from \citet{SmithMNRAS2012b}, three CSEDs from \citet{AndrewsMNRAS2017}, and the Spi4 template of \citet{PollettaApJ2007}. The SEDs from the literature have been normalised to the stellar luminosity of the highest-$L_\mathrm{bolo}$ template.}
\label{fig:sed_L_compa}
\end{figure*}

We have seen in Sect.~\ref{sec:lumi} that the stronger trend we found for  $f_\mathrm{abs}$ (and thus for the shape of the SED) is that with $L_\mathrm{bolo}$ for galaxies later than Sb. Fig.~\ref{fig:sed_L} shows the templates obtained after grouping these later type galaxies in the four bolometric luminosity bins defined in Table~\ref{tab:fabs_vs_L}. With respect to the templates for morphological bins, there is a slight improvement in the scatter, in particular for the dust emission spectrum, at least for the three brighter bins. The average templates are shown together in Fig.~\ref{fig:sed_L_compa}, after scaling them for the mean $L_\mathrm{bolo}$ in each bin. The trend is clear, with templates becoming redder in the optical-UV and thus more extinguished and with more emission in the FIR as the luminosity increases. For the three higher luminosity bins, the peak of thermal emission shifts to shorter wavelengths with increasing $L_\mathrm{bolo}$, with $T_\mathrm{peak}$ increasing from 21~K to 24~K.
The average template of galaxies in the lower $L_\mathrm{bolo}$ bin, instead, show a hotter temperature ($T_\mathrm{peak} = 29$~K) and a broader FIR peak.  Again with a caveat on the larger variety of objects in this bin (shown by the larger discrepancy between mean and median SED), the template is in agreement with the findings on dwarf galaxies, where the clumpier nature of the ISM and the reduced extinction result in hotter dust and a larger range of heating conditions \citep{RemyRuyerA&A2015}.

The template SEDs binned in $L_\mathrm{bolo}$ will be compared with those from samples at higher redshift in the next section.

\section{DustPedia galaxies and the evolution of $f_\mathrm{abs}$}
\label{sec:evol}

The EBL spectrum is the summation of the SED of all galaxies, integrated over cosmic time. By considering separately the wavelength ranges dominated by stellar and dust emission, $f_\mathrm{abs} \approx 50\%$ can be estimated \citep[using, e.g.\ , the fitting models in ][]{DriverMNRAS2016,FranceschiniA&A2017}. The value is analogous to a  {\em luminosity-weighted} average $f_\mathrm{abs}$  for the whole galactic population in the Universe. The $\langle f_\mathrm{abs}\rangle$ presented so far are unweighted averages. It is $\langle f_\mathrm{abs}\rangle \approx 20\%$ for our full sample. It raises to $\approx 28\%$ if we properly compute an $L_\mathrm{bolo}$-weighted value. The difference between the result on our local, DustPedia, sample and that from the EBL 
is yet another confirmation of the strong evolution in the fraction of stellar radiation reprocessed by dust \citep[for a recent review including {\em Herschel} studies, see][]{LutzARA&A2014}.  
In this section, we compare our results for $z<0.01$ with those of surveys at higher redshifts to highlight the physical properties of the high- $f_\mathrm{abs}$ objects which are underrepresented, or missing, in DustPedia.

\subsection{A simple evolutionary scenario}
\label{sec:simple}

\begin{figure}
\center
\includegraphics[height=8.5cm,clip=true,trim=5 15 0 0]{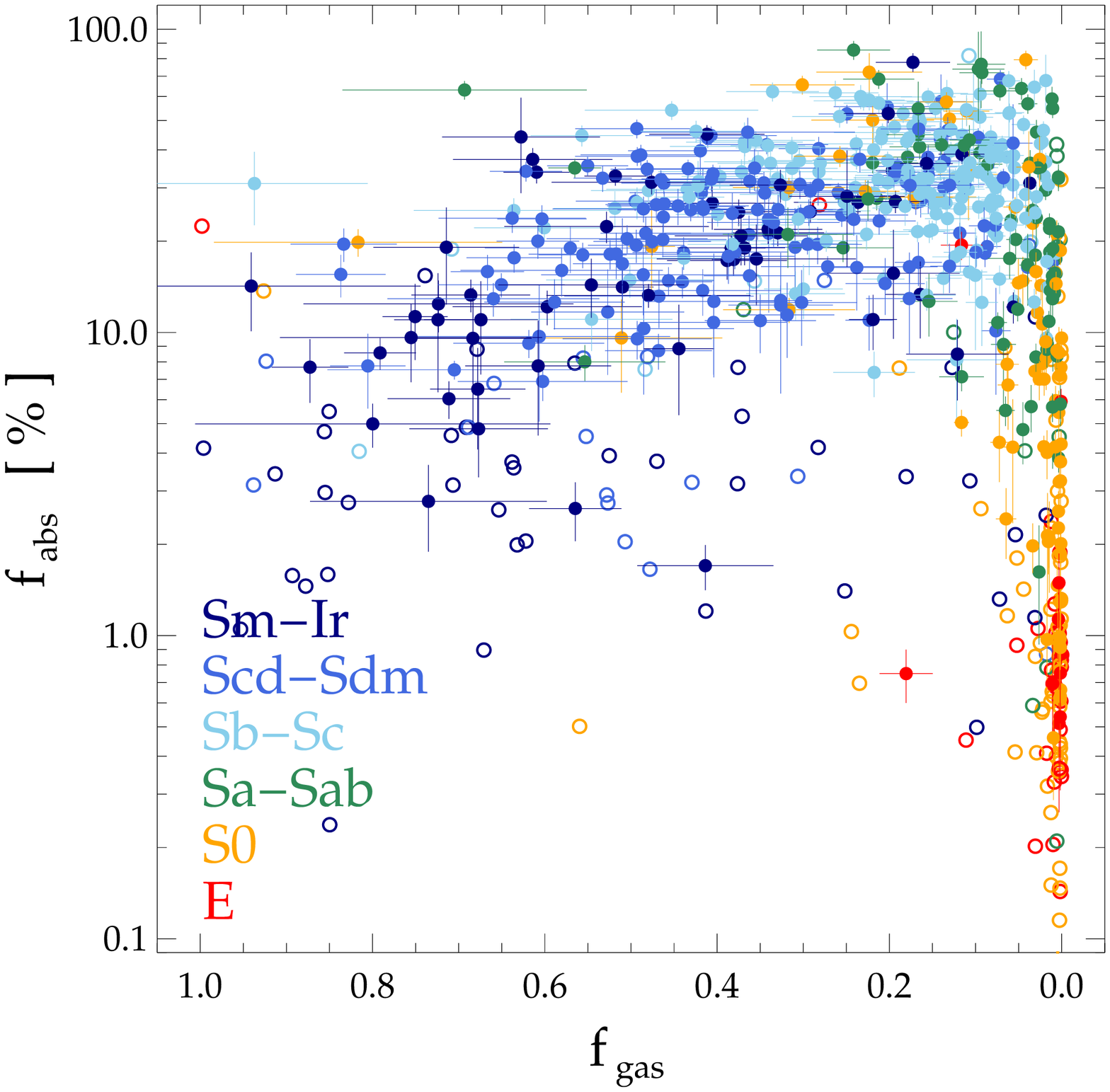}
\caption{
$f_\mathrm{abs}$ vs  $f_\mathrm{gas}$. Same symbol and colour conventions as in Fig.~\ref{fig:corr_lum}.}
\label{fig:corr_fgas}
\end{figure}

It has been shown that the evolution of the dust mass in a galaxy, relatively to the stellar \citep{CorteseA&A2012} or to the baryonic (gas+stars) mass \citep{ClarkMNRAS2015,DeVisMNRAS2017a}, can be broadly described by relatively simple analytical models, taking into account the formation of dust seeds in asymptotic giant branch stars and supernovae, their destruction along with the astration process, and possibly with the inclusion of grain growth in the ISM and gas inflows and outflows \citep{EdmundsMNRAS2001,RowlandsMNRAS2014,DeVisMNRAS2017b}. 
In these studies, the gas fraction, $f_\mathrm{gas}=M_\mathrm{\ion{H}{i}} / (M_\mathrm{star}+M_\mathrm{\ion{H}{i}})$ is used as a coarse indicator of the evolutionary stage of a galaxy. We investigate here if the evolution of the relative dust content with $f_\mathrm{gas}$ is reflected by an analogous trend for $f_\mathrm{abs}$, and if a simple evolutionary scenario can reconcile the difference between the local and EBL-derived $f_\mathrm{abs}$.

In Fig.~\ref{fig:corr_fgas}  we show $f_\mathrm{abs}$ vs $f_\mathrm{gas}$. Indeed a trend can be seen: objects where $f_\mathrm{gas}$ is high, and thus in the earlier stages of their star-formation cycles, are preferentially of later type and lower $f_\mathrm{abs}$; $f_\mathrm{abs}$ increases moving to earlier type spirals (Sb-Sc) with a smaller $f_\mathrm{gas}$. A peak in the average $f_\mathrm{abs}$  is reached at $f_\mathrm{gas} \approx 0.1$. 
When considering only the galaxies up to this limit, the (negative) correlation has  $\tau_\mathrm{K} = $-0.34.
The evolution with morphological type is also reflected by that with the  S\'ersic index (not shown): most of disk-dominated systems are along the trend, the number of bulge-dominated ones increasing as the peak is reached. Earlier types, bulge-dominated, objects have instead smaller $f_\mathrm{gas}$ and their $f_\mathrm{abs}$ dramatically reduces, though the abrupt change might be due to the neglect of molecular gas.
Since gas-rich galaxies follow a well defined trend in Fig.~\ref{fig:corr_fgas}, we used the selection $f_\mathrm{gas}> 0.1$ to try and improve the  definition of the correlation of  $f_\mathrm{abs}$ vs $L_\mathrm{bolo}$ shown in Sect.~\ref{sec:lumi}. We found $\tau_\mathrm{K} = $0.53, a value similar to those obtained by selecting galaxies later than Sb, or disk-dominated galaxies.

A similar evolution is found when studying the variation of the dust mass with respect to the baryon mass \citep{ClarkMNRAS2015,DeVisMNRAS2017a}. Objects with very high $f_\mathrm{gas}$ are found to have a much smaller dust content than that extrapolated from scaling laws derived from more evolved galaxies; they have a high SFR per unit $M_\mathrm{star}$ (specific star formation rate, sSFR); models suggest that their radiation is absorbed near the star formation sites, where the initial seeds for dust grains also form (\citealt{DeVisMNRAS2017b}; 
\citealt{RemyRuyerA&A2015}). 
As galaxies evolve and  $f_\mathrm{gas}$ decreases, the dust mass grows by accreting elements/molecules from the ISM onto the initial seed grains; the contribution of diffuse dust  to the extinction increases \citep{DeVisMNRAS2017b}. The dust mass, relative to the baryonic mass, reaches a peak at $f_\mathrm{gas}\approx 0.5$, and then decreases  when the dust destruction during the astration process dominates over the grain production/accretion  \citep{ClarkMNRAS2015,DeVisMNRAS2017a}. 

The delayed peak of the $f_\mathrm{abs}$ evolution at $f_\mathrm{gas}\approx 0.1$ reflects the correlation of $f_\mathrm{abs}$ with $L_\mathrm{bolo}$ and must be due to the dependence of absorption on the absolute dust content (which is higher for higher $L_\mathrm{bolo}$), and on its distribution with respect to the stars. Also, in high-$L_\mathrm{bolo}$, high-$f_\mathrm{abs}$ galaxies, the dust/star geometry might be more effective in absorption: an increasing contribution to absorption of the more pervasive diffuse medium is seen for objects with lower sSFR \citep{DeVisMNRAS2017a} which correspond, in our sample, to the more evolved objects.

Within this simple evolutionary picture, the precursor of current ETGs with $f_\mathrm{gas}\le 0.1$ must have passed  through a phase which considerably reduced their dust mass. As a numerical experiment, we tried to simulate the conditions {\em a step earlier} in galaxy evolution by assigning to each of the objects with $f_\mathrm{gas}\le 0.1$ a  value of  $f_\mathrm{abs}$ derived 
{
from the galaxies with $f_\mathrm{gas}>0.1$, assuming the trends of $f_\mathrm{abs}$ with either $f_\mathrm{gas}$ or $L_\mathrm{bolo}$ (and their scatter).
After the insertion of these mock values, the {\em luminosity-weighted} $\langle f_\mathrm{abs}\rangle$ of the sample rises from $\approx$28\% to 40-45\%. }
Thus, while a simple monolithical scenario can in principle describe the evolution in the dust content of galaxies, it still falls short in predicting the evolution of $f_\mathrm{abs}$:
a passive backward evolution from the DustPedia local sample does not explain the higher  $f_\mathrm{abs}$ level implied by the EBL spectrum.

\subsection{$f_\mathrm{abs}$, $M_\mathrm{star}$  and sSFR for local and high-z objects}
\label{sect:mstar}

Current models agree in assigning to star-forming disk galaxies the role of major contributors to the EBL; these objects undergo star-formation events driven by the infall of cold gas and dominates the star-formation history and infrared luminosity density for $z<1$; a minor contribution is expected from spheroids, whose peak in star-formation evolution - driven by merging events - occurs at higher redshifts \citep{FranceschiniA&A2008,FranceschiniA&A2010,DominguezMNRAS2011,DriverMNRAS2013,FranceschiniA&A2017,AndrewsMNRAS2018}. 

Large-area multi-wavelength surveys including the FIR/submm, such as  {\em Herschel}-ATLAS \citep{EalesPASP2010},  can be used to study the dependence of $f_\mathrm{abs}$ on other physical properties and  its evolution. \citet{SmithMNRAS2012b} produced MAGPHYS fits to {\em H}-ATLAS extragalactic point sources detected at 250 $\mu$m and derived template SEDs by median stacking. After the original publication, templates have been updated to include  the successive releases of  {\em H}-ATLAS data, totalling up to more than 12000 objects with redshift $z<0.35$ (D. J. B. Smith, private communications; see also \citealt{RowlandsMNRAS2014}).  By binning SEDs according to $M_\mathrm{stars}$ and sSFR, \citet{SmithMNRAS2012b} find that the SED has a weak dependence on the former, while the latter governs the most striking changes, with an increasing contribution of dust emission for larger sSFRs. We used their templates to derive $f_\mathrm{abs}$ and compare them to the values for the DustPedia sample.

\begin{figure*}
\center
\includegraphics[height=8.5cm,clip=true,trim=5 15 0 0]{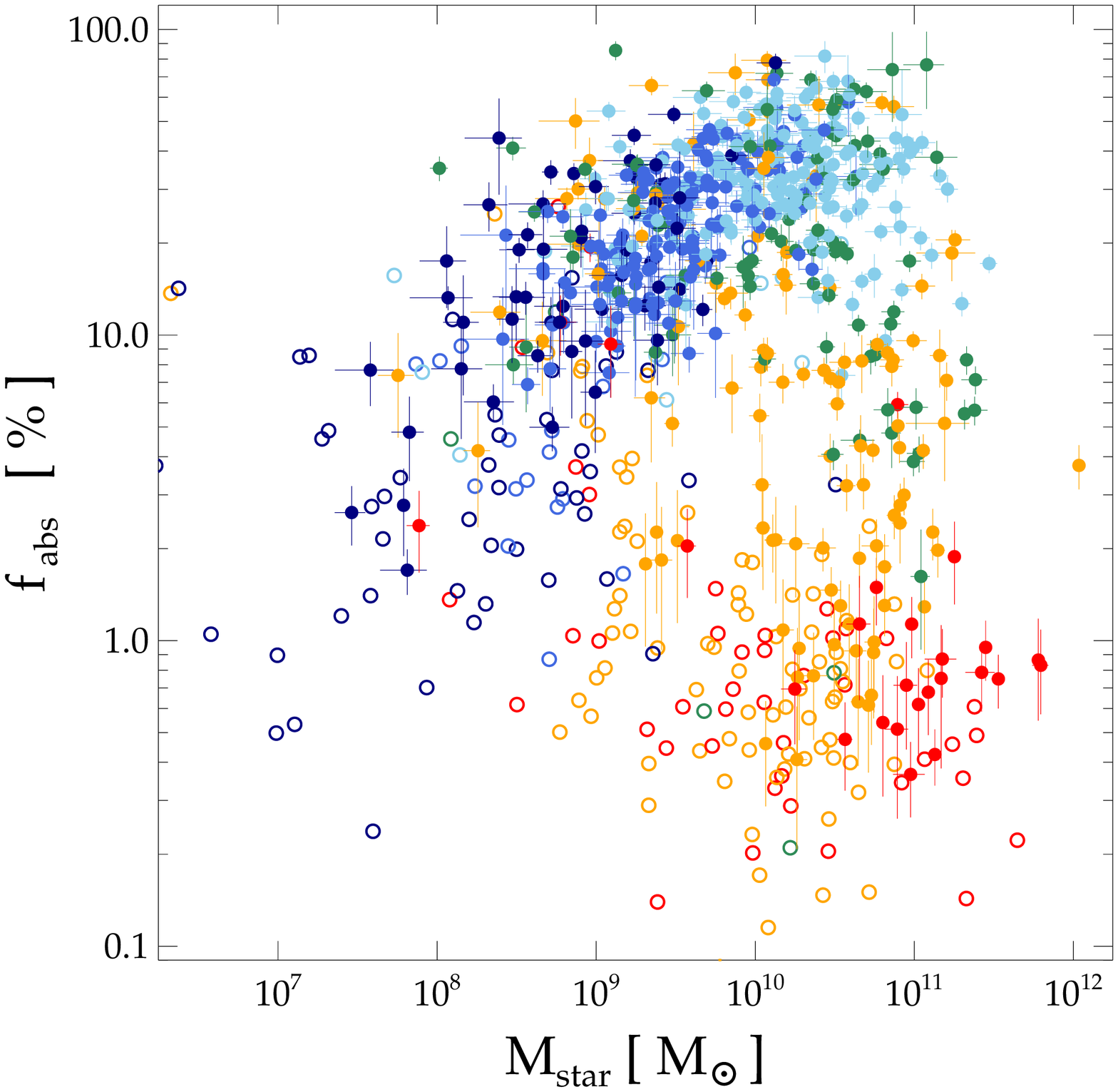}\includegraphics[height=8.5cm,clip=true,trim=82 15 0 0]{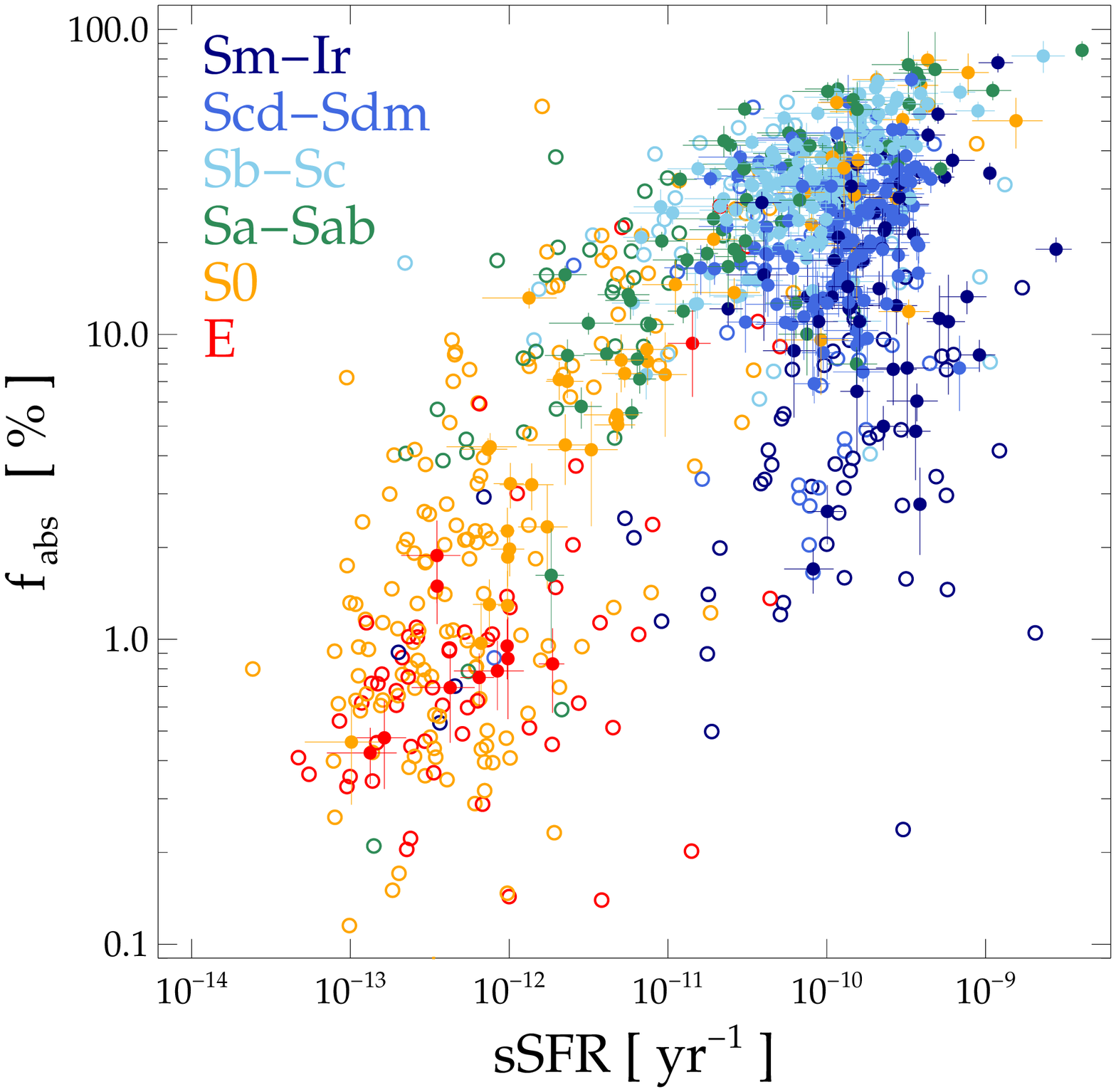}
\caption{
$f_\mathrm{abs}$ vs  $M_\mathrm{star}$ (left panel) and specific SFR (right panel). Same symbol and colour conventions as in Fig.~\ref{fig:corr_lum}.}
\label{fig:corr_mod}
\label{fig:corr_col}
\end{figure*}

The variation of $f_\mathrm{abs}$ as a function of  $M_\mathrm{star}$ and sSFR for DustPedia galaxies are shown in  Fig.~\ref{fig:corr_mod}.
Not surprisingly, the trend for $M_\mathrm{star}$ shown in the left plot is similar to that for $L_\mathrm{bolo}$ in Fig.~\ref{fig:corr_lum}, displaying a correlation for late-types  ($\tau_\mathrm{K} = $ 0.47 for objects later than Sb). 
A correlation of FIR/optical luminosity ratio with rotational velocity, and thus galactic mass, was already noted by \citet{WangApJ1996} using IRAS data.
\citet{SkibbaApJ2011} found a correlation with $M_\mathrm{star}$ for the KINGFISH late and dwarf galaxies. Instead,
\citet{ViaeneA&A2016} find none for HRS LTGs. The reason for this is in the smaller dynamic range in $M_\mathrm{stars}$ for HRS (complete for 
$M_\mathrm{stars} \ga 8\times10^8 M_\odot$; \citealt{EalesMNRAS2017}).
Indeed, if we select LTGs in common with HRS, we obtain $\tau_\mathrm{K} = 0.06$ (the only estimate, among those presented here, with a non-null probability that the data are uncorrelated, 22\%). 
Instead, the KINGFISH sample, despite the smaller number of objects, has a sufficient variety as to include several lower luminosity later-type galaxies. 

In the right panel of Fig.~\ref{fig:corr_col} we show $f_\mathrm{abs}$ vs the sSFR. 
There is a clear correlation between the two quantities for ETGs and spirals earlier than Scd, with $\tau_\mathrm{K} = $0.66. 
The correlation degrades to $\tau_\mathrm{K} = $0.48 when including the later type spirals and in particular the Sm-Ir galaxies, several of which show lower $f_\mathrm{abs}$ for their high sSFR. 
The correlation was noted by both \citet{SkibbaApJ2011}  and \citet{ViaeneA&A2016}, though the latter missed the deviant behaviour of the dwarf galaxies, again because of the HRS flux 
limits\footnote{These galaxies align with the rest of LTGs (and with ETGs) if the SFR is normalised by the amount of gas available for star formation (i.e.\ a {\em gas} sSFR = SFR/$M_\mathrm{\ion{H}{i}}$).  
Sm-Ir have a smaller {\em gas}-sSFR than the rest of LTGs (but higher than ETGs) and $\tau_\mathrm{K} = $0.49 for the full sample. We also experimented other correlations, such as $f_\mathrm{abs}$ vs $M_\mathrm{dust}$ ($\tau_\mathrm{K} = 0.48$ for types later than Sb).
}. 
Motivated by Fig.~\ref{fig:corr_mod}, for objects with sSFR $\ge 10^{-11}$ yr$^{-1}$ we derived $\tau_\mathrm{K} = $0.53 for $f_\mathrm{abs}$ vs $L_\mathrm{bolo}$, not significantly different than the correlations obtained with the other selections discussed so far.

The templates of \citet{SmithMNRAS2012b} do not show strong variations of $f_\mathrm{abs}$ with $M_\mathrm{star}$ until the largest masses, probably because of a selection bias (see later) and the reduced dynamic range  in  $M_\mathrm{stars}$ (about 2 orders of magnitude). For $10.5 \le \log(M_\mathrm{stars}/M_\odot) < 11 $, they have $f_\mathrm{abs} \approx 50\%$, considerably higher than what found in DustPedia for the same range, 33$\pm$3 \% (here and in the following,  we select objects later than Sb). This already implies an evolution in their sample, reaching $z\approx 0.35$. For the largest masses, $\log(M_\mathrm{stars}/M_\odot)\ge 11$ , their  $f_\mathrm{abs}$ reduces to $\approx$ 40\%. The reduction could be caused by the inclusion in the stacking of earlier type objects, which in DustPedia dominate over the range and have lower $f_\mathrm{abs}$.

As anticipated, the templates of \citet{SmithMNRAS2012b}  binned in sSFR show strong changes in $f_\mathrm{abs}$, from $\approx$ 30\% for objects with $\log(\mathrm{sSFR}/\mathrm{yr}^{-1}) < -10.5$ to $\ga 60\%$ for $\log(\mathrm{sSFR}/\mathrm{yr}^{-1})\ge -9.5$. In DustPedia, instead, we find $f_\mathrm{abs} = 28\pm 3\%$ for galaxies with  $-9.5 \le \log(\mathrm{sSFR}/\mathrm{yr}^{-1}) <-9$. The value is biased low by the presence of the low-$f_\mathrm{abs}$, high-sSFR, Sm-Ir objects. Values for $f_\mathrm{abs}\ga 60\%$ can also be obtained for $\log(L_\mathrm{dust}/L_\odot) \ge 11$
(from the templates of \citealt{SmithMNRAS2012b}  binned in this quantity). 
Actively star forming galaxies with these characteristics could align on the higher ends of the $f_\mathrm{abs}$ vs $L_\mathrm{dust}$ (or $L_\mathrm{bolo}$) trends of Fig.~\ref{fig:corr_lum}, and of the $f_\mathrm{abs}$ vs sSFR trends of  Fig.~\ref{fig:corr_col}. 
With a presumably high $f_\mathrm{gas}$, their position is instead expected to shift upward in the trend of Fig.~\ref{fig:corr_fgas}. These galaxies are under-represented (or absent) in DustPedia but are necessary to match the energy budget of the EBL.

The evolution with redshift is clearly revealed by comparing the templates of \citet{SmithMNRAS2012b} for local galaxies with $z<0.07$ and those for their highest redshift bin, $0.28<z<0.35$. They are shown in Fig.~\ref{fig:sed_L_compa} together with our templates binned in   $L_\mathrm{bolo}$.  If we exclude the unconstrained MIR range, the shape of their local template is remarkably close to that of our brightest bin, while that at higher redshift has a higher (and hotter) peak of thermal radiation; between these two redshifts, $f_\mathrm{abs}$ increases from $\approx40\%$ to $60\%$. The templates by \citet{SmithMNRAS2012b} might be biased to higher $f_\mathrm{abs}$ because of the  selection at 250 $\mu$m. Indeed fainter levels of FIR emission, for similar redshift bins, characterise the CSEDs of \citet{AndrewsMNRAS2017};  they were derived by summing-up MAGPHYS fits of about 300000 galaxies in the GAMA and COSMOS surveys, including objects detected and undetected in the FIR. Yet a progressive increase of $f_\mathrm{abs}$ with $z$ can be seen, from 38\% for $0.02<z<0.07$ (close to our brightest template, though again excluding the unconstrained MIR in their fits) to 45\% for  $0.28<z<0.36$, up to 60\% for $0.82<z<0.99$ (see Fig.~\ref{fig:sed_L_compa}).

As we have discussed, the increase in $f_\mathrm{abs}$ with $z$ might result from a larger fraction of luminous, star-forming galaxies - which are uncommon in DustPedia; this is consistent with the cosmic evolution of the SFR density and with the requirements set by the EBL. Apparently at odd with this, \citet{DominguezMNRAS2011} are able to predict the EBL with a sizeable contribution from normal star-forming galaxies. They use the pre-{\em Herschel} templates for local objects of \citet{PollettaApJ2007}\footnote{available at {\tt http://www.iasf-milano.inaf.it/ $\sim$polletta/templates/swire\_templates.html}}, generated by the spectral evolution code GRASIL \citep{SilvaApJprep1998}.
The templates are used to fit UV to MIR SEDs in a sample of $0.2<z<1$ galaxies, and to predict their FIR; most of the objects are fitted by templates for spirals, among which the most successful is one named {\tt spi4} (see Fig.~\ref{fig:sed_L_compa}). However, that template is not really representative of the galaxies sampled in DustPedia, since it has $f_\mathrm{abs}\approx 70\%$. Such high values are relatively rare in DustPedia, as there are just 14 galaxies with $f_\mathrm{abs} \ge 60\%$. Among them is the prototypical starburst M82 (NGC~3034) for which we measured $f_\mathrm{abs} = 68\pm5\%$.
This is a further confirmation that the EBL is dominated by objects under-represented in our analysis. 

Yet these $z<1$ objects do not represent the most extreme cases. Though not the dominant contributors to the local FIR background, submm galaxies (thought to be the precursors of local spheroid) have even higher values of sSFR and FIR luminosities, with  almost all of their radiation processed by dust: from the median template of submm-galaxies with  $z>1$ derived by \citet[][ see their Fig.~4]{RowlandsMNRAS2014},  we estimated $f_\mathrm{abs}\approx 95\%$.
  
\section{Summary and conclusions}
\label{sec:summary}

We used the code CIGALE to estimate the dust and bolometric luminosities, $L_\mathrm{dust}$ and $L_\mathrm{bolo}$, for 814 galaxies in the local Universe. The objects of our study constitute the vast majority (93\%) of the DustPedia sample, for which an extensive coverage of the SED from the UV to the submm is available. From the luminosities we have derived the fraction of radiation absorbed by dust, $f_\mathrm{abs}$, and studied its dependence on a galaxy's morphological type, luminosity and other physical quantities. From the SED fits, we obtained averaged templates as a function of morphology and luminosity. Our main findings are:

\begin{itemize}

\item For the full sample,  $\langle f_\mathrm{abs} \rangle \approx 19\%$, rising to 25\% if only LTGs are considered. The LTG average is in line with previous determinations, though somewhat smaller. We argue that this is the result of the larger number and diversity of objects in DustPedia.

\item A mild correlation is found between $f_\mathrm{abs}$  and $L_\mathrm{bolo}$, for galaxies of type later than Sb, disk-dominated, gas-rich and with high specific star-formation rates. None of these specifications resulted in a significant improvement in the correlation: the scatter is large and the Kendall's correlation measure $\tau_K \approx 0.5$ in all cases. A fit for objects later than Sb yields $f_\mathrm{abs}\sim L_\mathrm{bolo}^{1/3}$. Most of spheroid-dominated LTGs and of Sa-Sab galaxies align on the same trend, as well as about a quarter of lenticulars. 

\item Similar correlations are seen between $f_\mathrm{abs}$ and other physical quantities estimated by CIGALE, such as the mass of stars and the specific star-formation rate. Within our sample, a larger fraction of radiation is absorbed in LTGs where most of the build-up of stars and dust, and of the consumption of \ion{H}{i} (or conversion into H$_2$), has occurred.

\item No apparent change in $f_\mathrm{abs}$ with inclination is found, contrary to the predictions of RT models of edge-on galaxies.

\item While $f_\mathrm{abs}$ is higher for the higher luminosity LTGs in the sample ($\langle f_\mathrm{abs} \rangle \approx 39\%$ for $L_\mathrm{bolo}\approx 10^{11} L_\odot)$, the fraction of absorbed radiation is not sufficient to explain the energy budget of the EBL, which requires $\langle f_\mathrm{abs} \rangle \approx 50\%$. 
DustPedia is missing the high luminosity and specific-star-formation-rate galaxies found in higher redshift samples, in agreement with the evolutionary picture that has emerged from observations and EBL modelling.

\end{itemize}

While it is not unexpected that intrinsically brighter galaxies have a larger dust content, it remains unclear why the fraction of absorbed radiation scales with the bolometric luminosity. Certainly,  $f_\mathrm{abs}$ must depend on several galactic properties: the star-formation history, the dust mass,  the geometry of dust and stars, the relative proportion of radiation emitted in different environments, the optical properties of grains in each of them, to name a few. Indeed the scatter is large, yet correlations are clearly detected in our analysis. Their explanation requires a study of RT coupled with the evolution of a galaxy stellar, gas and dust content and structure.

Insights on these topics could be gained from RT models of large numbers of objects from cosmological simulations.  Recently, \citet{CampsMNRAS2016} have used the code SKIRT to include the effects of dust in present-day mock galaxies from the Evolution and Assembly of GaLaxies and their Environments (EAGLE; \citealt{SchayeMNRAS2015}) cosmological simulations; the general characteristics of local samples have been reproduced, both for what concern dust-extinguished optical radiation \citep{TrayfordMNRAS2017} and dust infrared emission \citep{CampsMNRAS2016}. Post-processed simulations for half a million EAGLE galaxies up to $z = 6$ have been made available by \citet{CampsApJS2018}. An analysis of mock SEDs from that database along the lines of what has been done in the current work is yielding promising results: the trend  $f_\mathrm{abs}$ vs $L_\mathrm{bolo}$ seen in DustPedia galaxies is reproduced (Tr\v{c}ka et al. in prep.). Hopefully, the detailed knowledge of the input parameters will help in understanding the key physical properties drawing the correlation, and in studying the evolution of dust absorption/emission with cosmic time.

For a few open questions, the current resolution of large-volume hydrodynamical simulations might still be insufficient to describe the full range of ISM scales (such as the thickness of LTG disks;  \citealt{TrayfordMNRAS2017});  in particular, a higher resolution might be needed to understand the apparent lack of anisotropy we found for $f_\mathrm{abs}$, which could result from local dust extinction and emission at the scale of star-formation regions. In an exploratory study, \citet{SaftlyA&A2015} analysed a few hydrodynamical simulations of MW class objects and  concluded that large-scale structure, such as spiral arms, might concur in {\em hiding} a considerable fraction of the dust mass from simple-geometry RT fits, an effect which is generally imputed to small-scale clumps \citep{PopescuA&A2000,BianchiA&A2008}. Modern higher resolution cosmological zoom-in simulations, such as those in the  Numerical Investigations of Hundred Astrophysical Objects \citep[NIHAO;][]{WangMNRAS2015}, or {\em Auriga} \citep{GrandMNRAS2017}, to name a few,  are already providing the ground to try to settle these issues and assess the combined effect of both small and large scales on a galaxy's dust emission and energy budget.

\begin{acknowledgements}
We thank M. Boquien and L. Ciesla for several discussions and help on the use of CIGALE, and an anonymous referee for helpful comments.
IDL gratefully acknowledge the support of the Research Foundation - Flanders (FWO Vlaanderen).
\end{acknowledgements}

\bibliographystyle{aa} 
\bibliography{/Users/sbianchi/Documents/tex/DUST}

\begin{appendix}

\section{Comparison between different $f_\mathrm{abs}$ estimates}
\label{app:compa}

\begin{figure*}
\center
\includegraphics[width=\hsize]{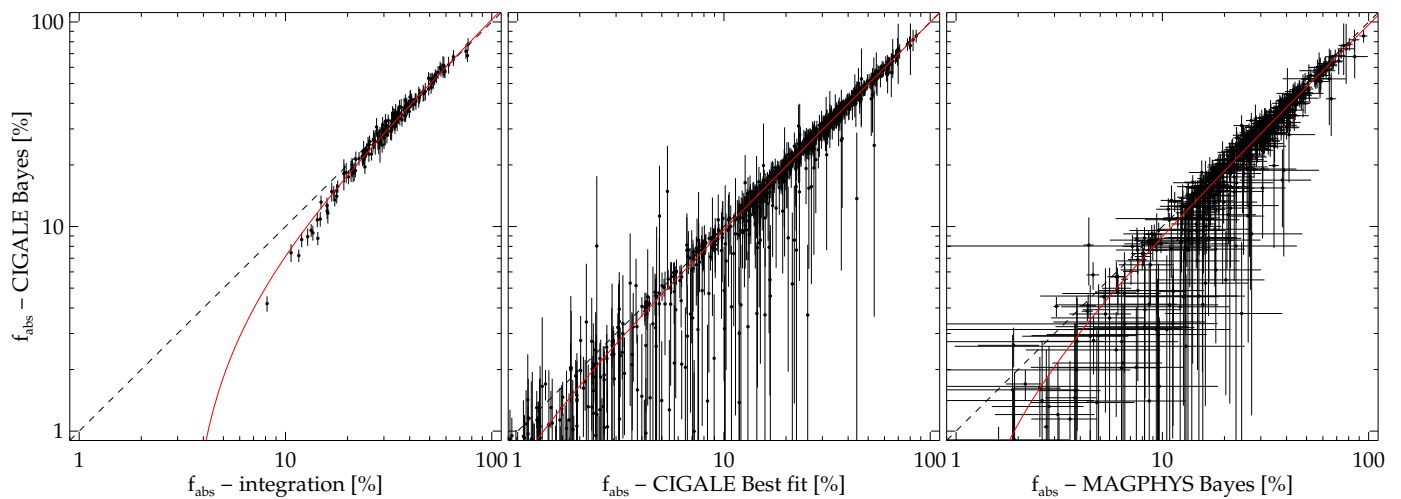}
\caption{Comparison between the $f_\mathrm{abs}$ estimated used in the main text and alternative estimates for the quantity. The dashed line shown the 1:1 relation. The red line is a linear fit between the estimates.}
\label{fig:corr_panel}
\end{figure*}

As described in Sect.~\ref{sec:cigale}, in this work we derive $f_\mathrm{abs}$ from the bayesian estimates for $L_\mathrm{bolo}$ and $L_\mathrm{dust}$ given by CIGALE. In CIGALE,  $L_\mathrm{bolo}$ and $L_\mathrm{dust}$ are computed for each model by integrating the SED over the entire wavelength grid, though we set as a lower limit $\lambda=0.0912$ $\mu$m (to exclude ionizing radiation); also, for the modules we used, emission at $\lambda> 1000$ $\mu\mathrm{m}$ constitutes a negligible contribution to the luminosities. The bayesian estimates are derived from all the models, while {\em best-fit} luminosities are derived from the model which minimise the $\chi^2$.

At earlier stages we have experimented a variety of other methods: a direct interpolation through the photometric datapoints; the use of CIGALE's best-fit luminosities; an independent estimate of the parameters using the MAGPHYS fitting code. We report here briefly on the main differences between these approaches. The three panels of Fig.~\ref{fig:corr_panel} show the comparison between the reference estimate of $f_\mathrm{abs}$ (on the y-axis) and each of alternative ones (on the x-axis), together with a linear fit to them.

\subsection{Direct integration}
\label{app:direct}
When the wavelength range dominated by non-ionizing stellar and dust radiation ($0.0912 < \lambda/\mu\mathrm{m} <1000$) is covered uniformly by a good number of good-quality photometric datapoints, it is possible to construct a  piecewise representation of the SED. $L_\mathrm{bolo}$ can be derived by integrating this SED over the whole wavelength range. For the estimate of $L_\mathrm{dust}$ one can assume as a lower integration limit a NIR wavelength where the dust emission starts to dominate over the stellar emission, or include  a correction to take into account the (declining) stellar contribution at longer wavelengths. Here we use the former method and derive $L_\mathrm{dust}$ by integrating the SED from the  WISE 3.4 $\mu$m band. The latter method was used in the analogous estimates by \citet{SkibbaApJ2011} on the KINGFISH sample. We have used a subsample of DustPedia galaxies for which photometry is available at $> 2\sigma$ level in all GALEX, SDSS, 2MASS, WISE and SPIRE wavebands, and in at least two PACS wavebands. Only 155 galaxies have at least this coverage. We extrapolated the SED to 912\AA\ by assuming $\lambda \times F_\lambda$ = constant for $0.0912 < \lambda / \mu\mathrm{m} < 0.15$, and to $1000  \mu$m, assuming a modified black-body of $T=20 K$ and an absorption cross section  $\kappa_\lambda \propto \lambda^{-\beta}$ with $\beta=1.6$ \citep[typical values for the diffuse cirrus in the MW; see, e.g., ][]{BianchiA&A2017}. These extrapolations, and in particular the extension down to the Lyman break, change the estimate of $f_\mathrm{abs}$ by less than 1\%. Of similar order are the effects on the choice of the limit wavelength in the estimate of $L_\mathrm{dust}$:  if instead of a sharp cut the stellar contribution at $\lambda \ge 3.4 \mu$m is assumed to follow a Rayleigh-Jeans spectrum,   $f_\mathrm{abs}$ changes by about 1.0 \%. 

The comparison between this method and the reference one is shown in the left panel of Fig.~\ref{fig:corr_panel}. The fit yields $f_\mathrm{abs} (\mathrm{bayes}) = 1.06 \times f_\mathrm{abs} (\mathrm{integration}) -3.5 \%$. For $f_\mathrm{abs} >20\%$ (integration) the agreement is good and within the average CIGALE error provided by the bayesian analysis, 3\% for the same sample of galaxies. For $f_\mathrm{abs} < 20\%$, the direct integration of the piecewise SED leads to higher values of $f_\mathrm{abs}$. This is mostly due to the straight-line nature of the piecewise SED over the MIR features, and over the   $25 \la \lambda/\mu\mathrm{m} \la 60$ range, where no data is available. An example of this behaviour is shown in Fig.~\ref{fig:ltgs_sedfits} for NGC~3254. The effect can bias the estimate of $f_\mathrm{abs}$ high, especially when the FIR peak is low.

\subsection{CIGALE best fits}
\label{app:bestfits}
In a second test, we used CIGALE best-fit luminosities to derive $f_\mathrm{abs}$. The comparison of the bayesian and best-fit estimates is shown in the central panel of Fig.~\ref{fig:corr_panel}. In general, there's a good agreement between the two estimates, with  $f_\mathrm{abs} (\mathrm{bayes}) = 1.0 \times f_\mathrm{abs} (\mathrm{best}) -0.3 \%$. For $f_\mathrm{abs} <10\%$, the bayesian estimate is slightly lower than the best-fit one, but still consistent with the average bayesian error estimate on the same sample, 1\%. In a few instances, though, the discrepancy is larger: they correspond, however, to objects where most of the (dust emission) flux densities are consistent with zero within a few $\sigma$, and the fit is poorly constrained. In particular for these cases, the bayesian-like approach provides a more conservative and robust estimate, less prone to the degeneracies in the SED fit. 

 \subsection{MAGPHYS}
 \label{app:magphys}
 
For late-type galaxies only, we also modelled the SED using the MAGPHYS\footnote{publicly available at: {\tt www.iap.fr/magphys}.} code of \citet{DaCunhaMNRAS2008}. MAGPHYS uses libraries of physically justified optical and IR models. The unattenuated stellar SED is computed by assuming a \citet{ChabrierPASP2003} initial mass function and using the 2007 version of the \citet{BruzualMNRAS2003} stellar population synthesis models \citep[see][]{BruzualProc2007}. The star formation history (SFH) is given by a continuous  exponentially declining SFH, with random bursts of star formation superimposed. The attenuation by dust is described by the prescription of \citet{CharlotApJ2000} and includes attenuation in birth clouds (i.e. molecular clouds where stars form) and in the ambient (i.e. diffuse) interstellar medium (ISM). The dust emission is made up of four different components:  polycyclic aromatic hydrocarbons (PAHs), small stochastically heated grains that produce mid-infrared continuum emission and warm and cold dust in thermal equilibrium with the radiation field. The warm and cold components are described using modified blackbody spectra 
and absorption cross section with spectral indices $\beta = 1.5$ and 2.0 respectively, normalised to $\kappa_{850 \mu\mathrm{m}} = 0.077 \ \rm{m^2 kg^{-1}}$ \citep{DunneMNRAS2000a,JamesMNRAS2002}. 
 
Within MAGPHYS, 50000 optical models and 50000 IR models are combined together maintaining energy balance, i.e. the energy absorbed between the attenuated and unattenuated SEDs in the optical model, is equal to the energy emitted in the IR model. The large number of templates obtained in this way is compared to the multi-wavelength photometry and a goodness-of-fit $\chi^2$ is calculated.  By running over each template, probability density functions (PDF) are made for the parameters used to build each template by weighting the value of that parameter by the probability $e^{-\chi^2}$ corresponding to that template. The median values of these PDF produce reliable estimates for each parameter \citep{HaywardMNRAS2015} and the corresponding uncertainties are the 16th and 84th percentiles of the PDF.  As in \citet{ViaeneA&A2016}, we use the PDF-derived values for $L_\mathrm{dust}$ (directly provided by code, after integrating the modelled SED over the code's grid), while we estimated $L_\mathrm{bolo}$ by integrating the best-fit SED in $0.0912 < \lambda/\mu\mathrm{m} <1000$. The uncertainty in $f_\mathrm{abs}$ is assumed to be entirely due to the uncertainty in $L_\mathrm{dust}$, as verified in CIGALE fits also.

We made adaptations to MAGPHYS in order to be able to consistently compare with CIGALE. First, we allow the use of the (few) negative fluxes, yet consistent with zero at the 2$\sigma$ level, present in the database. These negative fluxes still provide constrains on the data and can easily be incorporated in the $\chi^2$ calculation. Second, we extended the MAGPHYS infrared library to the temperature range 10 - 30 K, which encompasses better the properties of galaxies in our sample \citep[see also][]{ViaeneA&A2016,DeVisMNRAS2017a}. Third, CIGALE adds an extra source of error (10\% of the flux) to each photometric point to take into account uncertainties in the models (Boquien, \textit{private communication}). We add the same uncertainty to our photometry before fitting the SEDs with MAGPHYS. 

The right panel of Fig.~\ref{fig:corr_panel} shows the comparison between the CIGALE and MAGPHYS estimates (in this panel, we also plot on the x-axis the uncertainties derived from MAGPHYS). Again, there is a good correspondence between the estimates, with $f_\mathrm{abs} (\mathrm{CIGALE}) = 0.98 \times f_\mathrm{abs} (\mathrm{MAGPHYS}) -0.9 \%$. The average difference and scatter between CIGALE and MAGPHYS is $\Delta f_\mathrm{abs} = -2.2\pm 4.3$ \% for the whole late-type sample. The scatter is compatible with the average errors on $f_\mathrm{abs}$, which is 3.3\% for both methods, leading to an average error on the difference of 4.7\%. Given the diversity of the physical recipes and of the methods used by the codes for fitting the SED, and for deriving PDF-based estimates and uncertainties, it is difficult to understand the reason for the slightly higher MAGPHYS values for $f_\mathrm{abs}$. We note, however, that  MAGPHYS has a tendency to produce higher $f_\mathrm{abs}$ values when the SED coverage in the MIR region between 24 and 100 $\mu$m is insufficient: a large, unconstrained, component from the warm dust template might be present which instead is not seen in the physically-based emission templated of CIGALE.  An example of this unconstrained bump is shown by the MAGPHYS fit of UGC~5692 in Fig.~\ref{fig:ltgs_sedfits}.

\subsection{Differences in the results}
Despite the differences in the various method we experimented, they affect little the main conclusions of this work. For examples, for LTGs $\langle f_\mathrm{abs}\rangle = 24.9\pm 0.7\%$ using the reference model,  $25.6\pm0.7\%$ for CIGALE best-fit, and $27.2\pm 0.7$ for MAGPHYS; for galaxies later than Sb, $f_\mathrm{abs}$ vs $L_\mathrm{bolo}$ has a Kendall's correlation measure $\tau_K = 0.54$ for  CIGALE's bayesian estimate, 0.52 for CIGALE best-fit, and $0.5$ for MAGPHYS. Finally, there is little difference (at least in the determination of $f_\mathrm{abs}$) if the different dust properties of the {\tt dl14} dust emission module (see Sect.~\ref{sec:cigale}) are used: for LTGs, it is $\langle f_\mathrm{abs}\rangle = 25.1\pm 0.7\%$; for types later than Sb, $f_\mathrm{abs}$ vs $L_\mathrm{bolo}$ has $\tau_K = 0.54$.

\section{$f_\mathrm{abs}$ from models of edge-on galaxies}
\label{app:rt}

We show here that typical RT models of edge-on galaxies with prominent dust-extinction lanes imply a sizeable dependence of $f_\mathrm{abs}$ on inclination.  \citet{DeGeyterMNRAS2014} fitted the optical/NIR images of a sample of 12 edge-on galaxies with diffuse, homogeneous, geometrical components, using an automated routine, FitSKIRT \citep{DeGeyterA&A2013} based on the code SKIRT \citep{BaesAstrComp2015}. In order to study the inclination effects on our results, we simulated a galaxy with the average properties of the sample fitted by \citet[see their Table~4]{DeGeyterMNRAS2014}, composed of an exponential disk and a bulge for stars, and an exponential disk for dust. We used the code SKIRT, its built-in template for an Sc galaxy spectrum \citep[from][]{FiocA&A1997} and the optical properties of the THEMIS dust model. 

\begin{figure}
\center
\includegraphics[width=\hsize]{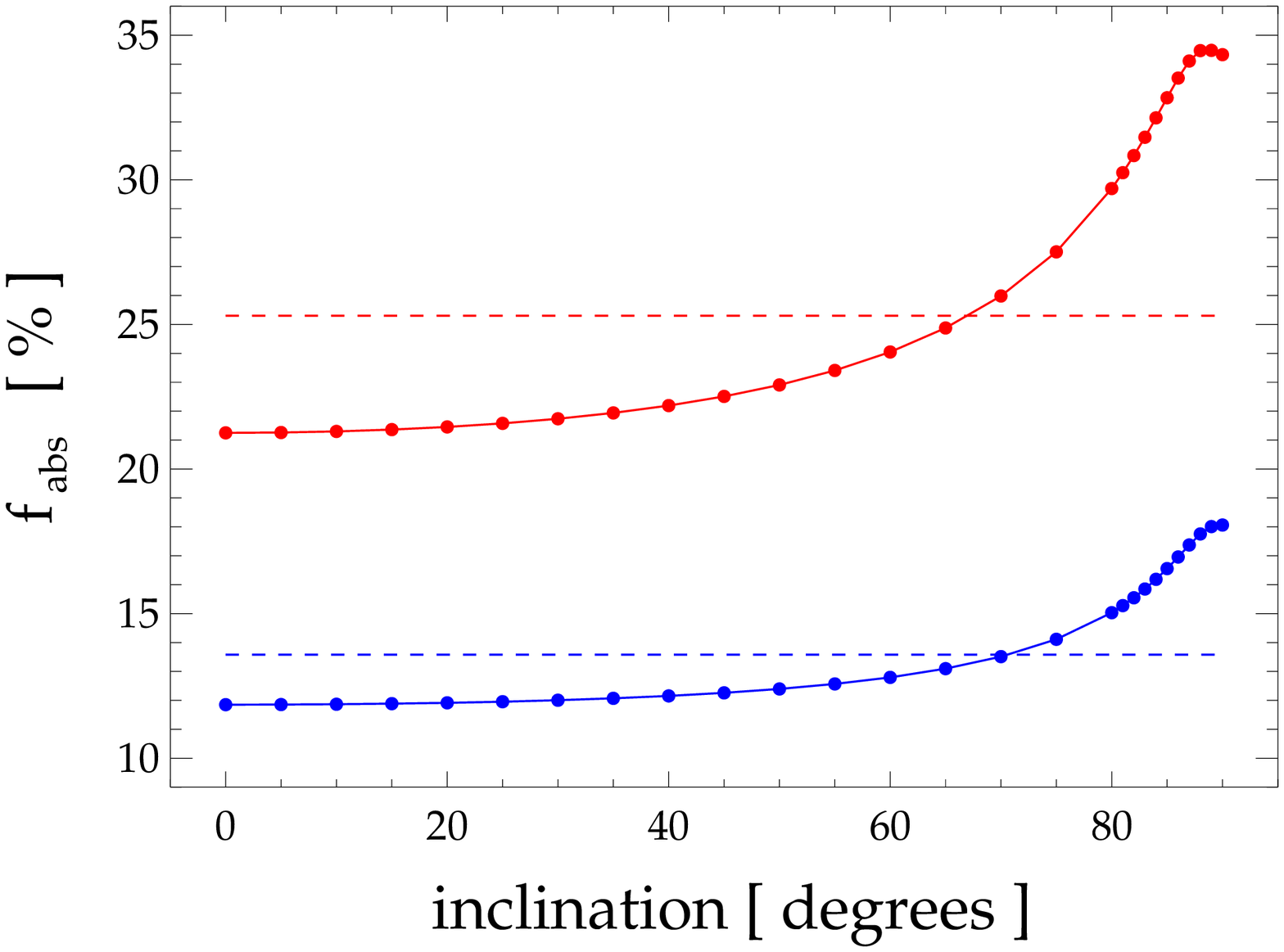}
\caption{$f_\mathrm{abs}$ vs inclination estimated from the SEDs of the {\em average} edge-on galaxy model of  \citet[][blue lines and symbols]{DeGeyterMNRAS2014}, and from the model with $3\times$ the dust mass (red lines and symbols). In both cases, dashed lines refers to the {\em true} $f_\mathrm{abs}$ derived from the RT energy budget.
}
\label{fig:fabs_RT}
\end{figure}

The models were produced for several galactic inclinations, and $f_\mathrm{abs}$ estimated from the mock SEDs in a way analogous to that adopted for real observations, integrating the output SEDs over  the range $0.0912 < \lambda/\mu\mathrm{m} <1000$ and removing the contribution of unattenuated starlight in the NIR  when deriving $L_\mathrm{dust}$ (see Sect.~\ref{app:compa}). The results are shown in Fig.~\ref{fig:fabs_RT} (blue symbols) and compared to the {\em true} $f_\mathrm{abs}$, i.e.\  the fraction of stellar radiation that is absorbed within the model globally,  directly derived from the RT  modelling. As expected, stellar emission is not isotropic: not only it is reduced at higher inclinations, because of dust extinction; it is also enhanced at face-on inclinations, because UV-optical radiation originally travelling along the galactic plane is scattered by dust in directions close to the galaxy axis. The dust emission, instead, is almost isotropic. As a consequence, $f_\mathrm{abs} \approx 12\%$ for the face-on case, smaller that {\em true} value obtained from the RT, 14\%. It increases to 18\% for the edge-on case. It is worth noting that the estimated $L_\mathrm{bolo}$ decreases by $\approx$ 32\% between the two inclination extremes.

It is well known that the average fit to edge-on galaxies is not able to reproduce the amount of radiation observed in the FIR: a factor $\sim 3$ higher opacity (or, for the same dust properties, a $\sim 3\times$ increase in the dust mass) is required by the energy conservation \citep[see, e.g. ][and the other references in their Introduction]{MosenkovA&A2018}. If we increase the dust mass of the average model by a factor 3 (red symbols in Fig.~\ref{fig:fabs_RT}), the {\em true} $f_\mathrm{abs}$ rises to 25\%, while the estimated  $f_\mathrm{abs}$  ranges from 21 to 34\% going from the face-on to the edge-on case;  $L_\mathrm{bolo}$ decreases by 38\%. In this second model also dust emission shows a  (moderate) anisotropy, due to the self-absorption of radiation emitted at shorter wavelength by non-thermal, stochastic, processes. 

Thus, RT models commonly fitted to edge-on galaxies suggest differences between the edge-on and face-on $f_\mathrm{abs}$ of $\approx$ 5-15\%.

\end{appendix}

\end{document}